\documentclass{article}

\usepackage{amssymb,amsfonts,amsmath}
\usepackage{cite,enumerate,float}
\usepackage{color}
\usepackage{ytableau}

\usepackage{mathrsfs}

\usepackage{tikz}
\usetikzlibrary{arrows}
\usetikzlibrary{arrows.meta}
\usetikzlibrary{positioning}
\usetikzlibrary{shapes,snakes}
\usetikzlibrary{fit}
\usetikzlibrary{decorations.pathmorphing,decorations.pathreplacing,decorations.markings}
\usetikzlibrary{calc}

\usepackage{xcolor}
\definecolor{burgundy}{rgb}{0.5, 0.0, 0.13}
\definecolor{olive}{rgb}{0.50, 0.50, 0.0}
\usepackage[linktocpage=true,colorlinks=true,linkcolor=burgundy,citecolor=black!20!blue,urlcolor=black!40!violet]{hyperref}

\usepackage[most]{tcolorbox}

\usepackage{pdflscape}
\usepackage{array, longtable}
\newcolumntype{C}{>{$}c<{$}}

\def\be{\begin{eqnarray}}
\def\ee{\end{eqnarray}}

\def\p{\partial}

\def\Tr{{\rm Tr}\,}

\definecolor{red}{rgb}{1,0,0}
\definecolor{orange}{rgb}{1,0.5,0}
\definecolor{violet}{rgb}{0.7,0,1}

\def\CF {{\cal F}}
\def\CG {{\cal G}}

\def\CI {{\cal I}}
\def\CJ {{\cal J}}

\def\CN {{\cal N}}

\def\CR {{\cal R}}

\def\CG {{\cal G}}

\def\CI {{{\cal I}}}

\def\CS {{\cal S}}

\def\IC{\mathbb{C}}

\def\IZ{{\mathbb{Z}}}

\def\fg{\mathfrak{g}}

\def\fl{\mathfrak{l}}
\def\fm{\mathfrak{m}}

\def\fS{\mathfrak{S}}

\def\lm{\limits}

\DeclareSymbolFont{bbsymbol}{U}{bbold}{m}{n}
\DeclareMathSymbol{\bbzero}{\mathbin}{bbsymbol}{"30}
\DeclareMathSymbol{\bbone}{\mathbin}{bbsymbol}{"31}
\DeclareMathSymbol{\bbtwo}{\mathbin}{bbsymbol}{"32}
\DeclareMathSymbol{\bbthree}{\mathbin}{bbsymbol}{"33}
\DeclareMathSymbol{\bbfour}{\mathbin}{bbsymbol}{"34}
\DeclareMathSymbol{\bbfive}{\mathbin}{bbsymbol}{"35}
\DeclareMathSymbol{\bbsix}{\mathbin}{bbsymbol}{"36}
\DeclareMathSymbol{\bbseven}{\mathbin}{bbsymbol}{"37}
\DeclareMathSymbol{\bbeight}{\mathbin}{bbsymbol}{"38}
\DeclareMathSymbol{\bbnine}{\mathbin}{bbsymbol}{"39}

\def\myY{\mathsf{Y}}
\newcommand\sqbox[1]{{
	\setbox0=\hbox{\mbox{$\Box$}}
	\setbox1=\hbox{\mbox{\raisebox{0.35ex}{\tiny #1}}}
	\mbox{\raisebox{-0.2ex}{\rlap{\hbox to \wd0{\hss{\box1}\hss}}\box0}}
}}

\def\stile{\begin{tikzpicture}[scale=0.15]
	\draw (0,0) -- (1,0) -- (1,-1) -- (0,-1) -- cycle;
\end{tikzpicture}}
\def\shtile{\begin{tikzpicture}[scale=0.15]
	\draw (0,0) -- (1,0) -- (0,-1) -- cycle;
\end{tikzpicture}}

\def\ntile{\begin{tikzpicture}[scale=0.2]
	\draw (0,0) -- (1,0) -- (1,-1) -- (0,-1) -- cycle;
\end{tikzpicture}}
\def\nhtile{\begin{tikzpicture}[scale=0.2]
	\draw (0,0) -- (1,0) -- (0,-1) -- cycle;
\end{tikzpicture}}

\tcbset{boxrule=0.6pt,colback=white!97!blue}

\hoffset= - 1.0in         

\begin{document}

\hfill MIPT/TH-14/23

\hfill ITEP/TH-17/23

\hfill IITP/TH-13/23

\vskip 1.5in
\begin{center}
	
	{\bf\Large{Super-Schur Polynomials for Affine Super Yangian $\mathsf{Y}(\widehat{\mathfrak{gl}}_{1|1})$}}
	\vskip 0.2in
	\renewcommand{\thefootnote}{\fnsymbol{footnote}}
	{Dmitry Galakhov$^{2,3,4,}$\footnote[2]{e-mail: galakhov@itep.ru},  Alexei Morozov$^{1,2,3,4,}$\footnote[3]{e-mail: morozov@itep.ru} and Nikita Tselousov$^{1,2,4,}$\footnote[4]{e-mail: tselousov.ns@phystech.edu}} 
	\vskip 0.2in 
	\renewcommand{\thefootnote}{\roman{footnote}}
	{\small{ 
			\textit{$^1$MIPT, 141701, Dolgoprudny, Russia}
			\vskip 0 cm 
			\textit{$^2$NRC “Kurchatov Institute”, 123182, Moscow, Russia}
			\vskip 0 cm 
			\textit{$^3$IITP RAS, 127051, Moscow, Russia}
			\vskip 0 cm 
			\textit{$^4$ITEP, Moscow, Russia}
	}}
\end{center}

\vskip 0.2in
\baselineskip 16pt

\centerline{ABSTRACT}

\bigskip

{\footnotesize
	We explicitly construct cut-and-join operators and their eigenfunctions --
	the Super-Schur functions --
	for the case of the affine super-Yangian $\mathsf{Y}(\widehat{\fg\fl}_{1|1})$.
	This is the simplest non-trivial (semi-Fock) representation, where eigenfunctions
	are labeled by the superanalogue of 2d Young diagrams,
	and depend on the supertime variables $(p_k,\theta_k)$.
	The action of other generators on diagrams is described by the analogue
	of the Pieri rule.
	As well we present generalizations of the hook formula for the measure on super-Young diagrams and of the Cauchy formula.
	Also a discussion of string theory origins for these relations is provided.
	
}

\bigskip

\bigskip

\tableofcontents

\section{Introduction}

Yangian and DIM symmetries,
which are the far-going generalizations of  the Lie algebraic ones,
are currently in the center of study in theoretical physics.
They appear in a variety of problems, from celestial amplitudes \cite{Dolan:2004ps, arkani-hamed_bourjaily_cachazo_goncharov_postnikov_trnka_2016}
to Seiberg-Witten-Nekrasov theory (low-energy sectors of brane models) \cite{Rapcak:2018nsl, Li:2020rij}.
And they have rich and interesting representation theory,
including Fock, MacMahon and triangular-times sectors.
In another direction, raised to Yangian and DIM can be arbitrary
simple Lie (super)algebra, and one can wonder on how representation
theory depends on this choice.

In the present paper we address one of initial questions on this route:
the semi-Fock representation of the simplest super-Yangian $\myY(\widehat{\fg\fl}_{1|1})$.
Our task is to describe the corresponding analogue of Young diagrams,
associated time-variables, commuting cut-and-join operators,
their common eigenfunctions (super-Schur/Jack functions)
and the action of other generators on them.
As usual, generators act by adding or subtracting boxes to the diagram
and/or by the differential operators in time variables.

We do not go into details and motivations of the definitions,
just cite a few selected papers, emphasising different aspects of the story: \cite{wang2020affine, cui2022jack, MT, Qschurs, Qschurs1, Qschurs2, MMZh, DIM1, DIM2}.
Instead we refer the reader to sec.2 of \cite{Morozov:2022ndt} and sec.3 of \cite{MT} for reminder of the properties of the ordinary Schur/Jack functions,
which are the eigenfunctions of the ordinary cut-and-join operators \cite{MMN},
associated with the ordinary Young diagrams and with the ordinary Yangian $\myY(\widehat{\fg\fl}_{1})$.\footnote{Yet we preferred a name ``super-Schur'' rather than ``super-Jack'' since super-Jack polynomials were introduced in \cite{sergeev2005generalised} in a different context, and we hope for much farther reaching generalization opportunities than discussed in this note.}
It is this construction that we are going to generalize to $\myY(\widehat{\fg\fl}_{1|1})$.
Generalizations to other representations, other Yangians and DIM algebras (their quantization)
are the three obvious directions for further developments --
which are beyond the scope of the present paper.

This note is organized as follows.
In sec.\ref{sec:AsY} we review a construction of affine super Yangian $\myY(\widehat{\fg\fl}_{1|1})$ and its semi-Fock representation, introduce super-partitions.
Sec.\ref{sec:boson} is the heart of this note. 
We present a construction of super-commuting super-time variables and derive in these terms a set of differential cut-and-join operators \eqref{super-cut-and-join}.
We define super-Schur polynomials as a set of cut-and-join eigenfunctions.
In sec.\ref{sec:conclude} we put remarks on this derivation as well as possible project that might be interesting to develop in the future.
In this note we were trying to concentrate on simple algebraic properties of novel cut-and-join operators and super-Schur functions, therefore we accumulate hints and reasoning from the string theory on this subject in app.\ref{app:BPS}.
Finally, in the very end of this note, in app.\ref{app:data}, we provide data (super-Schur functions, hook measures, cut-and-join eigenvalues) on all super-partitions up to level 9/2.

\section{Affine super-Yangian}\label{sec:AsY}

\subsection{Algebra: affine Yangian \texorpdfstring{$\myY(\widehat{\fg\fl}_{1|1})$}{Ygl11}}

Algebra $\myY(\widehat{\fg\fl}_{1|1})$ is an (shifted) affine Yangian algebra based on the affine Dynkin diagram:
\begin{equation}\label{gl_1_1_Dynkin}
	\widehat{\fg\fl}_{1|1}:\quad \begin{array}{c}
		\begin{tikzpicture}
			\draw (0,0) to[out=30,in=150] (2,0) (0,0) to[out=-30,in=-150] (2,0);
			\draw[fill=gray] (0,0) circle (0.1);
			\draw[fill=white!40!blue] (2,0) circle (0.1);
			\node[above] at (0,0.1) {$+$};
			\node[above] at (2,0.1) {$-$};
		\end{tikzpicture}
	\end{array}
\end{equation}
We would like to mark two nodes of this diagram by different colors, and we will distinguish those nodes as a $(+)$ node and a $(-)$ node.
$\myY(\widehat{\fg\fl}_{1|1})$ is generated by two families of Chevalley generators $e^{\pm}_{n}$, $f^{\pm}_{n}$, $\psi^{\pm}_{k}$, where $n\in\IZ_{\geq 0}$, $k\in\IZ$, for two nodes.
This is a superalgebra, we assign odd (fermionic) parity to raising $e$- and lowering $f$-generators, and even (bosonic) parity to Cartan $\psi$-generators.

In addition algebra $\myY(\widehat{\fg\fl}_{1|1})$ depends on two parameters we denote as $h_{1,2}$.
The (super-)commutation relations between generators read:
\begin{equation}\label{Yang_modes}
	\begin{split}
		&\left\{e^{a}_n,e^{a}_k\right\}=\left\{f^{a}_n,f^{a}_k\right\}=\left[\psi^{a}_n,e^{a}_k\right]=\left[\psi^{a}_n,f^{a}_k\right]=0,\quad a=\pm\,,\\
		&\left\{e_{n+2}^{+},e_{k}^{-}\right\}-2\left\{e_{n+1}^{+},e_{k+1}^{-}\right\}+\left\{e_{n}^{+},e_{k+2}^{-}\right\}-\frac{h_1^2+h_2^2}{2}\left\{e_{n}^{+},e_{k}^{-}\right\}+\frac{h_1^2-h_2^2}{2}\left[e_{n}^{+},e_{k}^{-}\right]=0\,,\\
		&\left\{f_{n+2}^{+},f_{k}^{-}\right\}-2\left\{f_{n+1}^{+},f_{k+1}^{-}\right\}+\left\{f_{n}^{+},f_{k+2}^{-}\right\}-\frac{h_1^2+h_2^2}{2}\left\{f_{n}^{+},f_{k}^{-}\right\}-\frac{h_1^2-h_2^2}{2}\left[f_{n}^{+},f_{k}^{-}\right]=0\,,\\
		&\left[\psi_{n+2}^{+},e_{k}^{-}\right]-2\left[\psi_{n+1}^{+},e_{k+1}^{-}\right]+\left[\psi_{n}^{+},e_{k+2}^{-}\right]-\frac{h_1^2+h_2^2}{2}\left[\psi_{n}^{+},e_{k}^{-}\right]+\frac{h_1^2-h_2^2}{2}\left\{\psi_{n}^{+},e_{k}^{-}\right\}=0\,,\\
		&\left[\psi_{n+2}^{+},f_{k}^{-}\right]-2\left[\psi_{n+1}^{+},f_{k+1}^{-}\right]+\left[\psi_{n}^{+},f_{k+2}^{-}\right]-\frac{h_1^2+h_2^2}{2}\left[\psi_{n}^{+},f_{k}^{-}\right]-\frac{h_1^2-h_2^2}{2}\left\{\psi_{n}^{+},f_{k}^{-}\right\}=0\,,\\
		&\left[\psi_{n+2}^{-},e_{k}^{+}\right]-2\left[\psi_{n+1}^{-},e_{k+1}^{+}\right]+\left[\psi_{n}^{-},e_{k+2}^{+}\right]-\frac{h_1^2+h_2^2}{2}\left[\psi_{n}^{-},e_{k}^{+}\right]-\frac{h_1^2-h_2^2}{2}\left\{\psi_{n}^{-},e_{k}^{+}\right\}=0\,,\\
		&\left[\psi_{n+2}^{-},f_{k}^{+}\right]-2\left[\psi_{n+1}^{-},f_{k+1}^{+}\right]+\left[\psi_{n}^{-},f_{k+2}^{+}\right]-\frac{h_1^2+h_2^2}{2}\left[\psi_{n}^{-},f_{k}^{+}\right]+\frac{h_1^2-h_2^2}{2}\left\{\psi_{n}^{-},f_{k}^{+}\right\}=0\,,\\
		&\left\{e^{a}_n,f^{b}_k\right\}=-\delta_{ab}\psi^{a}_{n+k},\quad a,b=\pm\,.
	\end{split}
\end{equation}

These relations are accompanied by quartic Serre relations:
\begin{equation}
	\begin{split}
		&\mathop{\rm Sym}\lm_{i,j}\mathop{\rm Sym}\lm_{k,l}\left\{e_i^+,\left[e_k^-,\left\{e_j^+,e_m^-\right\}\right]\right\}=0\,,\\
		&\mathop{\rm Sym}\lm_{i,j}\mathop{\rm Sym}\lm_{k,l}\left\{f_i^+,\left[f_k^-,\left\{f_j^+,f_m^-\right\}\right]\right\}=0\,.
	\end{split}
\end{equation}

In some contexts it turns out to be useful to introduce generating functions for the generators:
\begin{equation}
	e^{\pm}(z)=\sum\lm_{k=0}^{\infty}\frac{e_k^{\pm}}{z^{k+1}},\quad 	f^{\pm}(z)=\sum\lm_{k=0}^{\infty}\frac{f_k^{\pm}}{z^{k+1}},\quad 	\psi^{\pm}(z)=\sum\lm_{k=-\infty}^{\infty}\frac{\psi_k^{\pm}}{z^{k+1}}\,.
\end{equation}

In terms of these generating functions we could rewrite relations \eqref{Yang_modes} in the following form:
\begin{equation}\label{Yangian}
	\begin{split}
		&\left\{e^{a}(z),e^{a}(w)\right\}=\left\{f^{a}(z),f^{a}(w)\right\}=\left[\psi^{a}(z),e^{a}(w)\right]=\left[\psi^{a}(z),f^{a}(w)\right]=0,\quad a=\pm,\\
		&\left((z-w)^2-h_2^2\right)e^{+}(z)e^{-}(w)\simeq -\left((z-w)^2-h_1^2\right)e^{-}(w)e^{+}(z)\,,\\
		&\left((z-w)^2-h_1^2\right)f^{+}(z)f^{-}(w)\simeq -\left((z-w)^2-h_2^2\right)f^{-}(w)f^{+}(z)\,,\\
		&\left((z-w)^2-h_2^2\right)\psi^{+}(z)e^{-}(w)\simeq \left((z-w)^2-h_1^2\right)e^{-}(w)\psi^{+}(z)\,,\\
		&\left((z-w)^2-h_1^2\right)\psi^{-}(z)e^{+}(w)\simeq \left((z-w)^2-h_2^2\right)e^{+}(w)\psi^{-}(z)\,,\\
		&\left((z-w)^2-h_1^2\right)\psi^{+}(z)f^{-}(w)\simeq \left((z-w)^2-h_2^2\right)f^{-}(w)\psi^{+}(z)\,,\\
		&\left((z-w)^2-h_2^2\right)\psi^{-}(z)f^{+}(w)\simeq \left((z-w)^2-h_1^2\right)f^{+}(w)\psi^{-}(z)\,,\\
		&\left\{e^{a}(z),f^{b}(w)\right\}\simeq -\delta_{ab}\frac{\psi^{a}(z)-\psi^{a}(w)}{z-w},\quad a,b=\pm\,,
	\end{split}
\end{equation}
where sign $\simeq$ implies that we equate Laurent polynomials in $z^kw^m$ on both sides up to monomials $z^{k\geq 0}w^{m}$ and $z^{k}w^{m\geq 0}$.

Series of generators $\psi_k^{\pm}$ are bounded below, so that:
\begin{equation}
	\psi_{k<-M_+}^+=0,\quad 	\psi_{k<-M_-}^-=0\,,
\end{equation}
where $M_{\pm}$ account for shifts (see \cite{Galakhov:2021xum}).

\subsection{Semi-Fock and Fock representations}\label{sec:sF}

In this note we concentrate on a representation of $\myY(\widehat{\fg\fl}_{1|1})$ that we call \emph{semi-Fock} representation.
We will comment on this name in the end of this subsection.
It is an infinite representation where vectors in a module are labeled by molten 2d crystals, or super-Young diagrams.
We will define this representation and its properties in this subsection.

Let us consider a subset of an infinite bipartite graph forming a cone filling the bottom right quadrant of a 2d plane:
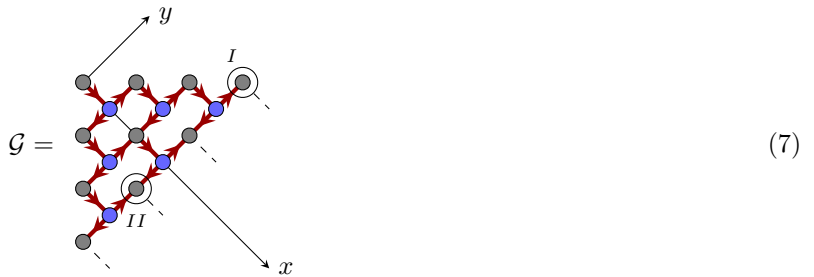
\begin{equation}\label{graph_G}
	\CG = \begin{array}{c}
		\begin{tikzpicture}[scale=0.5,rotate=-45]
			\draw[-stealth] (0,0) -- (7,0);
			\draw[-stealth] (0,0) -- (0,2.5);
			\node[right] at (7,0) {$x$};
			\node[right] at (0,2.5) {$y$};
			\tikzset{sty1/.style={fill=gray}}
			\tikzset{sty2/.style={fill=white!40!blue}}
			\tikzset{sty3/.style={ultra thick, black!40!red, postaction={decorate},
					decoration={markings, mark= at position 0.65 with {\arrow{stealth}}}}}
			\foreach \i in {0, 1, 2}
			{
				\foreach \j in {0, ..., \i}
				\draw[sty3] (\i, -\i + 2*\j) -- (\i + 1, -\i + 2*\j);
			}
			\foreach \i in {0, 1, 2}
			{
				\foreach \j in {0, ..., \i}
				{
					\draw[sty3] (\i+1, -\i + 2*\j) -- (\i+1, -\i + 2*\j + 1);
					\draw[sty3] (\i+1, -\i + 2*\j) -- (\i+1, -\i + 2*\j - 1);
				}
			}
			\foreach \i in{0, 1, 2, 3}
			{
				\draw[dashed] (3,-3 + 2 * \i) -- (4,-3+2*\i);
			}
			\foreach \i in {0, 1, 2, 3}
			{
				\foreach \j in {0, ..., \i}
				\draw[sty1] (\i, -\i + 2*\j) circle (0.2);
			}
			\foreach \i in {0, 1, 2}
			{
				\foreach \j in {0, ..., \i}
				\draw[sty2] (\i+1, -\i + 2*\j) circle (0.2);
			}
			\draw (3,3) circle (0.4);
			\node[above] at (2.6,3) {$\scriptstyle I$};
			\draw (3,-1) circle (0.4);
			\node[below] at (3.3,-1.3) {$\scriptstyle II$};
		\end{tikzpicture}
	\end{array}
\end{equation}
The nodes, we could call ``atoms'' by an analogy from chemistry, of lattice $\CG$ are of two types coinciding with the node types of $\widehat{\fg\fl}_{1|1}$ Dynkin diagram \eqref{gl_1_1_Dynkin}.
In depicting \eqref{graph_G} we have rotated the coordinate frame by $45^{\circ}$ in comparison to fig.\ref{fig:D-brane_Fock}, so that the crystal diagram is more similar to a Young diagram we discuss in the next subsection.
In \eqref{graph_G} we denoted nodes by the corresponding color code.
We define molten crystals in the following way.
Call a subset of nodes $\lambda\subset\CG$ a molten crystal if for any node $a\in\lambda$ any other node $b\in\CG$ is connected to $a$ by an arrow $b\to a$ then also $b\in\lambda$.

We place nodes in $\CG$ on an integral 2d lattice, so that nodes have integral coordinates $(x,y)$: the node in the tip of the cone has coordinates $(0,0)$, its nearest neighbor has coordinates $(1,0)$, the next nodes have coordinates $(1,\pm 1)$ and so on.
We weight $x$- and $y$-coordinates of a node with complex weights $h_1$ and $h_2$ parameterizing $\myY(\widehat{\fg\fl}_{1|1})$, so that a weight of node $a$ reads:
\begin{equation}
	\omega_a:=h_1 x_a+h_2 y_a\,.
\end{equation}

We would like to introduce some useful functions on crystals $\lambda$.
Let us denote by $\lambda^{\pm}$ a set of nodes of type $(+)$ or $(-)$ respectively in crystal $\lambda$.
We denote the numbers of nodes of respective types in $\lambda$ as:
\begin{equation}\label{scale_op}
	n_{\lambda}^{+}:=\left|\lambda^{+}\right|,\quad n_{\lambda}^{-}:=\left|\lambda^{-}\right|\,.
\end{equation}

Also we introduce net weights of $(+)$- and $(-)$-nodes in a crystal:
\begin{equation}\label{c-a-j_op}
	w_{\lambda}^+:=\sum\lm_{a\in\lambda^{+}}\omega_a,\quad 	w_{\lambda}^-:=\sum\lm_{a\in\lambda^{-}}\omega_a\,.
\end{equation}

In these terms we could write a generating function for the numbers of molten crystals in a compact form:
\begin{equation}\label{g_fun}
	\chi_{\frac{1}{2}}(q_1,q_2)=\sum\lm_{\lambda}q_1^{n^+_\lambda}q_2^{n^-_\lambda}=\prod\lm_{k=1}^{\infty}\frac{1+q_1^kq_2^{k-1}}{1-q_1^kq_2^{k}}\,.
\end{equation}

The semi-Fock representation is a crystal representation of $\myY(\widehat{\fg\fl}_{1|1})$ -- thus it falls into a class of representations of quiver BPS algebras \cite{Galakhov:2021xum}.
The vectors in the semi-Fock module are labeled by crystals $|\lambda\rangle$.
And we could use techniques of quiver BPS algebras to derive explicit matrix elements of  $\myY(\widehat{\fg\fl}_{1|1})$ generators $e_k^{\pm}$, $f_k^{\pm}$, $\psi_k^{\pm}$.
We derive explicit expressions in Appendix \ref{app:BPS}.

The semi-Fock representation is not unique.
The easiest way to observe this is to turn to the geometric picture in fig.\ref{fig:D-brane_Fock}.
We constructed the semi-Fock representation as a slice of atoms on the east face of the pyramid partition.
However we could have chosen different slices.
If a west side slice is chosen the resulting representation is equivalent to the semi-Fock module constructed in this subsection under switching signs of $h_{1,2}$. 
Let us denote these modules as $\frac{1}{2}\CF$.
Whereas if one chooses the north or south side with a simultaneous shift one step down along the pyramid the resulting representation $\widetilde{\frac{1}{2}\CF}$ is in involution with $\frac{1}{2}\CF$.
The involution permutes the nodes of the Dynkin diagram \eqref{gl_1_1_Dynkin} and the roles of $(+)-$ and $(-)$-generators respectively.
On $h_{1,2}$ it acts as $h_1\to -h_1$, $h_2\to h_2$, and the super-Young diagram is reflected with respect to the horizontal axis.

We could construct new crystal representations from old ones using a universal \emph{naive} tensor product defined in \cite{Galakhov:2022uyu}.
We define a Fock representation as a naive tensor product of two semi-Fock representations (therefore prefix ``semi''):
\begin{equation}
	\CF\,:=\,\mbox{$\frac{1}{2}\CF$}\,\otimes_{\rm naive}\,\mbox{$\widetilde{\frac{1}{2}\CF}$}\,.
\end{equation}
The character of the Fock representation is a product of characters:
\begin{equation}
	\chi_{\rm Fock}(q_1,q_2)=\chi_{\frac{1}{2}}(q_1,q_2)\chi_{\frac{1}{2}}(q_2,q_1)=\prod\lm_{k=1}^{\infty}\frac{\left(1+q_1^kq_2^{k-1}\right)\left(1+q_1^{k-1}q_2^{k}\right)}{\left(1-q_1^kq_2^{k}\right)^2}\,.
\end{equation}
This character coincides with \cite[eq.(3.15)]{Kolyaskin:2022tqi}, and we believe our Fock representation is isomorphic to a Fock representation constructed in \cite{Kolyaskin:2022tqi} in the context of a CFT $\beta\gamma$-system under embedding $\myY(\widehat{\fg\fl}_{1|1})$ in the corresponding conformal algebra.

In this note we will not use the Fock representation concentrating mostly on the semi-Fock one, however let us note that a problem of mapping our super-time variables onto CFT fields is intriguing.

\subsection{Super-Young diagrams for super-partitions}

We would like to note that molten crystals defined in the past subsection are in one-to-one correspondence with super-partitions and super-Young diagrams we define in this subsection.
We call by a super-partition $\lambda$ of a semi-integer number $u\in\IZ_{\geq 0}/2$ a sequence of semi-integer numbers:
\begin{equation}\label{parti}
	\lambda_1\geq \lambda_2\geq \lambda_3\geq\ldots\geq 0\,,
\end{equation}
such that $\sum\lm_{i}\lambda_i=u$ and if $\lambda_i$ is not integer then inequalities in sequence \eqref{parti} are strict: $\lambda_{i-1}>\lambda_i>\lambda_{i+1}$.

In a complete analogy with ordinary partitions we introduce for super-partitions super-Young diagrams.
The diagram is constructed as a filling of the bottom right plane quadrant with tiles, so that each next tile is supported on the top and left by a previous tile or a wall.
To the halves we assign triangular half-tiles.
The height of the tile column correspond to a number $\lambda_i$, for example:
\begin{equation}
	\left\{4,\frac{7}{2},2,2,\frac{3}{2},1\right\}=\begin{array}{c}
		\begin{tikzpicture}[scale=0.3]
			\foreach \i/\j in {0/-4, 0/-3, 0/-2, 0/-1, 0/0, 1/-3, 1/-2, 1/-1, 1/0, 2/-2, 2/-1, 2/0, 3/-2, 3/-1, 3/0, 4/-1, 4/0, 5/-1, 5/0}
			{
				\draw (\i,\j) -- (\i+1,\j);
			}
			\foreach \i/\j in {0/-3, 0/-2, 0/-1, 0/0, 1/-3, 1/-2, 1/-1, 1/0, 2/-2, 2/-1, 2/0, 3/-1, 3/0, 4/-1, 4/0, 5/0, 6/0}
			{
				\draw (\i,\j) -- (\i,\j-1);
			}
			\foreach \i/\j in {2/-3, 5/-1}
			{
				\draw (\i,\j) -- (\i-1,\j-1);
			}
		\end{tikzpicture}
	\end{array}\,.
\end{equation}

Let us select pairs of connected nodes $(+)\to(-)$ in crystal $\lambda$ (these pairs are highlighted by bubbles in fig.~\ref{fig:crystalYoung}).
Then we will find that some of pairs are complete, and in some pairs a $(-)$-node is missing.
We identify complete pairs with complete square tiles in the corresponding super-Young diagram and incomplete pairs with triangular half-tiles.
Then under such an identification a crystal diagram transforms into a super-Young diagram, see fig.~\ref{fig:crystalYoung}.\footnote{A similar diagram system for representations associated with crystal slices was introduced and developed in \cite{Noshita:2021dgj}. However here we do not split a complete square tile in two halves of different color.}
\begin{figure}[h!]
	\centering
	\begin{tikzpicture}
		\node(A) at (0,0) {$\begin{array}{c}
				\begin{tikzpicture}[scale=0.5, rotate=-45]
					\foreach \i/\j in {0/0, 1/1, 1/-1, 2/0, 2/2, 2/-2, 3/3, 3/1}
					{
						\begin{scope}[shift={(\i,\j)}]
							\draw[violet, thick] (0,0.4) to[out=180,in=90] (-0.4,0) to[out=270,in=180] (0,-0.4) -- (1,-0.4) to[out=0,in=270] (1.4,0) to[out=90,in=0] (1,0.4) -- cycle;
							\draw[violet, thick] (0.5,-0.4) -- (0.5,0.4);
						\end{scope}
					}
					\tikzset{sty1/.style={fill=gray}}
					\tikzset{sty2/.style={fill=white!40!blue}}
					\draw[thick] (0,0) -- (1,0) (1,-1) -- (1,1) (2,-2) -- (2,2) (1,1) -- (2,1) (1,-1) -- (2,-1) (2,2) -- (3,2) (2,2) -- (3,2) (2,0) -- (3,0) (2,-2) -- (3,-2) (3,0) -- (3,3);
					\draw[sty1] (0,0) circle (0.2) (1,1) circle (0.2) (1,-1) circle (0.2) (2,0) circle (0.2) (2,2) circle (0.2) (2,-2) circle (0.2) (3,3) circle (0.2) (3,1) circle (0.2);
					\draw[sty2] (1,0) circle (0.2) (2,1) circle (0.2) (2,-1) circle (0.2) (3,2) circle (0.2) (3,0) circle (0.2) (3,-2) circle (0.2);
				\end{tikzpicture}
			\end{array}$};
		\node(B) at (6,0) {$\begin{array}{c}
				\begin{tikzpicture}[scale=0.5]
					\draw[stealth-stealth] (-0.5,-3) -- (-0.5,0.5) -- (4.5,0.5);
					\node[below left] at (-0.5,-3) {$\scriptstyle \epsilon_1$};
					\node[right] at (4.5,0.5) {$\scriptstyle \epsilon_2$};
					\foreach \i/\j in {0/-3, 0/-2, 0/-1, 0/0, 1/-2, 1/-1, 1/0, 2/-1, 2/0, 3/0}
					{
						\draw[thick] (\i,\j) -- (\i+1,\j);
					}
					\foreach \i/\j in {0/-2, 0/-1, 0/0, 1/-2, 1/-1, 1/0, 2/-1, 2/0, 3/0}
					{
						\draw[thick] (\i,\j) -- (\i,\j-1);
					}
					\foreach \i/\j in {3/-1, 4/0}
					{
						\draw[thick] (\i,\j) -- (\i-1,\j-1);
					}
				\end{tikzpicture}
			\end{array}$};
		\node[right] at (B.east) {$=\left\{3,2,\frac{3}{2},\frac{1}{2}\right\}$};
		\path (A) edge[<->] (B);
	\end{tikzpicture}
	\caption{Crystal and diagrammatic labeling of $\myY(\widehat{\fg\fl}_{1|1})$ semi-Fock representation vectors.}
	\label{fig:crystalYoung}
\end{figure}

In terms of super-Young diagrams it is more convenient to introduce a different system of coordinates on the diagram plane.
We assume that a tile (or a half-tile) has coordinates $(i,j)$ where $i$ is the number of tile row and $j$ is the number of tile column, and the tile in the top left corner has coordinates $(0,0)$.

Each node of type $(+)$ corresponds to a tile or a half-tile, whereas a node of type $(-)$ corresponds to a complete tile.
Corresponding coordinate transformations are given by the following relations:
\begin{equation}
	(i,j)_+=\left(\frac{x_a-y_a}{2},\frac{x_a+y_a}{2}\right),\quad (i,j)_-=\left(\frac{x_a-y_a-1}{2},\frac{x_a+y_a-1}{2}\right)\,.
\end{equation}
Accordingly we introduce weights for the tiles in a diagram:
\begin{equation}
	\label{epsilons}
	\epsilon_1=h_1-h_2, \quad\epsilon_2 =h_1+h_2\,.
\end{equation}

In these terms we could redefine our functions \eqref{scale_op} and \eqref{c-a-j_op} in terms of diagrams:
\begin{equation}\label{eigen_values}
	\begin{split}
		n^+_\lambda=\sum\lm_{\stile,\shtile\in\lambda}1,\quad n^-_\lambda=\sum\lm_{\stile\in\lambda}1,\quad w^+_\lambda=\sum\lm_{\stile,\shtile\in\lambda}(\epsilon_1 i+\epsilon_2 j),\quad w^-_\lambda=\frac{\epsilon_1+\epsilon_2}{2}n_{\lambda}^-+\sum\lm_{\stile\in\lambda}(\epsilon_1 i+\epsilon_2 j)\,.
	\end{split}
\end{equation}

Finally, we would like to introduce a measure on super-Young diagrams.
Let us call a ``leg'' ${\bf leg}_{\lambda}(\ntile)$ of tile $\ntile$ in diagram $\lambda$ a string of all tiles (both complete and incomplete) below $\ntile$.
Similarly, by an ``arm'' ${\bf arm}_{\lambda}(\ntile)$ we call all the tiles located to the right (see fig.~\ref{fig:ArmLeg}).
We would like to distinguish \emph{even} and \emph{odd} arms/legs.
We say that a leg (an arm) is odd if the last tile in the string is a half-tile $\nhtile$, and even if the last tile is a complete tile $\ntile$.
For example, in fig.\ref{fig:ArmLeg} the leg for the highlighted red tile is odd, whereas the arm is even.
By $|\cdot|$ we denote a number of tiles in the leg (arm).
\begin{figure}[h!]
	\begin{center}
		\begin{tikzpicture}[scale=0.4]
			\foreach \i/\j in {0/-7, 0/-6, 0/-5, 0/-4, 0/-3, 0/-2, 0/-1, 0/0, 1/-7, 1/-6, 1/-5, 1/-4, 1/-3, 1/-2, 1/-1, 1/0, 2/-6, 2/-5, 2/-4, 2/-3, 2/-2, 2/-1, 2/0, 3/-5, 3/-4, 3/-3, 3/-2, 3/-1, 3/0, 4/-5, 4/-4, 4/-3, 4/-2, 4/-1, 4/0, 5/-4, 5/-3, 5/-2, 5/-1, 5/0, 6/-4, 6/-3, 6/-2, 6/-1, 6/0, 7/-4, 7/-3, 7/-2, 7/-1, 7/0, 8/-3, 8/-2, 8/-1, 8/0, 9/-3, 9/-2, 9/-1, 9/0, 10/-1, 10/0, 11/0}
			{
				\draw[white!40!gray] (\i,\j) -- (\i+1,\j);
			}
			\foreach \i/\j in {0/-6, 0/-5, 0/-4, 0/-3, 0/-2, 0/-1, 0/0, 1/-6, 1/-5, 1/-4, 1/-3, 1/-2, 1/-1, 1/0, 2/-6, 2/-5, 2/-4, 2/-3, 2/-2, 2/-1, 2/0, 3/-5, 3/-4, 3/-3, 3/-2, 3/-1, 3/0, 4/-4, 4/-3, 4/-2, 4/-1, 4/0, 5/-4, 5/-3, 5/-2, 5/-1, 5/0, 6/-3, 6/-2, 6/-1, 6/0, 7/-3, 7/-2, 7/-1, 7/0, 8/-3, 8/-2, 8/-1, 8/0, 9/-2, 9/-1, 9/0, 10/-2, 10/-1, 10/0, 11/0}
			{
				\draw[white!40!gray] (\i,\j) -- (\i,\j-1);
			}
			\foreach \i/\j in {3/-6, 6/-4, 12/0}
			{
				\draw[white!40!gray] (\i,\j) -- (\i-1,\j-1);
			}
			\draw[thick, fill=white!40!blue] (2,-3) -- (2,-7) -- (3,-6) -- (3,-3) -- cycle;
			\foreach \i in {-4, -5, -6}
			\draw[thick] (2,\i) -- (3,\i);
			\draw[thick, fill=orange] (3,-2) -- (3,-3) -- (10,-3) -- (10,-2) -- cycle;
			\foreach \i in {4, 5, 6, 7, 8, 9}
			\draw[thick] (\i,-2) -- (\i,-3);
			\draw[thick, fill=black!40!red] (2,-2) -- (2,-3) -- (3,-3) -- (3,-2) -- cycle;
			\begin{scope}[shift={(2,-3)}]
				\draw[thick, white!40!blue] (0,0) to[out=180,in=90] (-0.2,-0.2) -- (-0.2,-1.8) to[out=270,in=0] (-0.4,-2) to[out=0,in=90] (-0.2,-2.2) -- (-0.2,-3.8) to[out=270,in=180] (0,-4);
				\node[white!40!blue, rotate=90,anchor=center, fill=white] at (-1.1,-2) {Leg};
			\end{scope}
			\begin{scope}[shift={(3,-2)}]
				\draw[thick, black!10!orange] (0,0) to[out=90,in=180] (0.2,0.2) -- (3.3,0.2) to[out=0,in=270] (3.5,0.4) to[out=270,in=180] (3.7,0.2) -- (6.8,0.2) to[out=0, in=90] (7,0);
				\node[black!10!orange, fill=white] at (3.5,1.1) {Arm};
			\end{scope}
		\end{tikzpicture}
	\end{center}
	\caption{Hook, arm and leg in a super-Young diagram}\label{fig:ArmLeg}
\end{figure}

For tile $\ntile$ in Young diagram $\lambda$ we introduce a \emph{hook} potential $\upsilon_{\lambda}(\ntile)$ based on parity of ${\bf leg}_{\lambda}(\ntile)$ and ${\bf arm}_{\lambda}(\ntile)$ according to the following table:

\bigskip

\begin{equation}
	\begingroup
	\renewcommand*{\arraystretch}{2}
	\begin{array}{c|c|c}
		\mbox{Leg parity}& \mbox{Arm parity}& \upsilon_{\lambda}(\ntile) \\
		\hline
		\hline
		\mbox{even} & \mbox{even} &
		\Big(-\epsilon_1|{\bf leg}_{\lambda}(\ntile)|+\epsilon_2|{\bf arm}_{\lambda}(\ntile)|-\epsilon_1
		\Big)
		\cdot\Big (\epsilon_1|{\bf leg}_{\lambda}(\ntile)|-\epsilon_2|{\bf arm}_{\lambda}(\ntile)|-\epsilon_2\Big)
		\\
		\mbox{odd} & \mbox{even} &
		\Big(-\epsilon_1|{\bf leg}_{\lambda}(\ntile)|+\epsilon_2|{\bf arm}_{\lambda}(\ntile)|
		\Big)\cdot\Big(\epsilon_1|{\bf leg}_{\lambda}(\ntile)|-\epsilon_2|{\bf arm}_{\lambda}(\ntile)|-\epsilon_2\Big)
		\\
		\mbox{even} & \mbox{odd} &
		\Big(-\epsilon_1|{\bf leg}_{\lambda}(\ntile)|+\epsilon_2|{\bf arm}_{\lambda}(\ntile)|-\epsilon_1\Big)
		\cdot \Big(\epsilon_1|{\bf leg}_{\lambda}(\ntile)|-\epsilon_2|{\bf arm}_{\lambda}(\ntile)|\Big)  \\
		\mbox{odd} & \mbox{odd} &	1 \\
	\end{array}
	\endgroup
\end{equation}

\bigskip

We introduce a \emph{hook} measure (cf. \cite{nakajima1996jack, Nekrasov:2002qd, Nekrasov:2003rj, Losev:2003py}) on super-partitions according to the following:\footnote{We should note that the ordinary 2d Young diagrams are a subset of super-Young diagrams presented here. The hook norm reduces to a conventional expression \cite{nakajima1996jack} when there are no half-tiles in a diagram. Similarly expressions for the cut-and-join operators and super-Schur functions we will discuss in what follows become ordinary expressions for cut-and-join operators and Jack polynomials under a substitution $\theta_k\to 0$. A similar phenomenon was observed in \cite[sec.6]{Rapcak:2020ueh} where the algebra of generators induced by simultaneous adding/subtracting a pair of $(\pm)$-nodes is actually $\myY(\widehat{\fg\fl}_1)$.}
\begin{equation}\label{hook}
	\fm_{\lambda}:=\prod_{\stile\in\lambda} \upsilon_{\lambda}(\ntile)\,,
\end{equation}
where the product runs over only complete tiles.
The geometric meaning of this measure is that it coincides with the Euler class of the tangent space to the corresponding fixed point on the quiver variety \eqref{Euler}.
Also it defines a natural norm on semi-Fock module vectors (see \eqref{geom_norm} and \eqref{Cauchy}).

\section{``Bosonisation'' of semi-Fock representation}\label{sec:boson}

\subsection{Supertimes in semi-Fock representation}

Using \eqref{matrix_el} we define the following set of super-commuting operators $\theta_k$, $p_k$, $k\in\IZ_{\geq 1}$ from the Borel positive part of $\myY(\widehat{\fg\fl}_{1|1})$:
\begin{equation}\label{super-times}
	\begin{split}
		&\theta_1=e_0^{+}\,,\\
		&p_1=\left\{e_0^{-},e_0^{+}\right\}\,,\\
		&\theta_{k+1}=\frac{1}{k}\left[e_1^{+},\left\{e_0^{-},\theta_k\right\}\right]\,,\\
		&p_{k+1}=\frac{1}{k}\left\{e_0^{-},\left[e_1^{+},p_k\right]\right\}\,.
	\end{split}
\end{equation}

These operators super-commute (anti-commute for fermions and commute otherwise) as ordinary $p_k$ and Grassmann variables $\theta_k$.
It is natural to identify those with ``super-times'' -- analogs of time variables in the case of $\myY(\widehat{\fg\fl}_{1})$ and Jack/Schur polynomials -- and consider a polynomial ring:
\begin{equation}
	\CR=\IC[p_1,p_2,\ldots,\theta_1,\theta_2,\ldots]\,.
\end{equation}

Apparently, monomials in $\CR$ are in one-to-one correspondence with super-partitions: we simply identify an element $\lambda_i$ in $\lambda$ (see \eqref{parti}) with either $p_k$ or $\theta_k$ in a monomial product according to the following rule:
\begin{equation}
	\lambda_i\leftrightarrow \left\{\begin{array}{ll}
		p_{\lambda_i},& \mbox{ if }\lambda_i\mbox{ is integer}\,;\\
		\theta_{\lambda_i+\frac{1}{2}},&\mbox{ otherwise}\,.
	\end{array}\right.\,.
\end{equation}
One could introduce additive bosonic and fermionic degrees for time super-variables:
\begin{equation}
	{\rm deg}_b\,p_k=2k,\quad {\rm deg}_b\,\theta_k=2k-1,\quad {\rm deg}_f\,p_k=0,\quad {\rm deg}_f\,\theta_k=1\,.
\end{equation}
Then generating functions for monomial degrees and numbers of molten crystals/super-Young diagrams agree:
\begin{equation}\label{2dsgen}
	\begin{split}
		&\sum\lm_{\fm\in\CR}q^{{\rm deg}_b\fm}T^{{\rm deg}_f\fm}=\prod_{k=1}^{\infty}\frac{1+T q^{2k-1}}{1-q^{2k}}
		= \eqref{g_fun}\Big|_{q_1\to qT,\,q_2\to q/T}=\\
		&= 1 + Tq+q^2+2Tq^3+(2+T^2)q^4 + 4Tq^5+(3+2T^2)q^6+7Tq^7+(5+5T^2)q^8+(12T+T^3)q^9 + \ldots
	\end{split}
\end{equation}

The agreement between these generating functions suggests that in ring $\CR$ one could choose a \emph{selected} basis, so that vectors of this basis would correspond to molten crystals/super-Young diagrams under \eqref{super-times}.
In what follows we will argue that such basic polynomials exist and we call them super-Schur $\CS_{\lambda}$ polynomials by analogy with Jack/Schur polynomials in the case of $\myY(\widehat{\fg\fl_1})$.
An agreement between these bases may be expressed in the following relation:
\begin{tcolorbox}[ams equation]\label{basic_Jack_Yangian_rel}
	\CS_{\lambda}(p_1,p_2,\ldots,\theta_1,\theta_2,\ldots)|\varnothing\rangle =|\lambda\rangle
\end{tcolorbox}

Similarly we could have constructed  conjugate supertimes $\theta_{-k}$, $p_{-k}$, $k\in\IZ_{\geq 1}$ from the Borel negative part of $\myY(\widehat{\fg\fl}_{1|1})$:
\begin{equation}\label{super-co-times}
	\begin{split}
		&\theta_{-1}=f_0^{+}\,,\\
		&p_{-1}=\left\{f_0^{-},f_0^{+}\right\}\,,\\
		&\theta_{-k-1}=\frac{1}{k}\left[f_1^{+},\left\{f_0^{-},\theta_{-k}\right\}\right]\,\\
		&p_{-k-1}=\frac{1}{k}\left\{f_0^{-},\left[f_1^{+}, p_{-k}\right]\right\}\,.
	\end{split}
\end{equation}

Together these Borel positive and negative parts form a super-Heisenberg algebra:
\begin{equation}\label{Heisenberg}
	\begin{split}
		&\left\{\theta_{k},\theta_{m}\right\}=\delta_{k+m,0}(-1)^{k-1}\left(\epsilon_1\epsilon_2\right)^{2|k|-2}\cdot\bbone\,,\\
		&\left[\theta_k,p_m\right]=0\,,\\
		&\left[p_k,p_m\right]=\delta_{k+m,0}(-1)^{k-1}k\left(\epsilon_1\epsilon_2\right)^{2|k|-1}\cdot\bbone\,.
	\end{split}
\end{equation}

\subsection{Super-Cut-and-join operators and super-Jack/Schur polynomials}

The vectors of the semi-Fock module are eigenvectors in the basis of Cartan operators $\psi_k^{\pm}$.
Using relations \eqref{matrix_el} it is easy to derive that generating functions $\psi_k^{\pm}$ have the following expansions:\footnote{From this decomposition it becomes clear that the semi-Fock representation is defined for \emph{shifted} affine Yangian $\myY(\widehat{\fg\fl}_{1|1})$. We could have canceled the shift in $\psi^+(z)$ by introducing a new central element $C$ and redefining $\tilde\psi^+(z)=(z-C)\cdot \psi^+(z)$. However this simple trick is inapplicable to $\psi^-(z)$. To cancel the shift in $\psi^-(z)$ one has to introduce an additional pole leading to a new crystallization source and modifying the very representation to the Fock module. See \cite{Galakhov:2021xum} and app.\ref{app:stable} for details.}
\begin{equation}\label{psi-CaJ}
	\begin{split}
		&\psi^+(z)=\frac{1}{z}-\frac{\epsilon_1\epsilon_2}{z^3}\hat n^--\frac{2\epsilon_1\epsilon_2}{z^4}\hat w^-+O\left(\frac{1}{z^5}\right)\,,\\
		&\psi^-(z)=-z-\frac{\epsilon_1+\epsilon_2}{2}-\frac{\epsilon_1\epsilon_2}{z}\hat n^+-\frac{2\epsilon_1\epsilon_2}{z^2}\left(\hat w^++\frac{\epsilon_1+\epsilon_2}{4}\hat n^+\right)+O\left(\frac{1}{z^3}\right)\,,
	\end{split}
\end{equation}
where operators $\hat n^{\pm}$ and $\hat w^{\pm}$ are commuting operators with natural eigenvalues in the super-partition basis given by \eqref{eigen_values}:
\begin{equation}
	\hat n^{\pm}|\lambda\rangle=n^{\pm}_{\lambda}|\lambda\rangle,\quad\hat w^{\pm}|\lambda\rangle=w^{\pm}_{\lambda}|\lambda\rangle\,.
\end{equation}

Using identification \eqref{basic_Jack_Yangian_rel} we might wonder if these operators have a differential representation in polynomial ring $\CR$.
Operators $\hat n^{\pm}$ implement a function counting the number of boxes and half-boxes that is equivalent to counting degrees of corresponding polynomials.
Therefore we call these operators grading operator.
Naturally they acquire a form of the dilatation operator $x\frac{d}{dx}$:
\begin{equation}
	\begin{split}
		&\hat n^+=\sum\lm_{a=1}^{\infty}a\, p_a\frac{\p}{\p p_a}+\sum\lm_{a=1}^{\infty}a\,\theta_a\frac{\p}{\p \theta_a}\,,\\
		&\hat n^-=\sum\lm_{a=1}^{\infty}a\, p_a\frac{\p}{\p p_a}+\sum\lm_{a=1}^{\infty}(a-1)\,\theta_a\frac{\p}{\p \theta_a}\,.\\
	\end{split}
\end{equation}

Operators $\hat w^{\pm}$ have a more intricate form.
The closest analog to these operators in the case of $\myY(\widehat{\fg\fl}_1)$ is a cut-and-join operator \cite{MT}.
So we think of $\hat w^{\pm}$ as \emph{super-cut-and-join} operators:
\begin{tcolorbox}[ams equation]\label{super-cut-and-join}
	\begin{split}
		&\hat w^+=\frac{1}{2}\sum\lm_{a,b}\left[-\epsilon_1\epsilon_2(a+b)p_ap_b\frac{\p}{\p p_{a+b}}+ab\,p_{a+b}\frac{\p^2}{\p p_a\p p_b}\right]+\\
		&\quad+\sum\lm_{a,b}\left[-\epsilon_1\epsilon_2\,b\,p_a\theta_b\frac{\p}{\p\theta_{a+b}}+ab\,\theta_{a+b}\frac{\p^2}{\p p_a\p\theta_b}\right]+\sum\lm_a \frac{\epsilon_1+\epsilon_2}{2}a(a-1)\left(p_a\frac{\p}{\p p_a}+\theta_a\frac{\p}{\p\theta_a}\right)\,,\\
		&\hat w^-=\frac{1}{2}\sum\lm_{a,b}\left[-\epsilon_1\epsilon_2(a+b)p_ap_b\frac{\p}{\p p_{a+b}}+ab\,p_{a+b}\frac{\p^2}{\p p_a\p p_b}\right]+\\
		&\quad+\sum\lm_{a,b}\left[-\epsilon_1\epsilon_2\,(b-1)\,p_a\theta_b\frac{\p}{\p\theta_{a+b}}+a(b-1)\,\theta_{a+b}\frac{\p^2}{\p p_a\p\theta_b}\right]+\sum\lm_a \frac{\epsilon_1+\epsilon_2}{2}\left(a^2p_a\frac{\p}{\p p_a}+(a-1)^2\theta_a\frac{\p}{\p\theta_a}\right)\,.
	\end{split}
\end{tcolorbox}

We define the super-Schur polynomials $\CS_{\lambda}$ as eigen functions of these operators with respective eigen values:
\begin{equation}
	\begin{split}
		&\hat n^+\CS_{\lambda}(p_1,p_2,\ldots,\theta_1,\theta_2,\ldots)=n^+_{\lambda}\CS_{\lambda}(p_1,p_2,\ldots,\theta_1,\theta_2,\ldots)\,,\\
		&\hat n^-\CS_{\lambda}(p_1,p_2,\ldots,\theta_1,\theta_2,\ldots)=n^-_{\lambda}\CS_{\lambda}(p_1,p_2,\ldots,\theta_1,\theta_2,\ldots)\,,\\
		&\hat w^+\CS_{\lambda}(p_1,p_2,\ldots,\theta_1,\theta_2,\ldots)=w^+_{\lambda}\CS_{\lambda}(p_1,p_2,\ldots,\theta_1,\theta_2,\ldots)\,,\\
		&\hat w^-\CS_{\lambda}(p_1,p_2,\ldots,\theta_1,\theta_2,\ldots)=w^-_{\lambda}\CS_{\lambda}(p_1,p_2,\ldots,\theta_1,\theta_2,\ldots)\,.
	\end{split}
\end{equation}

There are no degeneracies, therefore these conditions are enough to obtain
all the polynomials from known eigenvalues \eqref{eigen_values}.
We present some explicit expressions for them in app.\ref{app:data}.
Also these polynomials satisfy generalized Pieri rule for addition and subtraction of boxes
to the diagram, and form a closed (generalized) Littlewood-Richardson algebra:
their products can be expanded in linear combinations of super-Schur polynomials:
\begin{equation}
	\CS_{\lambda}\cdot\CS_{\lambda'}=\CN_{\lambda\lambda'}^{\lambda''}\CS_{\lambda''}\,,
\end{equation}
with $|\lambda''|=|\lambda|+|\lambda'|$.

\subsection{Free field representation of semi-Fock modules}

To complete our discussion of super-times we would like to present a map inverse to \eqref{super-times} and \eqref{super-co-times}.

Using expansion \eqref{psi-CaJ} for the semi-Fock representation we derive that some modes of operators $\psi_k^{\pm}$ are simply central elements:
\begin{equation}
	\psi_0^+=1,\quad \psi_{-2}^{-}=-1,\quad \psi_{-1}^{-}=-\frac{\epsilon_1+\epsilon_2}{2}\,.
\end{equation}

Then using algebra relations \eqref{Yang_modes} we derive the following relation between modes:
\begin{equation}
	\begin{split}
		& \left[\psi^{\pm}_{1\pm 1},e_k^{\mp}\right]=-\epsilon_1\epsilon_2e_k^{\mp},\quad \left[\psi^{\pm}_{1\pm 1},f_k^{\mp}\right]=\epsilon_1\epsilon_2f_k^{\mp}\,,\\
		&\left[\psi^{\pm}_{2\pm 1},e_k^{\mp}\right]=-2\epsilon_1\epsilon_2\,e_{k+1}^{\mp}-(1\mp 1)\epsilon_1\epsilon_2\frac{\epsilon_1+\epsilon_2}{4}e_{k}^{\mp}\,,\\
		&\left[\psi^{\pm}_{2\pm 1},f_k^{\mp}\right]=2\epsilon_1\epsilon_2\,f_{k+1}^{\mp}+(1\mp 1)\epsilon_1\epsilon_2\frac{\epsilon_1+\epsilon_2}{4}f_{k}^{\mp}\,.
	\end{split}
\end{equation}

Then again translating these expressions with \eqref{psi-CaJ} one observes that cut-and-join operators $\hat w^{\pm}$ shift mode numbers of $e_k^{\pm}$, $f_k^{\pm}$:
\begin{equation}\label{mode-shift}
	e^{\pm}_{k+1}=\left[\hat w^{\pm},e_k^{\pm}\right],\quad f^{\pm}_{k+1}=-\left[\hat w^{\pm},f_k^{\pm}\right]\,.
\end{equation}
Thus, eventually, to reconstruct the whole $\myY(\widehat{\fg\fl}_{1|1})$ algebra we will need expressions only for $w^{\pm}$, $e_0^{\pm}$, $f_0^{\pm}$.
All the remaining generators may be re-constructed using \eqref{mode-shift} and the following relation from the algebra:
\begin{equation}
	\psi^{\pm}_{k+m}=-\left\{e_k^{\pm},f_m^{\pm}\right\}
\end{equation} 

Expressions for the zero modes of raising/lowering operators in terms of super-times may easily derived from \eqref{super-times} and \eqref{Heisenberg}:
\begin{tcolorbox}
	\begin{align}
		\begin{aligned}
			e_0^+&=\theta_1,&\hspace{10mm} f_0^+&=\frac{\p}{\p\theta_1},\\
			e_0^-&=\sum\lm_kp_k\frac{\p}{\p\theta_k},
			&\hspace{10mm} f_0^-&=\epsilon_1\epsilon_2\sum\lm_k k\,\theta_k\frac{\p}{\p p_k}\,.
		\end{aligned}
	\end{align}
\end{tcolorbox}

\subsection{Cauchy relation}

In this subsection we would like to confirm that super-Schur polynomials satisfy the following generalization of the Cauchy formula \cite{MorozovCauchy} for two sets of super-times $(p_i,\theta_i)$ and $(q_j,\xi_j)$:
\begin{equation}\label{Cauchy}
	\exp\left[\sum\lm_{k=1}^{\infty}\left(\frac{p_k q_k}{k}+\theta_k\xi_k\right)\right]=\sum\lm_{\lambda}\frac{(-\epsilon_1\epsilon_2)^{n_{\lambda}^-}}{{\fm}_{\lambda}}\CS_{\lambda}(p_i,\theta_i)\CS_{\lambda}(q_j,\xi_j)\,,
\end{equation}
where the summation runs over all super-Young diagrams, $n_{\lambda}^-$ is a number \eqref{eigen_values} of complete square boxes in $\lambda$, and, finally, $\fm_{\lambda}={\bf e}_{\lambda}$ is a geometrical hook measure on partitions (see \eqref{hook} and \eqref{geom_norm}).

\section{Conclusion and future directions}\label{sec:conclude}

In this paper we provided a detailed description of the semi-Fock representation
of the affine super Yangian $\myY(\widehat{\fg\fl}_{1|1})$.
This looks like a small step forward from $\myY(\widehat{\fg\fl}_{1})$,
which requires generalization in two complementary directions:
to arbitrary Yangians and to arbitrary representations.
However, we consider this consideration as quite important,
because it helps to emphasize non-conventional aspects of the story
and supports the possibility of a more ``physical'' approach to the problem.

Conventional approach to affine Yangian representations is to note that
they are associated with one or another substitute of Young diagrams,
and generators of the algebra add or subtract boxes.
The ``simple-root'' generators $e_\alpha$ and $f_\alpha$ act by
adding and subtracting single boxes.
The ``Cartan'' generators $\psi_{\alpha}$ do not change the diagrams, in this sense
diagrams can be considered as their common ``eigenfunctions''.
In this approach the only two essential questions concern the choice
of diagram system and the choice of coefficients in the action of generators.
The latter are usually described in terms of some ``charge'' function with poles
and zeroes, which are adjusted to the diagram system so that they allow
adding/subtracting only at the right places
(so that the diagram of a given type is deformed within the system).
There are still various unresolved problems with this approach,
including building the theory for solid (4d) partitions.
However, the approach is well-established and followed by the majority
of investigators in the field.

Unfortunately, it stops far from the needs of mathematical physics,
where we need much more:
a well-controlled set of special functions in convenient coordinates,
which can be used in description of correlators and partition functions
of the physically reasonable models.
This means that we need to associate with the diagrams concrete {\it states},
with the wave-functions which are eigenfunctions of $\psi_{\alpha}$, $e_{\alpha}$ and $f_{\alpha}$ acting as differential operators on them
(which become difference operators in the further lifting from Yangians
to DIM algebras).
These wave-functions should depend on appropriate variables and
are usually named time-dependent Schur functions.
In various examples the names can be different: Jacks, Halls, Macdonalds,
Shiraishi, Q-Schurs, 3d Shurs etc -- but we prefer and use Schurs as a common name.
The usual sequence of notions in {\it this} approach includes
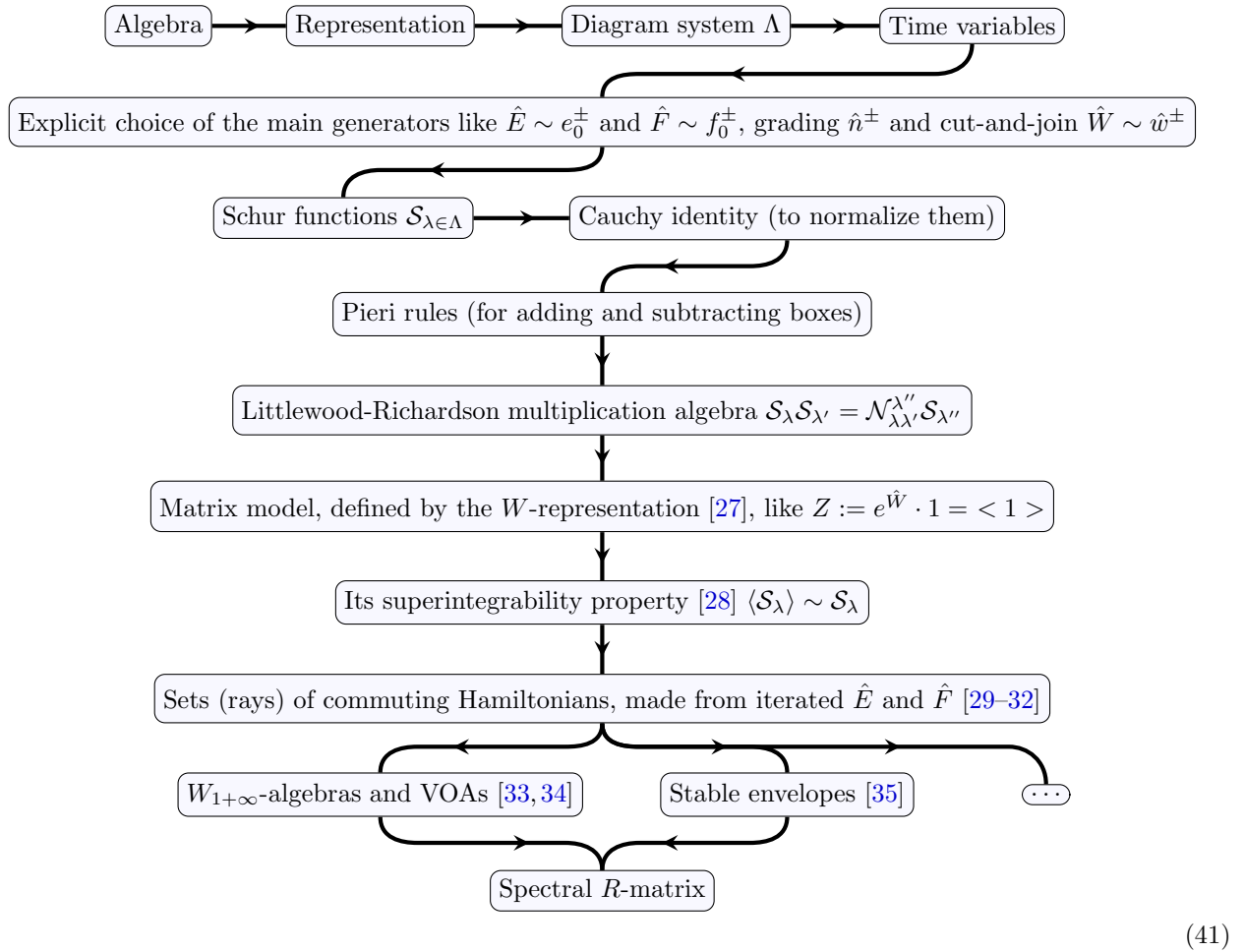
\begin{equation}\label{prog}
	\begin{array}{c}
		\begin{tikzpicture}[yscale=1.3]
			\tikzset{bbl/.style={draw, rounded corners, fill=white!97!blue}}
			\begin{scope}[shift={(-1,0)}]
				\node[bbl] (A) at (-5,0) {Algebra};
				\node[bbl] (B) at (-2,0) {Representation};
				\node[bbl] (C) at (2,0) {Diagram system $\Lambda$};
				\node[bbl] (D) at (6,0) {Time variables};
			\end{scope}
			\node[bbl] (E) at (0,-1) {Explicit choice of the main generators like $\hat E\sim e_0^{\pm}$ and $\hat F\sim f_0^{\pm}$, grading $\hat n^{\pm}$ and cut-and-join $\hat W\sim \hat w^{\pm}$ };
			\begin{scope}[shift={(-3.5,0)}]
				\node[bbl] (F) at (0,-2) {Schur functions ${\cal S}_{\lambda\in\Lambda}$};
				\node[bbl] (G) at (6,-2) {Cauchy identity (to normalize them)};
			\end{scope}
			\node[bbl] (H) at (0,-3) {Pieri rules (for adding and subtracting boxes)};
			\node[bbl] (I) at (0,-4) {Littlewood-Richardson multiplication algebra ${\cal S}_{\lambda}{\cal S}_{\lambda'} = {\cal N}_{\lambda\lambda'}^{\lambda''} {\cal S}_{\lambda''}$};
			\node[bbl] (J) at (0,-5) {Matrix model, defined by the $W$-representation \cite{Morozov:2009xk}, like
				$Z:=e^{\hat W}\cdot 1 =\ <1>$};
			\node[bbl] (K) at (0,-6) {Its superintegrability property \cite{Mironov:2022fsr} $\left\langle{\cal S}_{\lambda}\right\rangle \sim {\cal S}_{\lambda}$};
			\node[bbl] (L) at (0,-7) {Sets (rays) of commuting Hamiltonians, made from iterated $\hat E$ and $\hat F$ \cite{schiffmann2012cherednik, Smirnov:2021cyf, MMMP12, Mironov:2023wga}};
			\begin{scope}[shift={(-3,0)}]
				\node[bbl] (M) at (0,-8) {$W_{1+\infty}$-algebras and VOAs \cite{Prochazka:2023zdb, Gaiotto:2020dsq}};
				\node[bbl] (N) at (5.5,-8) {Stable envelopes \cite{Okounkov:2015spn}};
				\node[bbl] (O) at (9,-8) {$\ldots$};
			\end{scope}
			\node[bbl] (P) at (0,-9) {Spectral $R$-matrix};
			\tikzset{arr/.style={ultra thick, postaction={decorate},decoration={markings,
						mark= at position 0.65 with {\arrow{stealth}}}}}
			\path (A) edge[arr] (B) (B) edge[arr] (C) (C) edge[arr] (D) (F) edge[arr] (G) (H) edge[arr] (I) (I) edge[arr] (J) (J) edge[arr] (K) (K) edge[arr] (L);
			\path let \p1 = (D.south) in node (QQ) at ([shift={(-0.5,0)}]\x1,-0.5) {};
			\path let \p1 = (E.north) in node (PP) at ([shift={(0.5,0)}]\x1,-0.5) {};
			\draw[arr] (D.south) to[out=270,in=0] (QQ.center) -- (PP.center) to[out=180,in=90] (E.north);
			\path let \p1 = (E.south) in node (QQ) at ([shift={(-0.5,0)}]\x1,-1.5) {};
			\path let \p1 = (F.north) in node (PP) at ([shift={(0.5,0)}]\x1,-1.5) {};
			\draw[arr] (E.south) to[out=270,in=0] (QQ.center) -- (PP.center) to[out=180,in=90] (F.north);
			\path let \p1 = (G.south) in node (QQ) at ([shift={(-0.5,0)}]\x1,-2.5) {};
			\path let \p1 = (H.north) in node (PP) at ([shift={(0.5,0)}]\x1,-2.5) {};
			\draw[arr] (G.south) to[out=270,in=0] (QQ.center) -- (PP.center) to[out=180,in=90] (H.north);
			\path let \p1 = (L.south) in node (QQ) at ([shift={(-0.5,0)}]\x1,-7.5) {};
			\path let \p1 = (M.north) in node (PP) at ([shift={(0.5,0)}]\x1,-7.5) {};
			\draw[arr] (L.south) to[out=270,in=0] (QQ.center) -- (PP.center) to[out=180,in=90] (M.north);
			\path let \p1 = (L.south) in node (QQ) at ([shift={(0.5,0)}]\x1,-7.5) {};
			\path let \p1 = (N.north) in node (PP) at ([shift={(-0.5,0)}]\x1,-7.5) {};
			\draw[arr] (L.south) to[out=270,in=180] (QQ.center) -- (PP.center) to[out=0,in=90] (N.north);
			\path let \p1 = (L.south) in node (QQ) at ([shift={(0.5,0)}]\x1,-7.5) {};
			\path let \p1 = (O.north) in node (PP) at ([shift={(-0.5,0)}]\x1,-7.5) {};
			\draw[arr] (L.south) to[out=270,in=180] (QQ.center) -- (PP.center) to[out=0,in=90] (O.north);
			\path let \p1 = (M.south) in node (QQ) at ([shift={(0.5,0)}]\x1,-8.5) {};
			\path let \p1 = (P.north) in node (PP) at ([shift={(-0.5,0)}]\x1,-8.5) {};
			\draw[arr] (M.south) to[out=270,in=180] (QQ.center) -- (PP.center) to[out=0,in=90] (P.north);
			\path let \p1 = (N.south) in node (QQ) at ([shift={(-0.5,0)}]\x1,-8.5) {};
			\path let \p1 = (P.north) in node (PP) at ([shift={(0.5,0)}]\x1,-8.5) {};
			\draw[arr] (N.south) to[out=270,in=0] (QQ.center) -- (PP.center) to[out=180,in=90] (P.north);
		\end{tikzpicture}
	\end{array}
\end{equation}
While very well established for the ordinary Young diagrams
(i.e. for Fock representation of $\myY(\widehat{\fg\fl}_{1})$),
this program makes only first steps in various other directions:
\begin{itemize}
	
	\item{}to other diagram systems, like strict and odd partitions for \emph{Q-Schurs} \cite{Qschurs, Qschurs1, Qschurs2, MMZh},
	
	\item{}to quantized (finite-difference) case, i.e lifting to DIM \cite{DIM1, DIM2}
	
	\item{}to plane partitions (3d Schurs for triangular/true MacMahon representations of $\myY(\widehat{\fg\fl}_{1})$
	\cite{MT, Morozov:2022ndt}
	
	\item{} \ldots

\end{itemize}
In this sense any new non-trivial example is a big progress.
From this point of view our description of semi-Fock representations for $\myY(\widehat{\fg\fl}_{1|1})$
and their association with a new non-trivial generalization of Young diagrams seem important.
Even in this case we actually did only the first half of the job in \eqref{prog},
the second part remains for the future work.

Especially close to this particular story is the theory of $Q$-functions,
where some other super-Jacks also appeared \cite{sergeev2005generalised},
while the possibility of super-field description \cite{MMZh} remained under-investigated.
Another amusing parallel is with the theory of Nekrasov functions,
where parameters $\epsilon_1$ and $\epsilon_2$ also appear --
moreover, their hidden {\it raison d'etre (un secret de Polichinelle)} is the same:
they parameterize the  toric action on CY3.
These are also questions for further investigation.

Also we should mention another far reaching yet lying beyond the scope of this note aim of our journey -- a modern look at integrability as a property of a topological QFT moduli space.
This point of view allows one to interpret R-matrices intertwining tensor powers of representations as holonomies of a Berry connection on some TQFT moduli space \cite{Bullimore:2017lwu, Dedushenko:2022pem,  Galakhov:2020upa}.
This story has a continuation in a dominion of enumerative and algebraic geometry of quiver varieties in a form of stable envelopes \cite{2012arXiv1211.1287M, Aganagic:2016jmx, Rimanyi:2019zyi}.
Physically \cite{Dedushenko:2021mds, Bullimore:2021rnr, Dedushenko:2023qjq, Crew:2023tky, Galakhov:2022uyu, Galakhov:2023aev} a stable envelope is a transition matrix between various choices of brane boundary conditions for a string theory inhabiting a Calabi-Yau manifold whose crepant resolution defines the quiver in question.
In this context a family of toric Calabi-Yau 3-folds represented by generalized conifolds associated with affine Yangians $\myY(\widehat{\fg\fl}_{m|n})$ is interesting since it presents quiver varieties beyond the ones of the Nakajima type \cite{2009arXiv0905.0686G}, and a simple resolved conifold -- algebra $\myY(\widehat{\fg\fl}_{1|1})$ -- delivers the simplest example of a non-Nakajima quiver.
This quiver variety has no canonical symplectic structure pairing present in the Nakajima quiver varieties \cite{Okounkov:2015spn}, and, therefore, the standard enumerative geometry methods should be improved to capture the construction of stable envelopes for these varieties as well.
We believe that our construction of super-Schur polynomials and, especially, a similarity between languages of symmetric polynomials in this note and works \cite{Smirnov:2013hh,Smirnov:2014npa} on instanton R-matrices will open a new road towards this goal.

On the other hand, canonical constructions of the R-matrices for $\myY(\widehat{\fg\fl}_{m|n})$ become in the light of the present note no less exciting.
Unfortunately, canonical constructions in the case of $\myY(\widehat{\fg\fl}_{m|n})$ are rather involved for both purely algebraic \cite{2019arXiv191106666U, Bao:2022fpk, Bao:2022jhy} and CFT \cite{Kolyaskin:2022tqi} approaches.
We hope that combinatorial relations (hook formulae) for matrix coefficients of the semi-Fock representation would allow one to improve and simplify these techniques for a more open study of associated integrable models.

Surely, as the first step towards a completion of this program it would be interesting to construct a system of (semi-)Fock modules for the generic family of affine super Yangians $\myY(\widehat{\fg\fl}_{m|n})$ and, moreover, for quiver Yangians associated with generic toric Calabi-Yau 3-folds\cite{Li:2020rij}.
As a defining characteristic of such a representation we might assume an existence of an explicit simple cut-and-join representation.

\section*{Acknowledgments}

This work is supported by the grants of the Foundation for the Advancement of
Theoretical Physics “BASIS” (A.M., N.T.), by the RFBR grant 21-51-46010-ST-a (A.M., N.T.), and by the joint RFBR-MOST grant 21-52-52004-MNT-a (A.M.).

\appendix

\section{Semi-Fock representations of \texorpdfstring{$\myY(\widehat{\fg\fl}_{1|1})$}{Ygl11} from D-brane dynamics} \label{app:BPS}

\subsection{On smooth quiver varieties and Fock representations}

Attempts to construct free field representations for generic representations of affine Yangian algebras encounter certain difficulties \cite{MT}, and the resulting algorithms lack transparency and become rather involved quite rapidly.
In contrast a free field representation for the Fock representation of  $\myY(\widehat{\fg\fl}_1)$ is well-studied and is given in terms of Jack polynomials.\footnote{Other free field constructions (see e.g. \cite{Bezerra:2019dmp}) of $\myY(\widehat{\fg\fl}_{m|n})$ are rather involved and require a non-trivial extension making Hamiltonians $\psi_k^{\pm}$ non-commutative.}

This motivates us to search for related deformations of Jack polynomial families in a domain of Fock representations of affine Yangians.
However what representation should be promoted to a canonical Fock representation remains unclear, especially when we discuss super-algebras.
We will attempt to find an inspiration in the physical picture where the affine Yangian is a an algebra of scattering BPS states in a system of D-branes wrapping a toric Calabi-Yau 3-fold \cite{Li:2020rij,Galakhov:2020vyb}.

We will not attempt to cover this construction in details referring the interested reader to a concise \cite{Yamazaki:2022cdg} and a more detailed \cite{Li:2023zub} reviews.
For a family of super-algebras $\myY(\widehat{\fg\fl}_{m|n})$ the correspondence is given by the following recipe: the affine Dynkin diagram of $\widehat{\fg\fl}_{m|n}$ is translated to a quiver of the toric CY3 crepant resolution.
Each edge of the Dynkin diagram becomes a pair of counter-directed morphism, and each even node acquires a self-morphism, whereas an odd node does not.
Furthermore one puts a system of D0-D2-D4-D6 branes on the corresponding toric CY3, and the quiver describes its gauge-matter content in the IR description of this theory.
Corresponding quivers for $\myY(\widehat{\fg\fl}_1)$ and $\myY(\widehat{\fg\fl}_{1|1})$ are depicted in fig.~\ref{fig:D-brane_Fock}.

\begin{figure}[h!]
	\centering
	\begingroup
	\renewcommand*{\arraystretch}{2}
	$\begin{array}{c|c|c}
		\mbox{Algebra} & \myY(\widehat{\fg\fl}_1) & \myY(\widehat{\fg\fl}_{1|1}) \\
		\hline
		\mbox{4-cycle in CY3} & \begin{array}{c} 
			\begin{tikzpicture}[scale=0.7]
				\filldraw[fill=orange] (0,0) -- (1,0) decorate[decoration={snake, segment length=2mm, amplitude=0.5mm}] { -- (0,1)}  -- (0,0);
				\draw[ultra thick] (0,0) -- (1,0) (0,0) -- (0,1) (0,0) -- (-0.7,-0.7);
			\end{tikzpicture}
		\end{array} & \begin{array}{c} 
			\begin{tikzpicture}[scale=0.5]
				\filldraw[fill=orange] (0,0) -- (1,0) decorate[decoration={snake, segment length=2mm, amplitude=0.5mm}] { -- (0,1)}  -- (0,0);
				\draw[ultra thick] (0,0) -- (1,0) (0,0) -- (0,1) (0,0) -- (-0.4,-0.4) (-1.4,-0.4) -- (-0.4,-0.4) (-0.4,-1.4) -- (-0.4,-0.4);
			\end{tikzpicture}
		\end{array}  \\
		\hline
		\mbox{Quiver} &\begin{array}{c}
			\begin{tikzpicture}
				\draw[postaction={decorate},decoration={markings,
					mark= at position 0.75 with {\arrow{stealth}}}] (0,0) to[out=60,in=0] (0,1) to[out=180,in=120] (0,0);
				\node[above] at (0,1) {$\scriptstyle B_2$};
				\begin{scope}[rotate=90]
					\draw[postaction={decorate},decoration={markings,
						mark= at position 0.75 with {\arrow{stealth}}}] (0,0) to[out=60,in=0] (0,1) to[out=180,in=120] (0,0);
					\node[left] at (0,1) {$\scriptstyle B_1$};
				\end{scope}
				\begin{scope}[rotate=270]
					\draw[postaction={decorate},decoration={markings,
						mark= at position 0.75 with {\arrow{stealth}}}] (0,0) to[out=60,in=0] (0,1) to[out=180,in=120] (0,0);
					\node[right] at (0,1) {$\scriptstyle B_3$};
				\end{scope}
				\draw[postaction={decorate},decoration={markings,
					mark= at position 0.65 with {\arrow{stealth}}}] (0,-1.5) to[out=120,in=240] node[pos=0.5,left] {$\scriptstyle I$} (0,0);
				\draw[ultra thick,black!40!green,postaction={decorate},decoration={markings,
					mark= at position 0.65 with {\arrow{stealth}}}] (0,0) to[out=300,in=60] node[pos=0.5,right] {$\scriptstyle \color{black!40!green} J$} (0,-1.5);
				\begin{scope}[shift={(0,-1.5)}]
					\draw[fill=white] (-0.1,-0.1) -- (-0.1,0.1) -- (0.1,0.1) -- (0.1,-0.1) -- cycle;
				\end{scope}
				\draw[fill=black!30!red] (0,0) circle (0.1);
			\end{tikzpicture}
		\end{array}& \begin{array}{c}
			\begin{tikzpicture}
				\tikzstyle{col0} = [fill=white]
				\tikzstyle{col1} = [fill=gray]
				\tikzstyle{col2} = [fill=white!40!blue]
				\draw[postaction={decorate},
				decoration={markings, mark= at position 0.55 with {\arrow{stealth}}}]
				(1,-2) -- (0,0) node[pos=0.5,below left] {$\scriptstyle I$};
				\draw[ultra thick, black!40!green, postaction={decorate},
				decoration={markings, mark= at position 0.55 with {\arrow{stealth}}}]
				(2,0) -- (1,-2) node[pos=0.5,below right] {$\scriptstyle J$};
				\draw[postaction={decorate},
				decoration={markings, mark= at position 0.65 with {\arrow{stealth}}, mark= at position 0.45 with {\arrow{stealth}}}] (0,0) to[out=30,in=150] node[pos=0.5, above] {$\scriptstyle A_1, A_2$} (2,0);
				\draw[postaction={decorate},
				decoration={markings, mark= at position 0.65 with {\arrow{stealth}}, mark= at position 0.45 with {\arrow{stealth}}}] (2,0) to[out=210,in=330] node[pos=0.5, below] {$\scriptstyle B_1, B_2$} (0,0);
				\draw[col1] (0,0) circle (0.1);
				\draw[col2] (2,0) circle (0.1);
				\node[left] at (-0.1,0) {$+$};
				\node[right] at (2.1,0) {$-$};
				\begin{scope}[shift={(1,-2)}]
					\draw[col0] (-0.08,-0.08) -- (-0.08,0.08) -- (0.08,0.08) -- (0.08,-0.08) -- cycle;
				\end{scope}
			\end{tikzpicture}
		\end{array} \\
		\hline
		\mbox{Superpotential} & W = \Tr\left(B_1\left[B_2,B_3\right] + {\color{black!40!green} B_3IJ}\right)& W = \Tr\left(A_1B_1A_2B_2-A_1B_2A_2B_1 + {\color{black!40!green} A_2IJ}\right)\\
		\hline
		\mbox{Quiver variety} &\p_{B_3}W=\left[B_1,B_2\right]+IJ=0& \p_{A_2}W=B_2A_1B_1-B_1A_1B_2+IJ=0 \\
		\hline
		\mbox{Crystal slice} & \begin{array}{c}\begin{tikzpicture}
				\tikzset{style0/.style={thin, fill=black!30!red}}
				\tikzset{style1/.style={thin, fill=black!30!red}}
				\tikzset{style2/.style={thin, fill=white!70!red}}
				\begin{scope}[scale=0.35]
					\draw[style0] (-0.707107,2.85774) -- (-0.707107,3.67423) -- (0.,3.26599) -- (0.,2.44949) -- cycle;
					\draw[style0] (-0.707107,3.67423) -- (-0.707107,4.49073) -- (0.,4.08248) -- (0.,3.26599) -- cycle;
					\draw[style0] (-0.707107,4.49073) -- (-0.707107,5.30723) -- (0.,4.89898) -- (0.,4.08248) -- cycle;
					\draw[style0] (-0.707107,5.30723) -- (-0.707107,6.12372) -- (0.,5.71548) -- (0.,4.89898) -- cycle;
					\draw[style0] (-0.707107,6.12372) -- (-0.707107,6.94022) -- (0.,6.53197) -- (0.,5.71548) -- cycle;
					\draw[style0] (-0.707107,6.94022) -- (-0.707107,7.75672) -- (0.,7.34847) -- (0.,6.53197) -- cycle;
					\draw[style1] (0.,1.63299) -- (0.,2.44949) -- (0.707107,2.04124) -- (0.707107,1.22474) -- cycle;
					\draw[style1] (0.,2.44949) -- (0.,3.26599) -- (0.707107,2.85774) -- (0.707107,2.04124) -- cycle;
					\draw[style1] (0.,3.26599) -- (0.,4.08248) -- (0.707107,3.67423) -- (0.707107,2.85774) -- cycle;
					\draw[style1] (0.,4.08248) -- (0.,4.89898) -- (0.707107,4.49073) -- (0.707107,3.67423) -- cycle;
					\draw[style1] (0.,4.89898) -- (0.,5.71548) -- (0.707107,5.30723) -- (0.707107,4.49073) -- cycle;
					\draw[style1] (0.707107,0.408248) -- (0.707107,1.22474) -- (1.41421,0.816497) -- (1.41421,0.) -- cycle;
					\draw[style1] (0.707107,1.22474) -- (0.707107,2.04124) -- (1.41421,1.63299) -- (1.41421,0.816497) -- cycle;
					\draw[style1] (0.707107,2.04124) -- (0.707107,2.85774) -- (1.41421,2.44949) -- (1.41421,1.63299) -- cycle;
					\draw[style1] (0.707107,2.85774) -- (0.707107,3.67423) -- (1.41421,3.26599) -- (1.41421,2.44949) -- cycle;
					\draw[style1] (0.707107,3.67423) -- (0.707107,4.49073) -- (1.41421,4.08248) -- (1.41421,3.26599) -- cycle;
					\draw[style1] (1.41421,-0.816497) -- (1.41421,0.) -- (2.12132,-0.408248) -- (2.12132,-1.22474) -- cycle;
					\draw[style1] (1.41421,0.) -- (1.41421,0.816497) -- (2.12132,0.408248) -- (2.12132,-0.408248) -- cycle;
					\draw[style1] (1.41421,0.816497) -- (1.41421,1.63299) -- (2.12132,1.22474) -- (2.12132,0.408248) -- cycle;
					\draw[style1] (1.41421,1.63299) -- (1.41421,2.44949) -- (2.12132,2.04124) -- (2.12132,1.22474) -- cycle;
					\draw[style1] (2.12132,-2.04124) -- (2.12132,-1.22474) -- (2.82843,-1.63299) -- (2.82843,-2.44949) -- cycle;
					\draw[style1] (2.12132,-1.22474) -- (2.12132,-0.408248) -- (2.82843,-0.816497) -- (2.82843,-1.63299) -- cycle;
					\draw[style1] (2.12132,-0.408248) -- (2.12132,0.408248) -- (2.82843,0.) -- (2.82843,-0.816497) -- cycle;
					\draw[style1] (2.82843,-2.44949) -- (2.82843,-1.63299) -- (3.53553,-2.04124) -- (3.53553,-2.85774) -- cycle;
					\draw[style1] (3.53553,-2.85774) -- (3.53553,-2.04124) -- (4.24264,-2.44949) -- (4.24264,-3.26599) -- cycle;
					\draw[style2] (-1.41421,0.816497) -- (-1.41421,1.63299) -- (-0.707107,1.22474) -- (-0.707107,0.408248) -- cycle;
					\draw[style2] (-1.41421,1.63299) -- (-1.41421,2.44949) -- (-0.707107,2.04124) -- (-0.707107,1.22474) -- cycle;
					\draw[style2] (-0.707107,0.408248) -- (-0.707107,1.22474) -- (0.,0.816497) -- (0.,0.) -- cycle;
					\draw[style2] (0.,-0.816497) -- (0.,0.) -- (0.707107,-0.408248) -- (0.707107,-1.22474) -- cycle;
					\draw[style2] (-2.12132,-0.408248) -- (-2.12132,0.408248) -- (-1.41421,0.) -- (-1.41421,-0.816497) -- cycle;
					\draw[style2] (-1.41421,-0.816497) -- (-1.41421,0.) -- (-0.707107,-0.408248) -- (-0.707107,-1.22474) -- cycle;
					\draw[style2] (-0.707107,-2.04124) -- (-0.707107,-1.22474) -- (0.,-1.63299) -- (0.,-2.44949) -- cycle;
					\draw[style2] (0.,-2.44949) -- (0.,-1.63299) -- (0.707107,-2.04124) -- (0.707107,-2.85774) -- cycle;
					\draw[style2] (-2.12132,-2.04124) -- (-2.12132,-1.22474) -- (-1.41421,-1.63299) -- (-1.41421,-2.44949) -- cycle;
					\draw[style2] (-3.53553,-2.04124) -- (-3.53553,-1.22474) -- (-2.82843,-1.63299) -- (-2.82843,-2.44949) -- cycle;
					\draw[style0] (0.707107,6.12372) -- (0.707107,6.94022) -- (0.,6.53197) -- (0.,5.71548) -- cycle;
					\draw[style0] (0.707107,6.94022) -- (0.707107,7.75672) -- (0.,7.34847) -- (0.,6.53197) -- cycle;
					\draw[style1] (1.41421,4.89898) -- (1.41421,5.71548) -- (0.707107,5.30723) -- (0.707107,4.49073) -- cycle;
					\draw[style1] (2.12132,2.85774) -- (2.12132,3.67423) -- (1.41421,3.26599) -- (1.41421,2.44949) -- cycle;
					\draw[style1] (2.12132,3.67423) -- (2.12132,4.49073) -- (1.41421,4.08248) -- (1.41421,3.26599) -- cycle;
					\draw[style1] (2.82843,0.816497) -- (2.82843,1.63299) -- (2.12132,1.22474) -- (2.12132,0.408248) -- cycle;
					\draw[style1] (2.82843,1.63299) -- (2.82843,2.44949) -- (2.12132,2.04124) -- (2.12132,1.22474) -- cycle;
					\draw[style1] (3.53553,-1.22474) -- (3.53553,-0.408248) -- (2.82843,-0.816497) -- (2.82843,-1.63299) -- cycle;
					\draw[style1] (3.53553,-0.408248) -- (3.53553,0.408248) -- (2.82843,0.) -- (2.82843,-0.816497) -- cycle;
					\draw[style1] (4.94975,-2.85774) -- (4.94975,-2.04124) -- (4.24264,-2.44949) -- (4.24264,-3.26599) -- cycle;
					\draw[style2] (0.,1.63299) -- (0.,2.44949) -- (-0.707107,2.04124) -- (-0.707107,1.22474) -- cycle;
					\draw[style2] (0.707107,0.408248) -- (0.707107,1.22474) -- (0.,0.816497) -- (0.,0.) -- cycle;
					\draw[style2] (1.41421,-0.816497) -- (1.41421,0.) -- (0.707107,-0.408248) -- (0.707107,-1.22474) -- cycle;
					\draw[style2] (2.12132,-2.04124) -- (2.12132,-1.22474) -- (1.41421,-1.63299) -- (1.41421,-2.44949) -- cycle;
					\draw[style2] (0.,-0.816497) -- (0.,0.) -- (-0.707107,-0.408248) -- (-0.707107,-1.22474) -- cycle;
					\draw[style2] (1.41421,-2.44949) -- (1.41421,-1.63299) -- (0.707107,-2.04124) -- (0.707107,-2.85774) -- cycle;
					\draw[style2] (-0.707107,-2.04124) -- (-0.707107,-1.22474) -- (-1.41421,-1.63299) -- (-1.41421,-2.44949) -- cycle;
					\draw[style2] (-2.12132,-2.04124) -- (-2.12132,-1.22474) -- (-2.82843,-1.63299) -- (-2.82843,-2.44949) -- cycle;
					\draw[style0] (0.,8.16497) -- (0.707107,7.75672) -- (0.,7.34847) -- (-0.707107,7.75672) -- cycle;
					\draw[style1] (0.707107,6.12372) -- (1.41421,5.71548) -- (0.707107,5.30723) -- (0.,5.71548) -- cycle;
					\draw[style1] (1.41421,4.89898) -- (2.12132,4.49073) -- (1.41421,4.08248) -- (0.707107,4.49073) -- cycle;
					\draw[style1] (2.12132,2.85774) -- (2.82843,2.44949) -- (2.12132,2.04124) -- (1.41421,2.44949) -- cycle;
					\draw[style1] (2.82843,0.816497) -- (3.53553,0.408248) -- (2.82843,0.) -- (2.12132,0.408248) -- cycle;
					\draw[style1] (3.53553,-1.22474) -- (4.24264,-1.63299) -- (3.53553,-2.04124) -- (2.82843,-1.63299) -- cycle;
					\draw[style1] (4.24264,-1.63299) -- (4.94975,-2.04124) -- (4.24264,-2.44949) -- (3.53553,-2.04124) -- cycle;
					\draw[style2] (-0.707107,2.85774) -- (0.,2.44949) -- (-0.707107,2.04124) -- (-1.41421,2.44949) -- cycle;
					\draw[style2] (0.,1.63299) -- (0.707107,1.22474) -- (0.,0.816497) -- (-0.707107,1.22474) -- cycle;
					\draw[style2] (0.707107,0.408248) -- (1.41421,0.) -- (0.707107,-0.408248) -- (0.,0.) -- cycle;
					\draw[style2] (1.41421,-0.816497) -- (2.12132,-1.22474) -- (1.41421,-1.63299) -- (0.707107,-1.22474) -- cycle;
					\draw[style2] (-1.41421,0.816497) -- (-0.707107,0.408248) -- (-1.41421,0.) -- (-2.12132,0.408248) -- cycle;
					\draw[style2] (-0.707107,0.408248) -- (0.,0.) -- (-0.707107,-0.408248) -- (-1.41421,0.) -- cycle;
					\draw[style2] (0.,-0.816497) -- (0.707107,-1.22474) -- (0.,-1.63299) -- (-0.707107,-1.22474) -- cycle;
					\draw[style2] (0.707107,-1.22474) -- (1.41421,-1.63299) -- (0.707107,-2.04124) -- (0.,-1.63299) -- cycle;
					\draw[style2] (-2.12132,-0.408248) -- (-1.41421,-0.816497) -- (-2.12132,-1.22474) -- (-2.82843,-0.816497) -- cycle;
					\draw[style2] (-1.41421,-0.816497) -- (-0.707107,-1.22474) -- (-1.41421,-1.63299) -- (-2.12132,-1.22474) -- cycle;
					\draw[style2] (-2.82843,-0.816497) -- (-2.12132,-1.22474) -- (-2.82843,-1.63299) -- (-3.53553,-1.22474) -- cycle;
				\end{scope}
		\end{tikzpicture}\end{array} & \begin{array}{c} \begin{tikzpicture}[scale=0.25]
				\tikzset{sty1/.style={fill=white!40!gray}}
				\tikzset{sty2/.style={fill=white!70!blue}}
				\tikzset{sty3/.style={fill=gray}}
				\tikzset{sty4/.style={fill=white!40!blue}}
				\foreach \i/\j/\st in {5/6/sty2, 5/-6/sty2, 3/6/sty2, 3/-6/sty2, 1/6/sty2, 1/-6/sty2, -1/6/sty2, -1/-6/sty2, -3/6/sty2, -3/-6/sty2, -5/6/sty2, -5/-6/sty2, -7/6/sty2, -7/4/sty2, -7/2/sty2, -7/0/sty2, -7/-2/sty2, -7/-4/sty2, -7/-6/sty2, 4/6/sty1, 4/-6/sty1, 2/6/sty1, 2/-6/sty1, 0/6/sty1, 0/-6/sty1, -2/6/sty1, -2/-6/sty1, -4/6/sty1, -4/-6/sty1, -6/6/sty1, -6/4/sty1, -6/2/sty1, -6/0/sty1, -6/-2/sty1, -6/-4/sty1, -6/-6/sty1, 4/5/sty2, 4/-5/sty2, 2/5/sty2, 2/-5/sty2, 0/5/sty2, 0/-5/sty2, -2/5/sty2, -2/-5/sty2, -4/5/sty2, -4/-5/sty2, -6/5/sty2, -6/3/sty2, -6/1/sty2, -6/-1/sty2, -6/-3/sty2, -6/-5/sty2, 3/5/sty1, 3/-5/sty1, 1/5/sty1, 1/-5/sty1, -1/5/sty1, -1/-5/sty1, -3/5/sty1, -3/-5/sty1, -5/5/sty1, -5/3/sty1, -5/1/sty1, -5/-1/sty1, -5/-3/sty1, -5/-5/sty1, 3/4/sty2, 3/-4/sty2, 1/4/sty2, 1/-4/sty2, -1/4/sty2, -1/-4/sty2, -3/4/sty2, -3/-4/sty2, -5/4/sty2, -5/2/sty2, -5/0/sty2, -5/-2/sty2, -5/-4/sty2, 2/4/sty1, 2/-4/sty1, 0/4/sty1, 0/-4/sty1, -2/4/sty1, -2/-4/sty1, -4/4/sty1, -4/2/sty1, -4/0/sty1, -4/-2/sty1, -4/-4/sty1, 2/3/sty2, 2/-3/sty2, 0/3/sty2, 0/-3/sty2, -2/3/sty2, -2/-3/sty2, -4/3/sty2, -4/1/sty2, -4/-1/sty2, -4/-3/sty2, 1/3/sty1, 1/-3/sty1, -1/3/sty1, -1/-3/sty1, -3/3/sty1, -3/1/sty1, -3/-1/sty1, -3/-3/sty1, 1/2/sty2, 1/-2/sty2, -1/2/sty2, -1/-2/sty2, -3/2/sty2, -3/0/sty2, -3/-2/sty2, 0/2/sty1, 0/-2/sty1, -2/2/sty1, -2/0/sty1, -2/-2/sty1, 0/1/sty2, 0/-1/sty2, -2/1/sty2, -2/-1/sty2, -1/1/sty1, -1/-1/sty1, -1/0/sty2, 7/6/sty4, 7/4/sty4, 7/2/sty4, 7/0/sty4, 7/-2/sty4, 7/-4/sty4, 7/-6/sty4, 6/6/sty3, 6/4/sty3, 6/2/sty3, 6/0/sty3, 6/-2/sty3, 6/-4/sty3, 6/-6/sty3, 6/5/sty4, 6/3/sty4, 6/1/sty4, 6/-1/sty4, 6/-3/sty4, 6/-5/sty4, 5/5/sty3, 5/3/sty3, 5/1/sty3, 5/-1/sty3, 5/-3/sty3, 5/-5/sty3, 5/4/sty4, 5/2/sty4, 5/0/sty4, 5/-2/sty4, 5/-4/sty4, 4/4/sty3, 4/2/sty3, 4/0/sty3, 4/-2/sty3, 4/-4/sty3, 4/3/sty4, 4/1/sty4, 4/-1/sty4, 4/-3/sty4, 3/3/sty3, 3/1/sty3, 3/-1/sty3, 3/-3/sty3, 3/2/sty4, 3/0/sty4, 3/-2/sty4, 2/2/sty3, 2/0/sty3, 2/-2/sty3, 2/1/sty4, 2/-1/sty4, 1/1/sty3, 1/-1/sty3, 1/0/sty4, 0/0/sty3}
				{
					\draw[\st] (\i,\j) circle (0.7);
				}
				\begin{scope}[shift={(14,0)}]
					\draw[stealth-stealth] (-2,0) -- (2,0);
					\draw[stealth-stealth] (0,-2) -- (0,2);
					\node[left] at (-2,0) {$\scriptstyle A_2$};
					\node[right] at (2,0) {$\scriptstyle A_1$};
					\node[below] at (0,-2) {$\scriptstyle B_2$};
					\node[above] at (0,2) {$\scriptstyle B_1$};
				\end{scope}
		\end{tikzpicture}\end{array}\\
	\end{array}$
	\endgroup
	\caption{Comparing MacMahon (3d crystal) and (semi-)Fock (2d slice) representations for $\myY(\widehat{\fg\fl}_1)$ and $\myY(\widehat{\fg\fl}_{1|1})$.}
	\label{fig:D-brane_Fock}
\end{figure}

A canonical representation of the affine Yangian emerging in this setting is a MacMahon-like representation where vectors of a module are labeled by 3d crystals.
In the example of $\myY(\widehat{\fg\fl}_1)$ corresponding to $\IC^3$ these 3d crystals are simply plane partitions.
The Fock module emerges in this algebra when we try to confine D-branes inside a heavy D4-brane wrapping a real 4-cycle inside CY3, in this particular case any of three $\IC^2$-planes invariant with respect to the toric action inside $\IC^3$.
This deformation \cite{Nishinaka:2013mba, Galakhov:2021xum, Noshita:2021dgj} changes the quiver framing and modifies the superpotential (see fig.~\ref{fig:D-brane_Fock}), so that a specific field $X_1$ becomes selected and the superpotential acquires the following form:
\begin{equation}
	W={\rm Tr}\, X_1\,\left[f(X_2,X_3,\ldots)+IJ\right]\,,
\end{equation}
where function $f$ is independent of $X_1$, and $I$, $J$ are fields corresponding to the quiver framing.

We should note that in the cases of $\myY(\widehat{\fg\fl}_n)$ this procedure enhances the initial supersymmetry from 8 charges to 16, and corresponding fields playing the role of $X_1$ become ``Lagrange multipliers'', so that a quiver variety cut out by an F-term condition:
\begin{equation}
	\p_{X_1}W=f(X_2,X_3,\ldots)+IJ=0\,,
\end{equation}
is a Nakajima quiver variety known to be \emph{smooth} \cite{2009arXiv0905.0686G} compared to the initial quiver variety that is singular.
In the case of $\myY(\widehat{\fg\fl}_1)$ this is a well-known ADHM description of Hilbert scheme ${\rm Hilb}^*(\IC^2)$ \cite{nakajima1999lectures}:
\begin{equation}
	\left[B_1,B_2\right]+IJ=0\,.
\end{equation}

The supersymmetry of the quiver gauge theory corresponding to a superalgebra $\myY(\widehat{\fg\fl}_{m|n})$ can not be enhanced.
Yet we hope to acquire a smooth quiver variety by repeating the strategy described above.
We will concentrate on the case of $\myY(\widehat{\fg\fl}_{1|1})$ corresponding to a conifold $xy=zw$ resolution.
By choosing a 4-cycle we modify the quiver and the superpotential accordingly (see fig.~\ref{fig:D-brane_Fock}).
In this case the role of a Lagrange multiplier is played by field $A_2$, and the F-term condition describes the following quiver variety (see fig.~\ref{fig:D-brane_Fock}):
\begin{equation}\label{quiver_variety}
	B_2A_1B_1-B_1A_1B_2+IJ=0\,.
\end{equation}

This procedure results in representations that are slices of the canonical MacMahon representation -- a pyramid \cite{Szendroi} representation for $\myY(\widehat{\fg\fl}_{1|1})$.
We have chosen parameters in such a way that the slice goes over one-atom layer of the east side of the pyramid.

We describe explicitly the molten crystals for the semi-Fock representation in sec.\ref{sec:sF}.

Let us denote quiver dimensions as $d_{\pm}$ respectively and the moduli space of the quiver variety representation confined to surface \eqref{quiver_variety} as $\mathscr{M}(\vec d)$.
Crystal $\lambda$ denotes a fixed point on this space with respect to the equivariant action of the complexified gauge and flavor groups.
We denote the corresponding tangent space as $\mathsf{T}_{\lambda}\mathscr{M}(\vec d)$.
A BPS wave function of a D-brane system is a cohomology of the supercharge and corresponds to the Euler class of the tangent space:
\begin{equation}
	{\bf e}_{\lambda}:={\rm Euler}(\mathsf{T}_{\lambda}\mathscr{M}(\vec d))\,.
\end{equation}

A BPS algebra on the Hilbert space of BPS D-brane states is induced by processes of capturing/emitting an elementary brane of fractional charge, such that the quiver dimension vector is shifted by a unit vector $e_+=(1,0)$ or $e_-=(0,1)$.
Depending if the shift is positive (capture)  or negative (emission) we distinguish raising $\bf E$ and lowering $\bf F$ operators analogously to Borel positive and negative parts of $\myY({\fg\fl}_{1|1})$.
Practically these generators are constructed according to the following recipe \cite{Galakhov:2020vyb}(cf.~\cite{nakajima1999lectures,Rapcak:2018nsl}).
Let us denote $\mathscr{M}=\mathscr{M}(\vec d)$ and $\mathscr{M}'=\mathscr{M}(\vec d+\vec e_{\pm})$.
Consider a surface $\CI\subset \mathscr{M}\times \mathscr{M}'$ defined by a condition that there is a homomorphism of quiver representations $\zeta:\,\mathscr{M}'\to \mathscr{M}$.
Fixed points on $\CI$ are labeled by pairs $\lambda$, $\lambda'$ such that $\lambda'=\lambda+a$ contains an additional node $a$ in comparison to $\lambda$.
The matrix coefficients of the raising/lowering operators of the BPS algebra are calculated with the help of equivariant integration accordingly:
\begin{equation}\label{BPS_EF}
	{\bf E}_{\lambda}(a):=\frac{{\rm Euler}\left(\mathsf{T}_{\lambda}\mathscr{M}\right)}{{\rm Euler}\left(\mathsf{T}_{\lambda,\lambda'}\CI\right)},\quad {\bf F}_{\lambda}(a):=\frac{{\rm Euler}\left(\mathsf{T}_{\lambda'}\mathscr{M}'\right)}{{\rm Euler}\left(\mathsf{T}_{\lambda,\lambda'}\CI\right)}\,.
\end{equation}

\subsection{Explicit formulae for matrix coefficients and Euler class}

We find that the Euler class of the tangent space to a fixed point given by partition $\lambda$ coincides with the hook measure \eqref{hook} and could be represented in an integral form (similarly to \cite{Moore:1997dj, Moore:1998et, Nekrasov:2002qd, Nekrasov:2017cih, Nekrasov:2018xsb}), or as a residue of an integral measure:
\begin{equation}\label{Euler}
	{\bf e}_{\lambda}=\fm_{\lambda}=\left(\mathop{\rm res}_{\lambda}{\rm Mes}(\vec x,\vec y)^{-1}\right)^{-1}\,,
\end{equation}
where
\begin{equation}\label{Eul}
	\begin{split}
		{\rm Mes}(\vec x,\vec y)&=\frac{{\rm weight}(\mbox{degrees of freedom})}{{\rm weight}(\mbox{constraints})\times {\rm weight}(\mbox{gauge})}=\\
		&=\frac{\prod\lm_{i=1}^{d_1}\prod\lm_{j=1}^{d_2}(y_j-x_i-h_1)(x_i-y_j-h_2)(x_i-y_j+h_2)\times \prod\lm_{i=1}^{d_1}x_i\times\prod\lm_{j=1}^{d_2}(-y_j-h_1)}{\prod\lm_{i,i'=1}^{d_1}(x_i-x_{i'})\times \prod\lm_{j,j'=1}^{d_2}(y_j-y_{j'})\times \prod\lm_{i=1}^{d_1}\prod\lm_{j=1}^{d_2}(x_i-y_j-h_1)}\,.
	\end{split}
\end{equation}
and $d_{1,2}$ are quiver dimensions given by $n_{\lambda}^{\pm}$ respectively.
To calculate the residue in \eqref{Euler} we order nodes of crystal $\lambda$ according to the length of a path in the crystal connecting the node in consideration with the node located at $(0,0)$.
Then we change variables $x_{i(a)}\to \omega_a+t_a$ if node $a$ is of $(+)$-type, or $y_{j(a)}\to \omega_a+t_a$ if $a$ is of $(-)$-type.
Then we define:
\begin{equation}
	\mathop{\rm res}\lm_{\lambda}\,:=\,\ldots\mathop{\rm res}\lm_{t_4=0}\mathop{\rm res}\lm_{t_3=0}\mathop{\rm res}\lm_{t_2=0}\,\mathop{\rm res}\lm_{t_1=0}\,.
\end{equation}

The Euler class imposes a suitable norm on the semi-Fock representation:
\begin{equation}\label{geom_norm}
	\langle\lambda,\lambda'\rangle=\delta_{\lambda,\lambda'}{\bf e}_{\lambda}\,,
\end{equation}
so that operators $e_k^{\pm}$ and $f_k^{\pm}$ are conjugate to each other with respect to this norm.

We derive the following formulae for matrix coefficients of the semi-Fock representation of $\myY(\widehat{\fg\fl}_{1|1})$ (cf. \cite{Prochazka:2015deb}):
\begin{equation}\label{matrix_el}
	\begin{split}
		&e^{\pm}(z)|\lambda\rangle=\sum\lm_{k=0}^{\infty}\frac{e_k^{\pm}}{z^{k+1}}|\lambda\rangle=\sum\lm_{a\in{\rm Add}\,\lambda^{\pm}}\frac{{\bf E}_{\lambda}(a)}{z-\omega_{\sqbox{$a$}}}|\lambda+a\rangle\,,\\
		&f^{\pm}(z)|\lambda\rangle=\sum\lm_{k=0}^{\infty}\frac{f_k^{\pm}}{z^{k+1}}|\lambda\rangle=\sum\lm_{a\in{\rm Rem}\,\lambda^{\pm}}\frac{{\bf F}_{\lambda-a}(a)}{z-\omega_{\sqbox{$a$}}}|\lambda-a\rangle\,,\\
		&\psi^{\pm}(z)|\lambda\rangle=\sum\lm_{k}\frac{\psi^{\pm}}{z^{k+1}}|\lambda\rangle=\psi_{\lambda}^{a}(z)|\lambda\rangle\,,
	\end{split}
\end{equation}
where ${\rm Add}\,\lambda^{\pm}$ (${\rm Rem}\,\lambda^{\pm}$) are sets of nodes of a respective type $(+)$ or $(-)$ in $\CG$ such that they could be added to (removed from) $\lambda$ and a new set of nodes $\lambda\pm a$ is again a molten crystal.

Matrix coefficients and eigenvalues of operators are derived using \eqref{BPS_EF} and are given by the following combinatorial expressions (cf.~\cite{FeiginTsymbaliuk}):
\begin{equation}\label{mat_coeff}
	{\bf E}_{\lambda}(a)=A_a\times\prod\lm_{b\in\lambda}\eta_{\bar a,\bar b}(\omega_a-\omega_b),\quad {\bf F}_{\lambda}(a)=B_a\times\prod\lm_{b\in\lambda}\xi_{\bar a,\bar b}(\omega_a-\omega_b),\quad \psi^{\pm}_{\lambda}(z)=\psi^{\pm}_{\varnothing}(z)\times\prod\lm_{b\in\lambda}\varphi_{\pm,\bar b}(z-\omega_b)\,,
\end{equation}
where $\bar c$ denotes a node $c$ type, and coupling potentials are defined as:\footnote{We should warn the reader that bare generators derived using the BPS or Cohomological Hall algebra methods in the super case may not have a definite parity. One should switch signs of some matrix elements in a prescribed order \cite{Galakhov:2020vyb}. Here we simply switched the signs of $\xi_{ab}(z)$ form the equivariant calculation.}
\begin{equation}
	\begin{split}
		&\begin{array}{ll}
			\eta_{++}(z)=\eta_{--}(z)=z\,; & \xi_{++}(z)=\xi_{--}(z)=\dfrac{1}{z}\,;\\
			\eta_{+-}(z)=\left\{\begin{array}{ll}
				\dfrac{z-h_1}{2z(z+h_1)}, & \mbox{if }z=\pm h_2\,;\\
				\dfrac{1}{2z}, & \mbox{if }z=h_1\,;\\
				\dfrac{z-h_1}{z^2-h_2^2}, & \mbox{otherwise ;}
			\end{array}\right. &\xi_{+-}(z)=z+h_1\,;\\
			\eta_{-+}(z)=\left\{\begin{array}{ll}
				1, & \mbox{if }z=h_1\,;\\
				\dfrac{1}{z-h_1}, & \mbox{otherwise ;}
			\end{array}\right. & \xi_{-+}(z)=\dfrac{z^2-h_2^2}{z+h_1}\,.
		\end{array}
	\end{split}
\end{equation}
\begin{equation}
	\varphi_{\bar a,\bar b}(z)=-\frac{\eta_{\bar a,\bar b}(z)}{\eta_{\bar b,\bar a}(-z)}=-\frac{\xi_{\bar a,\bar b}(z)}{\xi_{\bar b,\bar a}(-z)}=\eta_{\bar a,\bar b}(z)\xi_{\bar a,\bar b}(z)=\left\{\begin{array}{ll}
		1, & \mbox{ if }\bar a=\bar b\,;\\
		\dfrac{z^2-h_1^2}{z^2-h_2^2}, &\mbox{ if }(\bar a,\bar b)=(+-)\,;\\
		\dfrac{z^2-h_2^2}{z^2-h_1^2}, &\mbox{ if }(\bar a,\bar b)=(-+)\,.\\
	\end{array}\right.
\end{equation}
and
\begin{equation}
	\begin{split}
		&A_a=\left\{\begin{array}{ll}
			\dfrac{2}{|x_a|+|y_a|}, &\mbox{if } \bar a=+\mbox{ and }a\mbox{ lies on the cone boundary as I in \eqref{graph_G}}\,;\\
			\dfrac{2h_1}{\omega_{a}}, &\mbox{if } \bar a=+\mbox{ and }a\mbox{ lies inside the cone as II in \eqref{graph_G}}\,;\\
			1, &\mbox{if }\bar a=-\,;
		\end{array}\right.\\
		&B_a=\left\{\begin{array}{ll}
			1, & \mbox{if }\bar a=+\,;\\
			-\omega_{a}-h_1, &\mbox{if }\bar a=-\,;
		\end{array}\right.\\
		&\psi^+_{\varnothing}(z) = \frac{1}{z},\quad \psi_{\varnothing}^-(z)=-z-h_1\,.
	\end{split}
\end{equation}

\subsection{Example of calculations}

In this subsection we present an explicit pedagogical calculation routine for matrix element ${\bf E}_{\lambda}(a)$ with:
\begin{equation}\label{examp_diags}
	\lambda=\begin{array}{c}
		\begin{tikzpicture}[scale=0.4]
			\foreach \i/\j in {0/-1, 0/0, 1/-1, 1/0}
			{
				\draw (\i,\j) -- (\i+1,\j);
			}
			\foreach \i/\j in {0/0, 1/0, 2/0}
			{
				\draw (\i,\j) -- (\i,\j-1);
			}
		\end{tikzpicture}
	\end{array}=\begin{array}{c}
		\begin{tikzpicture}[scale=0.7]
			\draw[thick] (0,0) -- (1,-1) (1,-1) -- (2,0) (2,0) -- (3,-1);
			\draw[fill=gray] (0,0) circle (0.25) (2,0) circle (0.25);
			\draw[fill=white!40!blue] (1,-1) circle (0.25) (3,-1) circle (0.25);
			\node[above] at (0,0.2) {$\scriptstyle 1_+$};
			\node[above] at (2,0.2) {$\scriptstyle 2_+$};
			\node[below] at (1,-1.2) {$\scriptstyle 1_-$};
			\node[below] at (3,-1.2) {$\scriptstyle 2_-$};
		\end{tikzpicture}
	\end{array}\quad\overset{{\bf E}_{\lambda}(a)}{\longrightarrow}\quad\lambda'=\begin{array}{c}
		\begin{tikzpicture}[scale=0.4]
			\foreach \i/\j in {0/-1, 0/0, 1/-1, 1/0}
			{
				\draw (\i,\j) -- (\i+1,\j);
			}
			\foreach \i/\j in {0/-1, 0/0, 1/0, 2/0}
			{
				\draw (\i,\j) -- (\i,\j-1);
			}
			\foreach \i/\j in {1/-1}
			{
				\draw (\i,\j) -- (\i-1,\j-1);
			}
			\node at (0.5,-0.5) {$\scriptstyle 1$};
			\node at (1.5,-0.5) {$\scriptstyle 2$};
		\end{tikzpicture}
	\end{array}=\begin{array}{c}
		\begin{tikzpicture}[scale=0.7]
			\draw[thick] (0,0) -- (1,-1) node[pos=0.5,below left=-0.2]{$\scriptstyle A_1$} (1,-1) -- (2,0) node[pos=0.5,above left=-0.2]{$\scriptstyle B_1$} (2,0) -- (3,-1) node[pos=0.5,above right=-0.2]{$\scriptstyle A_1$} (1,-1) -- (0,-2) node[pos=0.5,below right=-0.2]{$\scriptstyle B_2$};
			\draw[fill=gray] (0,0) circle (0.2) (2,0) circle (0.2) (0,-2) circle (0.2);
			\draw[fill=white!40!blue] (1,-1) circle (0.2) (3,-1) circle (0.2);
			\node[above] at (0,0.2) {$\scriptstyle 1_+$};
			\node[above] at (2,0.2) {$\scriptstyle 2_+$};
			\node[below right] at (1.1,-1.1) {$\scriptstyle 1_-$};
			\node[below right] at (3.1,-1.1) {$\scriptstyle 2_-$};
			\node[left] at (-0.2,-2) {$\scriptstyle a=3_+$};
			\draw[thick, black!40!red] (0,-2) circle (0.4);
		\end{tikzpicture}
	\end{array}
\end{equation}
In this process we add node $a$ marked by the red circle of type $(+)$.
In what follows we construct explicitly a tangent space to fixed point $\lambda'$, and the tangent space to fixed point $\lambda$ is constructed by a complete analogy.
We enumerate all the nodes according to \eqref{examp_diags}.
And construct vector spaces of the quiver representation:
\begin{equation}
	V_+'={\rm Span}\left\{|1_+\rangle,|2_+\rangle,|3_+\rangle\right\},\quad 	V_-'={\rm Span}\left\{|1_-\rangle,|2_-\rangle\right\}
\end{equation}

Vacuum fields are calculated according to the following rule: $I$ is a vector pointing towards the root atom, $J=0$, $A_2=0$ as the Lagrange multiplier, the rest of the fields acquire \emph{non-zero} vevs corresponding to the links of the crystal diagram:
\begin{equation}
	A_1'=\left(
	\begin{array}{ccc}
		1 & 0 & 0 \\
		0 & 1 & 0 \\
	\end{array}
	\right),\;B_1'=\left(
	\begin{array}{cc}
		0 & 0 \\
		1 & 0 \\
		0 & 0 \\
	\end{array}
	\right),\;B_2'=\left(
	\begin{array}{cc}
		0 & 0 \\
		0 & 0 \\
		1 & 0 \\
	\end{array}
	\right),\;I'=\left(
	\begin{array}{c}
		1 \\
		0 \\
		0 \\
	\end{array}
	\right),\;J'=\left(
	\begin{array}{cc}
		0 & 0 \\
	\end{array}
	\right)\,.
\end{equation}
This representation is a fixed point with respect to the action of the vector field induced by the complexified gauge and flavor symmetries:
\begin{equation}\label{vector}
	\mathscr{V}=\sum\lm_{a\to b\in\{{\rm quiver \; arrows}\}}\left(\Phi_bX_{a\to b}-X_{a\to b}\Phi_a-{\rm weight}(a\to b)\cdot X_{a\to b}\right)\frac{\p}{\p X_{a\to b}}\,.
\end{equation}
where expectation values of the complexified gauge fields are defined by node weights according to the following rule:
\begin{equation}
	V={\rm Span}\left\{|a\rangle,|b\rangle,|c\rangle,\ldots\right\}\;\Rightarrow\;\Phi_V={\rm diag}\left(\omega_a,\omega_b,\omega_c,\ldots\right)\,.
\end{equation}
In our case we have:
\begin{equation}
	\Phi_+={\rm diag}\left(0,\,h_1+h_2,\,h_1-h_2\right),\quad \Phi_-={\rm diag}\left(h_1,2h_1+h_2\right)\,.
\end{equation}

Generically the tangent space is parameterized by infinitesimal field deviations from the vacuum values:
\begin{equation}
	\begin{split}
		\delta A_1'=\left(
		\begin{array}{ccc}
			\alpha_{1;11}' & \alpha_{1;12}' & \alpha_{1;13}' \\
			\alpha_{1;21}' & \alpha_{1;22}' & \alpha_{1;23}' \\
		\end{array}
		\right),\quad \delta B_1'=\left(
		\begin{array}{cc}
			\beta_{1;11}' & \beta_{1;12}' \\
			\beta_{1;21}' & \beta_{1;22}' \\
			\beta_{1;31}' & \beta_{1;32}' \\
		\end{array}
		\right)\,,\\ \delta B_2'=\left(
		\begin{array}{cc}
			\beta_{2;11}' & \beta_{2;12}' \\
			\beta_{2;21}' & \beta_{2;22}' \\
			\beta_{2;31}' & \beta_{2;32}' \\
		\end{array}
		\right),\quad \delta I'=\left(
		\begin{array}{c}
			\iota_{1}' \\
			\iota_{2}' \\
			\iota_{3}' \\
		\end{array}
		\right),\quad \delta J'=\left(
		\begin{array}{cc}
			\upsilon_{1}' & \upsilon_{2}' \\
		\end{array}
		\right)
	\end{split}
\end{equation}
We would like to parameterize the infinitesimal scale by $t$ and work up to $O(t^2)$ in what follows.

The action of the complexified gauge groups is linearized to the action of the corresponding Lie algebra on the tangent space:
\begin{equation}
	\mbox{for }\fg_a=((g_{ij}^a))\in\fg\fl(d_a,\IC):\quad\delta X_{a\to b}\to \delta X_{a\to b}+\fg_{b}X_{a\to b}-X_{a\to b}\fg_a\,.
\end{equation}

For example, for $\delta A_1$ we have:
\begin{equation}
	\left(
	\begin{array}{ccc}
		\alpha_{1;11}' & \alpha_{1;12}' & \alpha_{1;13}' \\
		\alpha_{1;21}' & \alpha_{1;22}' & \alpha_{1;23}' \\
	\end{array}
	\right)\to\left(
	\begin{array}{ccc}
		\alpha_{1;11}' & \alpha_{1;12}' & \alpha_{1;13}' \\
		\alpha_{1;21}' & \alpha_{1;22}' & \alpha_{1;23}' \\
	\end{array}
	\right)+\left(
	\begin{array}{ccc}
		g_{11}^--g_{11}^+ & g_{12}^--g_{12}^+ & -g_{13}^+ \\
		g_{21}^--g_{21}^+ & g_{22}^--g_{22}^+ & -g_{2,3}^+ \\
	\end{array}
	\right)\,.
\end{equation}
To derive the gauge-invariant moduli of the quiver representation we simply subtract from the vector space spanned by all the degrees of freedom its vector subspace spanned by the gauge degrees of freedom.
As a result the gauge invariant parameterization reads:
\begin{equation}
	\begin{split}
		\delta A_1'=\left(
		\begin{array}{ccc}
			0 & 0 & 0 \\
			0 & 0 & 0 \\
		\end{array}
		\right),\;\delta B_1'=\left(
		\begin{array}{cc}
			0 & \beta _{1;12}' \\
			0 & \beta _{1;22}'\\
			0 & \beta _{1;32}' \\
		\end{array}
		\right),\;\delta B_2'=\left(
		\begin{array}{cc}
			\beta _{2;11}' & \beta _{2;12}' \\
			\beta _{2;21}' & \beta _{2;22}' \\
			0 & \beta _{2;32}' \\
		\end{array}
		\right),\;\delta I'=\left(
		\begin{array}{c}
			0 \\
			0 \\
			0 \\
		\end{array}
		\right),\delta J'=\left(
		\begin{array}{cc}
			\upsilon _{1}' & \upsilon _{2}' \\
		\end{array}
		\right)
	\end{split}
\end{equation}
We project these degrees of freedom on hypersurface \eqref{quiver_variety}:
\begin{equation}
	\left(B_2+t\,\delta B_2\right)\left(A_1+t\,\delta A_1\right)\left(B_1+t\,\delta B_1\right)-\left(B_1+t\,\delta B_1\right)\left(A_1+t\,\delta A_1\right)\left(B_2+t\,\delta B_2\right)+\left(I+t\,\delta I\right)\left(J+t\,\delta J\right)=O(t^2)\,.
\end{equation}
That results in a set of relations:
\begin{equation}
	\beta _{2;12}'+\upsilon _{1}'=0,\;\upsilon_{2}'=0,\;\beta_{2;22}'-\beta _{2;11}'=0,\;-\beta _{2;12}'=0,\;\beta _{2;32}'=0,\;\beta _{1;12}'=0\,.
\end{equation}
That further restricts our tangent space $\mathsf{T}_{\lambda'}\mathscr{M}'$ to the following form:
\begin{equation}
	\begin{split}
		\delta A_1'=\left(
		\begin{array}{ccc}
			0 & 0 & 0 \\
			0 & 0 & 0 \\
		\end{array}
		\right),\; \delta B_1'=\left(
		\begin{array}{cc}
			0 & 0 \\
			0 & \beta _{1;22}' \\
			0 & \beta _{1;32}' \\
		\end{array}
		\right),\; 
		\delta B_2'=\left(
		\begin{array}{cc}
			\beta _{2;22}' & 0 \\
			\beta _{2;21}' & \beta _{2;22}' \\
			0 & 0 \\
		\end{array}
		\right),\; \delta I'=\left(
		\begin{array}{c}
			0 \\
			0 \\
			0 \\
		\end{array}
		\right),\; \delta J'=\left(
		\begin{array}{cc}
			0 & 0 \\
		\end{array}
		\right)\,.
	\end{split}
\end{equation}
The equivariant weights of these degrees of freedom are defined by the $\mathscr{V}$-action:
\begin{equation}
	w\left(\beta_{1;22}\right)=-h_1-h_2,\quad w\left(\beta_{1;32}'\right)=-h_1-3h_2,\quad \left(\beta_{2;21}'\right)=2 h_2,\quad w\left(\beta _{2;22}'\right)=h_2-h_1\,.
\end{equation}
The resulting Euler class reads:
\begin{equation}
	\begin{split}
		&{\bf e}_{\lambda'}=-2 \left(h_1-h_2\right) h_2 \left(h_1+h_2\right) \left(h_1+3 h_2\right)=\\
		&=-\epsilon _1 \left(\epsilon _1-2 \epsilon _2\right) \left(\epsilon _1-\epsilon _2\right) \epsilon _2=\underbrace{\left(-\epsilon_1+\epsilon_2\right)\left(\epsilon_1-2\epsilon_2\right)}_{\upsilon_{\lambda'}(1)} \times \underbrace{\left(-\epsilon_1\right)\left(-\epsilon_2\right)}_{\upsilon_{\lambda'}(2)}\,,
	\end{split}
\end{equation}
where we split the last equality in a product of two hook contributions \eqref{hook} with hook corners located in tiles marked in \eqref{examp_diags} correspondingly.

Similarly, for fixed point $\lambda$ on $\mathscr{M}$ we have:
\begin{equation}
	\begin{split}
		&A_1=\left(
		\begin{array}{cc}
			1 & 0 \\
			0 & 1 \\
		\end{array}
		\right), \quad B_1=\left(
		\begin{array}{cc}
			0 & 0 \\
			1 & 0 \\
		\end{array}
		\right),\quad B_2=\left(
		\begin{array}{cc}
			0 & 0 \\
			0 & 0 \\
		\end{array}
		\right),\quad I=\left(
		\begin{array}{c}
			1 \\
			0 \\
		\end{array}
		\right),\quad J=\left(
		\begin{array}{cc}
			0 & 0 \\
		\end{array}
		\right)\,,\\
		&\delta A_1=\left(
		\begin{array}{cc}
			0 & 0 \\
			0 & 0 \\
		\end{array}
		\right),\;\delta B_1=\left(
		\begin{array}{cc}
			0 & \beta _{1;12} \\
			0 & \beta _{1;22} \\
		\end{array}
		\right),\;\delta B_2=\left(
		\begin{array}{cc}
			\beta _{2;22} & 0 \\
			\beta _{2;21} & \beta _{2;22} \\
		\end{array}
		\right),\;\delta I=\left(
		\begin{array}{c}
			0 \\
			0 \\
		\end{array}
		\right),\;\delta J=\left(
		\begin{array}{cc}
			0 & 0 \\
		\end{array}
		\right)\,,\\
		& w\left(\beta _{1;12}\right)=-2 h_1-2 h_2,\quad w\left(\beta _{1;22}\right)=-h_1-h_2,\quad w\left(\beta_{2;21}\right)=2 h_2,\quad w\left(\beta _{2;22}\right)=h_2-h_1\,.
	\end{split}
\end{equation}

Finally, the homomorphism of a quiver representation is a singular gauge map acting in quiver nodes so that the following diagram commutes (in our case up to $O(t^2)$):
\begin{equation}
	\begin{array}{c}
		\begin{tikzpicture}
			\node (A) at (0,0) {$V_a'$};
			\node (B) at (4,0) {$V_b'$};
			\node (C) at (0,-1.5) {$V_a$};
			\node (D) at (4,-1.5) {$V_b$};
			\path (A) edge[->] node[above] {$\scriptstyle X_{a\to b}'+t\,\delta X_{a\to b}'$}(B) (C) edge[->] node[above] {$\scriptstyle X_{a\to b}+t\,\delta X_{a\to b}$}(D) (A) edge[->] node[left] {$\scriptstyle \tau_a$}(C) (B) edge[->] node[right] {$\scriptstyle \tau_b$}(D);
		\end{tikzpicture}
	\end{array}
\end{equation}

In our case we derive:
\begin{equation}
	\tau_+=\left(
	\begin{array}{ccc}
		1 & 0 & 0 \\
		0 & 1 & 0 \\
	\end{array}
	\right),\quad\tau_-=\left(
	\begin{array}{cc}
		1 & 0 \\
		0 & 1 \\
	\end{array}
	\right)\,,
\end{equation}
and acquire an additional constraint on cases when such a homomorphism exists:
\begin{equation}
	\beta _{1;12}=0,\quad \beta_{1;22}-\beta _{1;22}'=0,\quad \beta _{2;21}-\beta_{2;21}'=0,\quad \beta _{2;22}-\beta _{2;22}'=0\,.
\end{equation}
By resolving those constraints in terms of $\beta$ and $\beta'$ variables we find the tangent space to the incidence locus:
\begin{equation}
	\mathsf{T}_{\lambda,\lambda'}\CI={\rm Span}\,\left\{\beta _{1;22}',\beta _{1;32}',\beta _{2;21}',\beta _{2;22}'\right\}\,.
\end{equation}
Thus for the matrix coefficient we have (cf.~\eqref{mat_coeff}):
\begin{equation}
	\begin{split}
		&{\bf E}_{\lambda}(a)=\frac{2 \left(h_1+h_2\right)}{h_1+3h_2}=\\
		&=\underbrace{1}_{A(3_+)}\times \underbrace{(h_1-h_2)}_{\eta_{++}(\omega_{3_+}-\omega_{1_+})}\times \underbrace{(-2h_2)}_{\eta_{++}(\omega_{3_+}-\omega_{2_+})}\times \underbrace{\frac{-h_2-h_1}{2(-h_2)(-h_2+h_1)}}_{\eta_{+-}(\omega_{3_+}-\omega_{1_-})}\times \underbrace{\frac{-2h_2-2h_1}{(-3h_2-h_1)(-h_2-h_1)}}_{\eta_{+-}(\omega_{3_+}-\omega_{2_-})}\,.
	\end{split}
\end{equation}

\subsection{Towards stable envelopes for \texorpdfstring{$\myY(\widehat{\fg\fl}_{1|1})$}{Ygl11}}\label{app:stable}

In this subsection we would like to perform a slight elaboration of scheme \eqref{prog} in applications to integrability.
Integrability is an intriguing subject especially in the light of its relation to quantum field theories and non-perturbative phenomena (see e.g. \cite{Nekrasov:2009ui}).
Yangian algebras are a natural source of non-trivial solutions to the Yang-Baxter equations -- $R$-matrices.
The resulting $R$-matrix is a function of 4 indices labeling vectors of representation and a complex spectral parameter.

In the context of the Yangian $R$-matrix is an intertwining operator for a \emph{non-diagonal} co-product structure (see \cite[sec.3]{Galakhov:2022uyu} for details and relations between naive and non-diagonal co-product structures):
\begin{equation}
	R_{12}\Delta_{12}=\Delta_{21}R_{12}\,.
\end{equation}
A difficulty of this construction form the side of a D-brane system description is that instead of the semi-Fock representations, that correspond to a shifted Yangian, and we have concentrated in this note on, one should use unshifted representations.
And the simplest unshifted representation is the Fock one.
Switching a representation modifies quiver framing and superpotential:
\begin{equation}
	\begin{array}{c}
		\begin{tikzpicture}
			\tikzstyle{col0} = [fill=white]
			\tikzstyle{col1} = [fill=gray]
			\tikzstyle{col2} = [fill=white!40!blue]
			\draw[postaction={decorate},
			decoration={markings, mark= at position 0.55 with {\arrow{stealth}}}]
			(0,-2) -- (0,0) node[pos=0.5,below left] {$\scriptstyle I_1$};
			\draw[postaction={decorate},
			decoration={markings, mark= at position 0.75 with {\arrow{stealth}}}]
			(2,0) -- (0,-2) node[pos=0.7,above left] {$\scriptstyle J_1$};
			\draw[postaction={decorate},
			decoration={markings, mark= at position 0.75 with {\arrow{stealth}}}]
			(0,0) -- (2,-2) node[pos=0.7,above right] {$\scriptstyle J_2$};
			\draw[postaction={decorate},
			decoration={markings, mark= at position 0.55 with {\arrow{stealth}}}]
			(2,-2) -- (2,0) node[pos=0.5,below right] {$\scriptstyle I_2$};
			\draw[postaction={decorate},
			decoration={markings, mark= at position 0.65 with {\arrow{stealth}}, mark= at position 0.45 with {\arrow{stealth}}}] (0,0) to[out=30,in=150] node[pos=0.5, above] {$\scriptstyle A_1, A_2$} (2,0);
			\draw[postaction={decorate},
			decoration={markings, mark= at position 0.65 with {\arrow{stealth}}, mark= at position 0.45 with {\arrow{stealth}}}] (2,0) to[out=210,in=330] node[pos=0.5, below] {$\scriptstyle B_1, B_2$} (0,0);
			\draw[col1] (0,0) circle (0.1);
			\draw[col2] (2,0) circle (0.1);
			\node[left] at (-0.1,0) {$1$};
			\node[right] at (2.1,0) {$2$};
			\begin{scope}[shift={(0,-2)}]
				\draw[col0] (-0.08,-0.08) -- (-0.08,0.08) -- (0.08,0.08) -- (0.08,-0.08) -- cycle;
			\end{scope}
			\begin{scope}[shift={(2,-2)}]
				\draw[col0] (-0.08,-0.08) -- (-0.08,0.08) -- (0.08,0.08) -- (0.08,-0.08) -- cycle;
			\end{scope}
		\end{tikzpicture}
	\end{array},\quad W=\Tr\left(A_1B_1A_2B_2-A_1B_2A_2B_1+ A_2I_1J_1+B_2I_2J_2\right)\,,
\end{equation}
so that the resulting quiver variety is not smooth anymore.
It is natural to expect that in this case there are no closed combinatorial expressions for the measure \eqref{Euler} or matrix coefficients.
The argument is purely dimensional.
Since all the variables in \eqref{Euler} and weights have a dimension of the mass, a closed combinatorial expression for the Euler class would allow one to derive a fixed expression for the dimension of the tangent space for given quiver dimensions $d_\pm$.
However we could observe just in a few examples that the dimension of the tangent space of a singular quiver variety jumps from one fixed point to another even for $d_{\pm}$ fixed.
Yet we expect to acquire some closed expressions using the algebraic approach discussed in sec.\ref{sec:sF}.
We will implement this strategy elsewhere.

Another problem related to a canonical construction of the co-product implemented in purely algebraic terms \cite{2019arXiv191106666U, Bao:2022jhy} and with the help of CFT $\beta\gamma$-system \cite{Kolyaskin:2022tqi} is that both require a rather involved construction of higher raising/lowering operators $e_{\alpha,k}$/$f_{\alpha,k}$ depending on a non-simple affine root and an extra mode number.
In the case of CFT system these operators are required to construct a map from the Yangian to the conformal or a $W$-algebra \cite{Bao:2022jhy}.
In principle, one needs the co-product only for the generating basis:
\begin{equation}
	\begin{split}
		&\Delta(e_0^{\pm})=e_0^{\pm}\otimes 1+1\otimes e_0^{\pm},\quad 	\Delta(f_0^{\pm})=f_0^{\pm}\otimes 1+1\otimes f_0^{\pm}\,,\\
		&\Delta(\psi_2^{\pm})=\psi_2^{\pm}\otimes 1+\psi_1^{\pm}\otimes\psi_0^{\pm}+\psi_0^{\pm}\otimes\psi_1^{\pm}+1\otimes \psi_2^{\pm}+\CJ^{\pm}
	\end{split}
\end{equation}
the rest of generators can be reconstructed via the co-product homomorphism.
Here $\CJ^{\pm}$ are non-trivial corrections to a diagonal co-product expected to have the following form:
\begin{equation}
	\CJ^{\pm}\sim\sum \left(f_{\alpha_1}\cdot f_{\alpha_2}\cdot\ldots\right)\otimes\left(e_{\beta_1}\cdot e_{\beta_2}\cdot\ldots\right)\,.
\end{equation}
And expectations of \cite{Bao:2022jhy} suggest that this higher order correction factorizes as $\sum\lm_{\alpha,k}f_{\alpha,k}\otimes e_{\alpha,k}$, however our calculation of the correction second level:
\begin{equation}
	\begin{split}
		&\CJ^{\pm}=\mp2(h_1^2-h_2^2)f_0^{\mp}\otimes e_0^{\mp}-\\
		&-2\bigg(\left\{f_0^{\pm},f_0^{\mp}\right\}\otimes \left\{e_0^{\pm},e_1^{\mp}\right\}+\left\{f_0^{\pm},f_1^{\mp}\right\}\otimes \left\{e_0^{\pm},e_0^{\mp}\right\}-\\
		&-\left\{f_0^{\pm},f_0^{\mp}\right\}\otimes \left\{e_1^{\mp},e_0^{\pm}\right\}-\left\{f_1^{(a)},f_0^{(b)}\right\}\otimes \left\{e_0^{\mp},e_0^{\pm}\right\}\bigg)+\ldots\,.
	\end{split}
\end{equation}
reveals no apparent factorization.
This might indicate that the case of $\myY(\widehat{\fg\fl}_{1|1})$ has certain stand alone hidden complications.

An alternative approach rather popular in the modern literature to a construction of a solution to the Yang-Baxter equation intertwines TQFTs with enumerative geometry and sends us to the realm of stable envelopes.
The easiest way to notice the stable envelope entity is to work with the mirror dual frame of a quiver sigma-model -- a Landau-Ginzburg model of vortex disorder operators \cite{Hori:2013ika}.
In this case if one is aiming for a disk partition function of such a theory a choice of brane boundary conditions is in order, and a natural choice ensuring a convergence of the path integral is a set of Lagrangian submanifolds of Lefschetz thimbles.
The ``body'' of a Lefschetz thimble is generated from a fixed point associated with some diagram $\lambda\in\Lambda$ by a gradient flow with respect to a Morse height function $h={\rm Re}\,W$, where $W$ is an effective superpotetnial in the Landau-Ginzburg model.
This structure induces an ordering on diagrams $\Lambda$ by the value of $h$ in each $\lambda$, so that the gradient flow may go from $\lambda$ to $\lambda'$ only if $\lambda<\lambda'$.

In this setting a non-perturbative disk partition function is a wave function and has the following expansion:
\begin{equation}
	\Psi_{\lambda}=\sum\lm_{\lambda'\geq\lambda}e^{-S_{\rm soliton}(\lambda\to\lambda')+\ldots}\psi_{\lambda'}\,,
\end{equation}
where $\psi_{\lambda}$ is a perturbative Gaussian wave function in the neighborhood of vacuum $\lambda$, and the expansion coefficient is given by a Euclidean path integral around a gradient flow soliton trajectory connecting vacua $\lambda$ and $\lambda'$.
A potentially non-trivial form of a Lefschetz thimble that could intersect greater vacua forces us to introduce a transition matrix $\fS_{\lambda,\lambda'}$ between non-perturbative and perturbative bases we call a \emph{stable envelope}.

If one considers a chain of diagrams $\vec \lambda$ embedded in the weight space, the form of Lefschetz thimble may depend on some moduli of the problem, whereas the perturbative basis remains ignorant.
This allows one to define an $R$-matrix as a holonomy of the non-perturbative bases on a permutation of moduli and factorize it a product of stable envelopes:
\begin{equation}\label{R-factor}
	R_{\vec\lambda,\vec\lambda''}(12)=\fS_{\vec\lambda,\vec\lambda'}(12)\cdot\fS_{\vec\lambda',\vec\lambda''}(21)^{-1}\,.
\end{equation}

The construction of the stable envelopes in physical terms is rather involved.
And enumerative geometry saves the day by mapping supersymmetric wave functions into elements of supercharge cohomology.

Mathematically literature \cite{Okounkov:2015spn} defines of a stable envelope as a \emph{unique} function on fixed points $\lambda$ satisfying the following rules:
\begin{enumerate}
	\item $\fS_{\lambda,\lambda'}=0$ for all $\lambda>\lambda'$.
	\item $\fS_{\lambda,\lambda}={\bf e}_{\lambda}$
	\item A Nakajima quiver variety has a specific symplectic structure. An edge of the Dynkin diagram is resolved in a symplectic pair of counter-oriented morphisms with equivariant weights $w$ and $\hbar-w$ for some fixed $\hbar$. Then $\fS_{\lambda,\lambda'}$ is divisible by $\hbar$.
\end{enumerate}
Two first rules are natural physical requirements.
The first one follows from the gradient flow directness discussed above.
The second one is a natural normalization condition for a choice of D-brane boundary conditions in the gauged linear sigma model phase of the theory -- the wave functions become simply the Euler cohomology classes.
And the last rule restricts the application of this construction to Nakajima quiver varieties only.
Therefore a generalization to our case of non-Nakajima conifold resolution associated with $\myY(\widehat{\fg\fl}_{1|1})$ is especially intriguing.

If one performs a phenomenological attempt of analyzing the results known in the literature for $\myY(\widehat{\fg\fl}_n)$ \cite{dinkins2021elliptic} one finds a direct resemblance between the stable envelope expression and an Euler class measure expression like \eqref{Eul}.
If the class is represented as product of weights, then the stable envelope is a product of weights in certain degrees:
\begin{equation}\label{stab_pheno}
	{\bf e}_{\lambda}\sim\prod\lm_{w(\lambda)}w(\lambda),\quad \fS_{\lambda,\lambda'}\sim\prod\lm_{w(\lambda')}w(\lambda')^{\langle w(\lambda),w(\lambda')\rangle}\,,
\end{equation}
where pairing $\langle\star,\star\rangle$ is allowed to take values $0$ and $1$ and is a function of moduli.
We hope that our construction of super-Schur polynomials and hook formulae for the Euler class in the case of semi-Fock representations will allow one to produce similar expressions for Fock representations and guess a phenomenological expression \`a la \eqref{stab_pheno}.
So that a new $R$-matrix in the form of \eqref{R-factor} for $\myY(\widehat{\fg\fl}_{1|1})$ is constructed.

\bibliographystyle{utphys}
\bibliography{biblio}

\providecommand{\href}[2]{#2}\begingroup\raggedright\begin{thebibliography}{10}

\bibitem{Dolan:2004ps}
L.~Dolan, C.~R. Nappi, and E.~Witten,
  \href{http://dx.doi.org/10.1142/9789812702340_0036}{``{Yangian symmetry in D
  = 4 superconformal Yang-Mills theory},''} in {\em {3rd International
  Symposium on Quantum Theory and Symmetries}}, pp.~300--315.
\newblock 1, 2004.
\newblock \href{http://arxiv.org/abs/hep-th/0401243}{{\ttfamily
  arXiv:hep-th/0401243}}.

\bibitem{arkani-hamed_bourjaily_cachazo_goncharov_postnikov_trnka_2016}
N.~Arkani-Hamed, J.~Bourjaily, F.~Cachazo, A.~Goncharov, A.~Postnikov, and
  J.~Trnka, \href{http://dx.doi.org/10.1017/CBO9781316091548}{{\em Grassmannian
  Geometry of Scattering Amplitudes}}.
\newblock Cambridge University Press, 2016.

\bibitem{Rapcak:2018nsl}
M.~Rapcak, Y.~Soibelman, Y.~Yang, and G.~Zhao, ``{Cohomological Hall algebras,
  vertex algebras and instantons},''
  \href{http://dx.doi.org/10.1007/s00220-019-03575-5}{{\em Commun. Math. Phys.}
  {\bfseries 376} no.~3, (2019) 1803--1873},
  \href{http://arxiv.org/abs/1810.10402}{{\ttfamily arXiv:1810.10402
  [math.QA]}}.

\bibitem{Li:2020rij}
W.~Li and M.~Yamazaki, ``{Quiver Yangian from Crystal Melting},''
  \href{http://dx.doi.org/10.1007/JHEP11(2020)035}{{\em JHEP} {\bfseries 11}
  (2020) 035}, \href{http://arxiv.org/abs/2003.08909}{{\ttfamily
  arXiv:2003.08909 [hep-th]}}.

\bibitem{wang2020affine}
N.~Wang, ``Affine yangian and 3-schur functions,'' {\em Nuclear Physics B}
  {\bfseries 960} (2020) 115173.

\bibitem{cui2022jack}
Z.~Cui, Y.~Bai, N.~Wang, and K.~Wu, ``Jack polynomials and affine yangian,''
  {\em Nuclear Physics B} {\bfseries 984} (2022) 115986.

\bibitem{MT}
A.~Morozov and N.~Tselousov, ``{3-Schurs from explicit representation of
  Yangian $Y(\hat{\mathfrak{gl}}_1)$. Levels 1-5},''
  \href{http://arxiv.org/abs/2305.12282}{{\ttfamily arXiv:2305.12282
  [hep-th]}}.

\bibitem{Qschurs}
A.~Mironov and A.~Morozov, ``{Superintegrability of Kontsevich matrix model},''
  \href{http://dx.doi.org/10.1140/epjc/s10052-021-09030-x}{{\em Eur. Phys. J.
  C} {\bfseries 81} no.~3, (2021) 270},
  \href{http://arxiv.org/abs/2011.12917}{{\ttfamily arXiv:2011.12917
  [hep-th]}}.

\bibitem{Qschurs1}
A.~D. Mironov and A.~Morozov, ``{Generalized Q-functions for GKM},''
  \href{http://dx.doi.org/10.1016/j.physletb.2021.136474}{{\em Phys. Lett. B}
  {\bfseries 819} (2021) 136474},
  \href{http://arxiv.org/abs/2101.08759}{{\ttfamily arXiv:2101.08759
  [hep-th]}}.

\bibitem{Qschurs2}
A.~D. Mironov, A.~Y. Morozov, S.~M. Natanzon, and A.~Y. Orlov, ``{Around spin
  Hurwitz numbers},'' \href{http://dx.doi.org/10.1007/s11005-021-01457-3}{{\em
  Lett. Math. Phys.} {\bfseries 111} no.~5, (2021) 124},
  \href{http://arxiv.org/abs/2012.09847}{{\ttfamily arXiv:2012.09847
  [math-ph]}}.

\bibitem{MMZh}
A.~Mironov, A.~Morozov, and A.~Zhabin, ``{Spin Hurwitz theory and Miwa
  transform for the Schur Q-functions},''
  \href{http://dx.doi.org/10.1016/j.physletb.2022.137131}{{\em Phys. Lett. B}
  {\bfseries 829} (2022) 137131},
  \href{http://arxiv.org/abs/2111.05776}{{\ttfamily arXiv:2111.05776
  [hep-th]}}.

\bibitem{DIM1}
H.~Awata, H.~Kanno, A.~Mironov, A.~Morozov, K.~Suetake, and Y.~Zenkevich,
  ``{$(q,t)$-KZ equations for quantum toroidal algebra and Nekrasov partition
  functions on ALE spaces},''
  \href{http://dx.doi.org/10.1007/JHEP03(2018)192}{{\em JHEP} {\bfseries 03}
  (2018) 192}, \href{http://arxiv.org/abs/1712.08016}{{\ttfamily
  arXiv:1712.08016 [hep-th]}}.

\bibitem{DIM2}
H.~Awata, H.~Kanno, T.~Matsumoto, A.~Mironov, A.~Morozov, A.~Morozov,
  Y.~Ohkubo, and Y.~Zenkevich, ``{Explicit examples of DIM constraints for
  network matrix models},''
  \href{http://dx.doi.org/10.1007/JHEP07(2016)103}{{\em JHEP} {\bfseries 07}
  (2016) 103}, \href{http://arxiv.org/abs/1604.08366}{{\ttfamily
  arXiv:1604.08366 [hep-th]}}.

\bibitem{Morozov:2022ndt}
A.~Morozov and N.~Tselousov, ``{Hunt for 3-Schur polynomials},''
  \href{http://dx.doi.org/10.1016/j.physletb.2023.137887}{{\em Phys. Lett. B}
  {\bfseries 840} (2023) 137887},
  \href{http://arxiv.org/abs/2211.14956}{{\ttfamily arXiv:2211.14956
  [hep-th]}}.

\bibitem{MMN}
A.~Mironov, A.~Morozov, and S.~Natanzon, ``{Complete Set of Cut-and-Join
  Operators in Hurwitz-Kontsevich Theory},''
  \href{http://dx.doi.org/10.1007/s11232-011-0001-6}{{\em Theor. Math. Phys.}
  {\bfseries 166} (2011) 1--22},
  \href{http://arxiv.org/abs/0904.4227}{{\ttfamily arXiv:0904.4227 [hep-th]}}.

\bibitem{sergeev2005generalised}
A.~Sergeev and A.~Veselov, ``Generalised discriminants, deformed
  calogero--moser--sutherland operators and super-jack polynomials,'' {\em
  Advances in Mathematics} {\bfseries 192} no.~2, (2005) 341--375.

\bibitem{Galakhov:2021xum}
D.~Galakhov, W.~Li, and M.~Yamazaki, ``{Shifted quiver Yangians and
  representations from BPS crystals},''
  \href{http://dx.doi.org/10.1007/JHEP08(2021)146}{{\em JHEP} {\bfseries 08}
  (2021) 146}, \href{http://arxiv.org/abs/2106.01230}{{\ttfamily
  arXiv:2106.01230 [hep-th]}}.

\bibitem{Galakhov:2022uyu}
D.~Galakhov, W.~Li, and M.~Yamazaki, ``{Gauge/Bethe correspondence from quiver
  BPS algebras},'' \href{http://dx.doi.org/10.1007/JHEP11(2022)119}{{\em JHEP}
  {\bfseries 11} (2022) 119}, \href{http://arxiv.org/abs/2206.13340}{{\ttfamily
  arXiv:2206.13340 [hep-th]}}.

\bibitem{Kolyaskin:2022tqi}
D.~Kolyaskin, A.~Litvinov, and A.~Zhukov, ``{R-matrix formulation of affine
  Yangian of gl\textasciicircum{}(1|1)},''
  \href{http://dx.doi.org/10.1016/j.nuclphysb.2022.116023}{{\em Nucl. Phys. B}
  {\bfseries 985} (2022) 116023},
  \href{http://arxiv.org/abs/2206.01636}{{\ttfamily arXiv:2206.01636
  [hep-th]}}.

\bibitem{Noshita:2021dgj}
G.~Noshita and A.~Watanabe, ``{Shifted quiver quantum toroidal algebra and
  subcrystal representations},''
  \href{http://dx.doi.org/10.1007/JHEP05(2022)122}{{\em JHEP} {\bfseries 05}
  (2022) 122}, \href{http://arxiv.org/abs/2109.02045}{{\ttfamily
  arXiv:2109.02045 [hep-th]}}.

\bibitem{nakajima1996jack}
H.~Nakajima, ``Jack polynomials and hilbert schemes of points on surfaces,''
  1996.

\bibitem{Nekrasov:2002qd}
N.~A. Nekrasov, ``{Seiberg-Witten prepotential from instanton counting},''
  \href{http://dx.doi.org/10.4310/ATMP.2003.v7.n5.a4}{{\em Adv. Theor. Math.
  Phys.} {\bfseries 7} no.~5, (2003) 831--864},
  \href{http://arxiv.org/abs/hep-th/0206161}{{\ttfamily arXiv:hep-th/0206161}}.

\bibitem{Nekrasov:2003rj}
N.~Nekrasov and A.~Okounkov, ``{Seiberg-Witten theory and random partitions},''
  \href{http://dx.doi.org/10.1007/0-8176-4467-9_15}{{\em Prog. Math.}
  {\bfseries 244} (2006) 525--596},
  \href{http://arxiv.org/abs/hep-th/0306238}{{\ttfamily arXiv:hep-th/0306238}}.

\bibitem{Losev:2003py}
A.~S. Losev, A.~Marshakov, and N.~A. Nekrasov, ``{Small instantons, little
  strings and free fermions},'' in {\em {From Fields to Strings:
  Circumnavigating Theoretical Physics: A Conference in Tribute to Ian Kogan}},
  pp.~581--621.
\newblock 2, 2003.
\newblock \href{http://arxiv.org/abs/hep-th/0302191}{{\ttfamily
  arXiv:hep-th/0302191}}.

\bibitem{Rapcak:2020ueh}
M.~Rapcak, Y.~Soibelman, Y.~Yang, and G.~Zhao, ``{Cohomological Hall algebras
  and perverse coherent sheaves on toric Calabi-Yau 3-folds},''
  \href{http://arxiv.org/abs/2007.13365}{{\ttfamily arXiv:2007.13365
  [math.QA]}}.

\bibitem{MorozovCauchy}
A.~Morozov, ``{Cauchy formula and the character ring},''
  \href{http://dx.doi.org/10.1140/epjc/s10052-019-6598-6}{{\em Eur. Phys. J. C}
  {\bfseries 79} no.~1, (2019) 76},
  \href{http://arxiv.org/abs/1812.03853}{{\ttfamily arXiv:1812.03853
  [hep-th]}}.

\bibitem{Morozov:2009xk}
A.~Morozov and S.~Shakirov, ``{Generation of Matrix Models by W-operators},''
  \href{http://dx.doi.org/10.1088/1126-6708/2009/04/064}{{\em JHEP} {\bfseries
  04} (2009) 064}, \href{http://arxiv.org/abs/0902.2627}{{\ttfamily
  arXiv:0902.2627 [hep-th]}}.

\bibitem{Mironov:2022fsr}
A.~Mironov and A.~Morozov, ``{Superintegrability summary},''
  \href{http://dx.doi.org/10.1016/j.physletb.2022.137573}{{\em Phys. Lett. B}
  {\bfseries 835} (2022) 137573},
  \href{http://arxiv.org/abs/2201.12917}{{\ttfamily arXiv:2201.12917
  [hep-th]}}.

\bibitem{schiffmann2012cherednik}
O.~Schiffmann and E.~Vasserot, ``Cherednik algebras, w algebras and the
  equivariant cohomology of the moduli space of instantons on {$A^2$},'' 2012.

\bibitem{Smirnov:2021cyf}
A.~Smirnov, ``{Quantum differential and difference equations for
  $\mathrm{Hilb}^{n}(\mathbb{C}^2)$},''
  \href{http://arxiv.org/abs/2102.10726}{{\ttfamily arXiv:2102.10726
  [math.AG]}}.

\bibitem{MMMP12}
A.~Mironov, V.~Mishnyakov, A.~Morozov, and A.~Popolitov, ``{Commutative
  families in $W_\infty$, integrable many-body systems and hypergeometric
  $\tau$-functions},'' \href{http://arxiv.org/abs/2306.06623}{{\ttfamily
  arXiv:2306.06623 [hep-th]}}.

\bibitem{Mironov:2023wga}
A.~Mironov, V.~Mishnyakov, A.~Morozov, and A.~Popolitov, ``{Commutative
  subalgebras from Serre relations},''
  \href{http://arxiv.org/abs/2307.01048}{{\ttfamily arXiv:2307.01048
  [hep-th]}}.

\bibitem{Prochazka:2023zdb}
T.~Proch\'azka and A.~Watanabe, ``{On Bethe equations of 2d conformal field
  theory},'' \href{http://arxiv.org/abs/2301.05147}{{\ttfamily arXiv:2301.05147
  [hep-th]}}.

\bibitem{Gaiotto:2020dsq}
D.~Gaiotto and M.~Rapcak, ``{Miura operators, degenerate fields and the M2-M5
  intersection},'' \href{http://dx.doi.org/10.1007/JHEP01(2022)086}{{\em JHEP}
  {\bfseries 01} (2022) 086}, \href{http://arxiv.org/abs/2012.04118}{{\ttfamily
  arXiv:2012.04118 [hep-th]}}.

\bibitem{Okounkov:2015spn}
A.~Okounkov, ``{Lectures on K-theoretic computations in enumerative
  geometry},'' \href{http://arxiv.org/abs/1512.07363}{{\ttfamily
  arXiv:1512.07363 [math.AG]}}.

\bibitem{Bullimore:2017lwu}
M.~Bullimore, H.-C. Kim, and T.~Lukowski, ``{Expanding the Bethe/Gauge
  Dictionary},'' \href{http://dx.doi.org/10.1007/JHEP11(2017)055}{{\em JHEP}
  {\bfseries 11} (2017) 055}, \href{http://arxiv.org/abs/1708.00445}{{\ttfamily
  arXiv:1708.00445 [hep-th]}}.

\bibitem{Dedushenko:2022pem}
M.~Dedushenko, ``{Remarks on Berry Connection in QFT, Anomalies, and
  Applications},'' \href{http://arxiv.org/abs/2211.15680}{{\ttfamily
  arXiv:2211.15680 [hep-th]}}.

\bibitem{Galakhov:2020upa}
D.~Galakhov, ``{On supersymmetric interface defects, brane parallel transport,
  order-disorder transition and homological mirror symmetry},''
  \href{http://dx.doi.org/10.1007/JHEP10(2022)076}{{\em JHEP} {\bfseries 22}
  (2020) 076}, \href{http://arxiv.org/abs/2105.07602}{{\ttfamily
  arXiv:2105.07602 [hep-th]}}.

\bibitem{2012arXiv1211.1287M}
D.~{Maulik} and A.~{Okounkov}, ``{Quantum Groups and Quantum Cohomology},''
  \href{http://arxiv.org/abs/1211.1287}{{\ttfamily arXiv:1211.1287 [math.AG]}}.

\bibitem{Aganagic:2016jmx}
M.~Aganagic and A.~Okounkov, ``{Elliptic stable envelopes},''
  \href{http://dx.doi.org/10.1090/jams/954}{{\em J. Am. Math. Soc.} {\bfseries
  34} no.~1, (2021) 79--133}, \href{http://arxiv.org/abs/1604.00423}{{\ttfamily
  arXiv:1604.00423 [math.AG]}}.

\bibitem{Rimanyi:2019zyi}
R.~Rim\'anyi, A.~Smirnov, Z.~Zhou, and A.~Varchenko, ``{Three-Dimensional
  Mirror Symmetry and Elliptic Stable Envelopes},''
  \href{http://dx.doi.org/10.1093/imrn/rnaa389}{{\em Int. Math. Res. Not.}
  {\bfseries 2022} no.~13, (2022) 10016--10094},
  \href{http://arxiv.org/abs/1902.03677}{{\ttfamily arXiv:1902.03677
  [math.AG]}}.

\bibitem{Dedushenko:2021mds}
M.~Dedushenko and N.~Nekrasov, ``{Interfaces and Quantum Algebras, I: Stable
  Envelopes},'' \href{http://arxiv.org/abs/2109.10941}{{\ttfamily
  arXiv:2109.10941 [hep-th]}}.

\bibitem{Bullimore:2021rnr}
M.~Bullimore and D.~Zhang, ``{3d $\mathcal{N}=4$ Gauge Theories on an Elliptic
  Curve},'' \href{http://dx.doi.org/10.21468/SciPostPhys.13.1.005}{{\em SciPost
  Phys.} {\bfseries 13} no.~1, (2022) 005},
  \href{http://arxiv.org/abs/2109.10907}{{\ttfamily arXiv:2109.10907
  [hep-th]}}.

\bibitem{Dedushenko:2023qjq}
M.~Dedushenko and N.~Nekrasov, ``{Interfaces and Quantum Algebras, II: Cigar
  Partition Function},'' \href{http://arxiv.org/abs/2306.16434}{{\ttfamily
  arXiv:2306.16434 [hep-th]}}.

\bibitem{Crew:2023tky}
S.~Crew, D.~Zhang, and B.~Zhao, ``{Boundaries \& Localisation with a
  Topological Twist},'' \href{http://arxiv.org/abs/2306.16448}{{\ttfamily
  arXiv:2306.16448 [hep-th]}}.

\bibitem{Galakhov:2023aev}
D.~Galakhov, ``{BPS States Meet Generalized Cohomology},''
  \href{http://arxiv.org/abs/2303.05538}{{\ttfamily arXiv:2303.05538
  [hep-th]}}.

\bibitem{2009arXiv0905.0686G}
V.~{Ginzburg}, ``{Lectures on Nakajima's Quiver Varieties},''
  \href{http://arxiv.org/abs/0905.0686}{{\ttfamily arXiv:0905.0686 [math.RT]}}.

\bibitem{Smirnov:2013hh}
A.~Smirnov, ``{On the Instanton R-matrix},''
  \href{http://dx.doi.org/10.1007/s00220-016-2686-8}{{\em Commun. Math. Phys.}
  {\bfseries 345} no.~3, (2016) 703--740},
  \href{http://arxiv.org/abs/1302.0799}{{\ttfamily arXiv:1302.0799 [math.AG]}}.

\bibitem{Smirnov:2014npa}
A.~Smirnov, ``{Polynomials associated with fixed points on the instanton moduli
  space},'' \href{http://arxiv.org/abs/1404.5304}{{\ttfamily arXiv:1404.5304
  [math-ph]}}.

\bibitem{2019arXiv191106666U}
M.~{Ueda}, ``{Affine Super Yangian},''
  \href{http://arxiv.org/abs/1911.06666}{{\ttfamily arXiv:1911.06666
  [math.RT]}}.

\bibitem{Bao:2022fpk}
J.~Bao, ``{A note on quiver Yangians and R-matrices},''
  \href{http://dx.doi.org/10.1007/JHEP08(2022)219}{{\em JHEP} {\bfseries 08}
  (2022) 219}, \href{http://arxiv.org/abs/2206.06186}{{\ttfamily
  arXiv:2206.06186 [hep-th]}}.

\bibitem{Bao:2022jhy}
J.~Bao, ``{Quiver Yangians and -algebras for generalized conifolds},''
  \href{http://dx.doi.org/10.1088/1751-8121/acd037}{{\em J. Phys. A} {\bfseries
  56} no.~22, (2023) 225203}, \href{http://arxiv.org/abs/2208.13395}{{\ttfamily
  arXiv:2208.13395 [hep-th]}}.

\bibitem{Bezerra:2019dmp}
L.~Bezerra and E.~Mukhin, ``{Quantum toroidal algebra associated with
  $\mathfrak{gl}_{m|n}$},''
\href{http://arxiv.org/abs/1904.07297}{{\ttfamily arXiv:1904.07297 [math.QA]}}.

\bibitem{Galakhov:2020vyb}
D.~Galakhov and M.~Yamazaki, ``{Quiver Yangian and Supersymmetric Quantum
  Mechanics},'' \href{http://dx.doi.org/10.1007/s00220-022-04490-y}{{\em
  Commun. Math. Phys.} {\bfseries 396} no.~2, (2022) 713--785},
  \href{http://arxiv.org/abs/2008.07006}{{\ttfamily arXiv:2008.07006
  [hep-th]}}.

\bibitem{Yamazaki:2022cdg}
M.~Yamazaki, ``{Quiver Yangians and crystal meltings: A concise summary},''
  \href{http://dx.doi.org/10.1063/5.0089785}{{\em J. Math. Phys.} {\bfseries
  64} no.~1, (2023) 011101}, \href{http://arxiv.org/abs/2203.14314}{{\ttfamily
  arXiv:2203.14314 [hep-th]}}.

\bibitem{Li:2023zub}
W.~Li, ``{Quiver algebras and their representations for arbitrary quivers},''
  \href{http://arxiv.org/abs/2303.05521}{{\ttfamily arXiv:2303.05521
  [hep-th]}}.

\bibitem{Nishinaka:2013mba}
T.~Nishinaka, S.~Yamaguchi, and Y.~Yoshida, ``{Two-dimensional crystal melting
  and D4-D2-D0 on toric Calabi-Yau singularities},''
  \href{http://dx.doi.org/10.1007/JHEP05(2014)139}{{\em JHEP} {\bfseries 05}
  (2014) 139}, \href{http://arxiv.org/abs/1304.6724}{{\ttfamily arXiv:1304.6724
  [hep-th]}}.

\bibitem{nakajima1999lectures}
H.~Nakajima, {\em Lectures on Hilbert schemes of points on surfaces}.
\newblock No.~18. American Mathematical Soc., 1999.

\bibitem{Szendroi}
B.~Szendr{\H{o}}i, ``Non-commutative {D}onaldson-{T}homas invariants and the
  conifold,'' {\em Geom. Topol.} {\bfseries 12} no.~2, (2008) 1171--1202,
  \href{http://arxiv.org/abs/0705.3419}{{\ttfamily arXiv:0705.3419 [math.AG]}}.

\bibitem{Moore:1997dj}
G.~W. Moore, N.~Nekrasov, and S.~Shatashvili, ``{Integrating over Higgs
  branches},'' \href{http://dx.doi.org/10.1007/PL00005525}{{\em Commun. Math.
  Phys.} {\bfseries 209} (2000) 97--121},
  \href{http://arxiv.org/abs/hep-th/9712241}{{\ttfamily arXiv:hep-th/9712241}}.

\bibitem{Moore:1998et}
G.~W. Moore, N.~Nekrasov, and S.~Shatashvili, ``{D particle bound states and
  generalized instantons},''
  \href{http://dx.doi.org/10.1007/s002200050016}{{\em Commun. Math. Phys.}
  {\bfseries 209} (2000) 77--95},
  \href{http://arxiv.org/abs/hep-th/9803265}{{\ttfamily arXiv:hep-th/9803265}}.

\bibitem{Nekrasov:2017cih}
N.~Nekrasov, ``{Magnificent four},''
  \href{http://dx.doi.org/10.4310/ATMP.2020.v24.n5.a4}{{\em Adv. Theor. Math.
  Phys.} {\bfseries 24} no.~5, (2020) 1171--1202},
  \href{http://arxiv.org/abs/1712.08128}{{\ttfamily arXiv:1712.08128
  [hep-th]}}.

\bibitem{Nekrasov:2018xsb}
N.~Nekrasov and N.~Piazzalunga, ``{Magnificent Four with Colors},''
  \href{http://dx.doi.org/10.1007/s00220-019-03426-3}{{\em Commun. Math. Phys.}
  {\bfseries 372} no.~2, (2019) 573--597},
  \href{http://arxiv.org/abs/1808.05206}{{\ttfamily arXiv:1808.05206
  [hep-th]}}.

\bibitem{Prochazka:2015deb}
T.~Proch\'azka, ``{$ \mathcal{W} $ -symmetry, topological vertex and affine
  Yangian},'' \href{http://dx.doi.org/10.1007/JHEP10(2016)077}{{\em JHEP}
  {\bfseries 10} (2016) 077}, \href{http://arxiv.org/abs/1512.07178}{{\ttfamily
  arXiv:1512.07178 [hep-th]}}.

\bibitem{FeiginTsymbaliuk}
B.~L. Feigin and A.~I. Tsymbaliuk, ``{Equivariant $K$-theory of Hilbert schemes
  via shuffle algebra},'' {\em Kyoto Journal of Mathematics} {\bfseries 51}
  no.~4, (2011) 831 -- 854, \href{http://arxiv.org/abs/0904.1679}{{\ttfamily
  arXiv:0904.1679 [math.RT]}}.

\bibitem{Nekrasov:2009ui}
N.~A. Nekrasov and S.~L. Shatashvili, ``{Quantum integrability and
  supersymmetric vacua},'' \href{http://dx.doi.org/10.1143/PTPS.177.105}{{\em
  Prog. Theor. Phys. Suppl.} {\bfseries 177} (2009) 105--119},
  \href{http://arxiv.org/abs/0901.4748}{{\ttfamily arXiv:0901.4748 [hep-th]}}.

\bibitem{Hori:2013ika}
K.~Hori and M.~Romo, ``{Exact Results In Two-Dimensional (2,2) Supersymmetric
  Gauge Theories With Boundary},''
  \href{http://arxiv.org/abs/1308.2438}{{\ttfamily arXiv:1308.2438 [hep-th]}}.

\bibitem{dinkins2021elliptic}
H.~Dinkins, ``Elliptic stable envelopes of affine type $a$ quiver varieties,''
  \href{http://arxiv.org/abs/2107.09569}{{\ttfamily arXiv:2107.09569
  [math.AG]}}.

\end{thebibliography}\endgroup

\newpage

\addtolength{\textheight}{1cm} 
\begin{landscape}
	\centering
	\section{Data on super-partitions up to level 9/2}\label{app:data}
	\vspace{-0.35cm}
	\begingroup
	\renewcommand*{\arraystretch}{1.85}
	\begin{longtable}{|C|C|C|C|C|C|C|C|}
		\hline
		\ell & \lambda & \CS_{\lambda} & \fm_{\lambda} & n_{\lambda}^+ & n_{\lambda}^- & w_{\lambda}^+ & w_{\lambda}^-\\
		\hline
		\hline
		0 & \varnothing & 1 & \scriptstyle 1& \scriptstyle  0& \scriptstyle 0& \scriptstyle 0 & \scriptstyle 0\\
		\hline
		\frac{1}{2} &
		\begin{array}{c}
			\begin{tikzpicture}[scale=0.15]
				\foreach \i/\j in {0/0}
				{
					\draw (\i,\j) -- (\i+1,\j);
				}
				\foreach \i/\j in {0/0}
				{
					\draw (\i,\j) -- (\i,\j-1);
				}
				\foreach \i/\j in {1/0}
				{
					\draw (\i,\j) -- (\i-1,\j-1);
				}
		\end{tikzpicture}\end{array}&
		\scalebox{0.95}{$\begin{array}{c}
				\theta _1
				
			\end{array}$}&
		{\scriptstyle1}&
		{\scriptstyle 1}&
		{\scriptstyle 0}&
		{\scriptstyle 0}&
		{\scriptstyle 0}\\
		\hline
		1 &
		\begin{array}{c}
			\begin{tikzpicture}[scale=0.15]
				\foreach \i/\j in {0/-1, 0/0}
				{
					\draw (\i,\j) -- (\i+1,\j);
				}
				\foreach \i/\j in {0/0, 1/0}
				{
					\draw (\i,\j) -- (\i,\j-1);
				}
		\end{tikzpicture}\end{array}&
		\scalebox{0.95}{$\begin{array}{c}
				p_1
				
			\end{array}$}&
		{\scriptstyle\epsilon _1 \epsilon _2}&
		{\scriptstyle 1}&
		{\scriptstyle 1}&
		{\scriptstyle 0}&
		{\scriptstyle \frac{\epsilon _1}{2}+\frac{\epsilon _2}{2}}\\
		\hline
		\frac{3}{2} &
		\begin{array}{c}
			\begin{tikzpicture}[scale=0.15]
				\foreach \i/\j in {0/-1, 0/0}
				{
					\draw (\i,\j) -- (\i+1,\j);
				}
				\foreach \i/\j in {0/-1, 0/0, 1/0}
				{
					\draw (\i,\j) -- (\i,\j-1);
				}
				\foreach \i/\j in {1/-1}
				{
					\draw (\i,\j) -- (\i-1,\j-1);
				}
		\end{tikzpicture}\end{array}&
		\scalebox{0.95}{$\begin{array}{c}
				\theta _1 p_1-\frac{\theta _2}{\epsilon _2}
				
			\end{array}$}&
		{\scriptstyle-\epsilon _1 \left(\epsilon _1-\epsilon _2\right)}&
		{\scriptstyle 2}&
		{\scriptstyle 1}&
		{\scriptstyle \epsilon _1}&
		{\scriptstyle \frac{\epsilon _1}{2}+\frac{\epsilon _2}{2}}\\
		\cline{2-8}
		&
		\begin{array}{c}
			\begin{tikzpicture}[scale=0.15]
				\foreach \i/\j in {0/-1, 0/0, 1/0}
				{
					\draw (\i,\j) -- (\i+1,\j);
				}
				\foreach \i/\j in {0/0, 1/0}
				{
					\draw (\i,\j) -- (\i,\j-1);
				}
				\foreach \i/\j in {2/0}
				{
					\draw (\i,\j) -- (\i-1,\j-1);
				}
		\end{tikzpicture}\end{array}&
		\scalebox{0.95}{$\begin{array}{c}
				\theta _1 p_1-\frac{\theta _2}{\epsilon _1}
				
			\end{array}$}&
		{\scriptstyle-\epsilon _2 \left(\epsilon _2-\epsilon _1\right)}&
		{\scriptstyle 2}&
		{\scriptstyle 1}&
		{\scriptstyle \epsilon _2}&
		{\scriptstyle \frac{\epsilon _1}{2}+\frac{\epsilon _2}{2}}\\
		\hline
		2 &
		\begin{array}{c}
			\begin{tikzpicture}[scale=0.15]
				\foreach \i/\j in {0/-2, 0/-1, 0/0}
				{
					\draw (\i,\j) -- (\i+1,\j);
				}
				\foreach \i/\j in {0/-1, 0/0, 1/-1, 1/0}
				{
					\draw (\i,\j) -- (\i,\j-1);
				}
		\end{tikzpicture}\end{array}&
		\scalebox{0.95}{$\begin{array}{c}
				p_1^2-\frac{p_2}{\epsilon _2}
				
			\end{array}$}&
		{\scriptstyle-2 \epsilon _1^2 \left(\epsilon _1-\epsilon _2\right) \epsilon _2}&
		{\scriptstyle 2}&
		{\scriptstyle 2}&
		{\scriptstyle \epsilon _1}&
		{\scriptstyle 2 \epsilon _1+\epsilon _2}\\
		\cline{2-8}
		&
		\begin{array}{c}
			\begin{tikzpicture}[scale=0.15]
				\foreach \i/\j in {0/-1, 0/0, 1/0}
				{
					\draw (\i,\j) -- (\i+1,\j);
				}
				\foreach \i/\j in {0/-1, 0/0, 1/0}
				{
					\draw (\i,\j) -- (\i,\j-1);
				}
				\foreach \i/\j in {1/-1, 2/0}
				{
					\draw (\i,\j) -- (\i-1,\j-1);
				}
		\end{tikzpicture}\end{array}&
		\scalebox{0.95}{$\begin{array}{c}
				\frac{\theta _1 \theta _2}{\epsilon _1 \epsilon _2}
				
			\end{array}$}&
		{\scriptstyle1}&
		{\scriptstyle 3}&
		{\scriptstyle 1}&
		{\scriptstyle \epsilon _1+\epsilon _2}&
		{\scriptstyle \frac{\epsilon _1}{2}+\frac{\epsilon _2}{2}}\\
		\cline{2-8}
		&
		\begin{array}{c}
			\begin{tikzpicture}[scale=0.15]
				\foreach \i/\j in {0/-1, 0/0, 1/-1, 1/0}
				{
					\draw (\i,\j) -- (\i+1,\j);
				}
				\foreach \i/\j in {0/0, 1/0, 2/0}
				{
					\draw (\i,\j) -- (\i,\j-1);
				}
		\end{tikzpicture}\end{array}&
		\scalebox{0.95}{$\begin{array}{c}
				p_1^2-\frac{p_2}{\epsilon _1}
				
			\end{array}$}&
		{\scriptstyle-2 \epsilon _1 \epsilon _2^2 \left(\epsilon _2-\epsilon _1\right)}&
		{\scriptstyle 2}&
		{\scriptstyle 2}&
		{\scriptstyle \epsilon _2}&
		{\scriptstyle \epsilon _1+2 \epsilon _2}\\
		\hline
		\frac{5}{2} &
		\begin{array}{c}
			\begin{tikzpicture}[scale=0.15]
				\foreach \i/\j in {0/-2, 0/-1, 0/0}
				{
					\draw (\i,\j) -- (\i+1,\j);
				}
				\foreach \i/\j in {0/-2, 0/-1, 0/0, 1/-1, 1/0}
				{
					\draw (\i,\j) -- (\i,\j-1);
				}
				\foreach \i/\j in {1/-2}
				{
					\draw (\i,\j) -- (\i-1,\j-1);
				}
		\end{tikzpicture}\end{array}&
		\scalebox{0.95}{$\begin{array}{c}
				\theta _1 p_1^2-\frac{2 \theta _2 p_1}{\epsilon _2}-\frac{\theta _1 p_2}{\epsilon _2}+\frac{2 \theta _3}{\epsilon _2^2}
				
			\end{array}$}&
		{\scriptstyle2 \epsilon _1^2 \left(\epsilon _1-\epsilon _2\right) \left(2 \epsilon _1-\epsilon _2\right)}&
		{\scriptstyle 3}&
		{\scriptstyle 2}&
		{\scriptstyle 3 \epsilon _1}&
		{\scriptstyle 2 \epsilon _1+\epsilon _2}\\
		\cline{2-8}
		&
		\begin{array}{c}
			\begin{tikzpicture}[scale=0.15]
				\foreach \i/\j in {0/-2, 0/-1, 0/0, 1/0}
				{
					\draw (\i,\j) -- (\i+1,\j);
				}
				\foreach \i/\j in {0/-1, 0/0, 1/-1, 1/0}
				{
					\draw (\i,\j) -- (\i,\j-1);
				}
				\foreach \i/\j in {2/0}
				{
					\draw (\i,\j) -- (\i-1,\j-1);
				}
		\end{tikzpicture}\end{array}&
		\scalebox{0.95}{$\begin{array}{c}
				\theta _1 p_1^2-\frac{\theta _2 p_1}{\epsilon _1}-\frac{\theta _1 p_2}{\epsilon _2}+\frac{\theta _3}{\epsilon _1 \epsilon _2}
				
			\end{array}$}&
		{\scriptstyle\epsilon _1 \left(\epsilon _1-\epsilon _2\right) \epsilon _2 \left(\epsilon _2-2 \epsilon _1\right)}&
		{\scriptstyle 3}&
		{\scriptstyle 2}&
		{\scriptstyle \epsilon _1+\epsilon _2}&
		{\scriptstyle 2 \epsilon _1+\epsilon _2}\\
		\cline{2-8}
		&
		\begin{array}{c}
			\begin{tikzpicture}[scale=0.15]
				\foreach \i/\j in {0/-1, 0/0, 1/-1, 1/0}
				{
					\draw (\i,\j) -- (\i+1,\j);
				}
				\foreach \i/\j in {0/-1, 0/0, 1/0, 2/0}
				{
					\draw (\i,\j) -- (\i,\j-1);
				}
				\foreach \i/\j in {1/-1}
				{
					\draw (\i,\j) -- (\i-1,\j-1);
				}
		\end{tikzpicture}\end{array}&
		\scalebox{0.95}{$\begin{array}{c}
				\theta _1 p_1^2-\frac{\theta _2 p_1}{\epsilon _2}-\frac{\theta _1 p_2}{\epsilon _1}+\frac{\theta _3}{\epsilon _1 \epsilon _2}
				
			\end{array}$}&
		{\scriptstyle\epsilon _1 \left(\epsilon _1-2 \epsilon _2\right) \epsilon _2 \left(\epsilon _2-\epsilon _1\right)}&
		{\scriptstyle 3}&
		{\scriptstyle 2}&
		{\scriptstyle \epsilon _1+\epsilon _2}&
		{\scriptstyle \epsilon _1+2 \epsilon _2}\\
		\cline{2-8}
		&
		\begin{array}{c}
			\begin{tikzpicture}[scale=0.15]
				\foreach \i/\j in {0/-1, 0/0, 1/-1, 1/0, 2/0}
				{
					\draw (\i,\j) -- (\i+1,\j);
				}
				\foreach \i/\j in {0/0, 1/0, 2/0}
				{
					\draw (\i,\j) -- (\i,\j-1);
				}
				\foreach \i/\j in {3/0}
				{
					\draw (\i,\j) -- (\i-1,\j-1);
				}
		\end{tikzpicture}\end{array}&
		\scalebox{0.95}{$\begin{array}{c}
				\theta _1 p_1^2-\frac{2 \theta _2 p_1}{\epsilon _1}-\frac{\theta _1 p_2}{\epsilon _1}+\frac{2 \theta _3}{\epsilon _1^2}
				
			\end{array}$}&
		{\scriptstyle2 \epsilon _2^2 \left(\epsilon _2-\epsilon _1\right) \left(2 \epsilon _2-\epsilon _1\right)}&
		{\scriptstyle 3}&
		{\scriptstyle 2}&
		{\scriptstyle 3 \epsilon _2}&
		{\scriptstyle \epsilon _1+2 \epsilon _2}\\
		\hline
		3 &
		\begin{array}{c}
			\begin{tikzpicture}[scale=0.15]
				\foreach \i/\j in {0/-3, 0/-2, 0/-1, 0/0}
				{
					\draw (\i,\j) -- (\i+1,\j);
				}
				\foreach \i/\j in {0/-2, 0/-1, 0/0, 1/-2, 1/-1, 1/0}
				{
					\draw (\i,\j) -- (\i,\j-1);
				}
		\end{tikzpicture}\end{array}&
		\scalebox{0.95}{$\begin{array}{c}
				-\frac{3 p_2 p_1}{\epsilon _2}+\frac{2 p_3}{\epsilon _2^2}+p_1^3
				
			\end{array}$}&
		{\scriptstyle6 \epsilon _1^3 \left(\epsilon _1-\epsilon _2\right) \left(2 \epsilon _1-\epsilon _2\right) \epsilon _2}&
		{\scriptstyle 3}&
		{\scriptstyle 3}&
		{\scriptstyle 3 \epsilon _1}&
		{\scriptstyle \frac{9 \epsilon _1}{2}+\frac{3 \epsilon _2}{2}}\\
		\cline{2-8}
		&
		\begin{array}{c}
			\begin{tikzpicture}[scale=0.15]
				\foreach \i/\j in {0/-2, 0/-1, 0/0, 1/0}
				{
					\draw (\i,\j) -- (\i+1,\j);
				}
				\foreach \i/\j in {0/-2, 0/-1, 0/0, 1/-1, 1/0}
				{
					\draw (\i,\j) -- (\i,\j-1);
				}
				\foreach \i/\j in {1/-2, 2/0}
				{
					\draw (\i,\j) -- (\i-1,\j-1);
				}
		\end{tikzpicture}\end{array}&
		\scalebox{0.95}{$\begin{array}{c}
				\frac{\theta _1 \theta _2 p_1}{\epsilon _1 \epsilon _2}-\frac{\theta _1 \theta _3}{\epsilon _1 \epsilon _2^2}
				
			\end{array}$}&
		{\scriptstyle-\epsilon _1 \left(\epsilon _1-\epsilon _2\right)}&
		{\scriptstyle 4}&
		{\scriptstyle 2}&
		{\scriptstyle 3 \epsilon _1+\epsilon _2}&
		{\scriptstyle 2 \epsilon _1+\epsilon _2}\\
		\cline{2-8}
		&
		\begin{array}{c}
			\begin{tikzpicture}[scale=0.15]
				\foreach \i/\j in {0/-2, 0/-1, 0/0, 1/-1, 1/0}
				{
					\draw (\i,\j) -- (\i+1,\j);
				}
				\foreach \i/\j in {0/-1, 0/0, 1/-1, 1/0, 2/0}
				{
					\draw (\i,\j) -- (\i,\j-1);
				}
		\end{tikzpicture}\end{array}&
		\scalebox{0.95}{$\begin{array}{c}
				-\frac{p_2 p_1 \left(\epsilon _1+\epsilon _2\right)}{\epsilon _1 \epsilon _2}+\frac{p_3}{\epsilon _1 \epsilon _2}+p_1^3
				
			\end{array}$}&
		{\scriptstyle\epsilon _1^2 \left(\epsilon _1-2 \epsilon _2\right) \epsilon _2^2 \left(\epsilon _2-2 \epsilon _1\right)}&
		{\scriptstyle 3}&
		{\scriptstyle 3}&
		{\scriptstyle \epsilon _1+\epsilon _2}&
		{\scriptstyle \frac{5 \epsilon _1}{2}+\frac{5 \epsilon _2}{2}}\\
		\cline{2-8}
		&
		\begin{array}{c}
			\begin{tikzpicture}[scale=0.15]
				\foreach \i/\j in {0/-1, 0/0, 1/-1, 1/0, 2/0}
				{
					\draw (\i,\j) -- (\i+1,\j);
				}
				\foreach \i/\j in {0/-1, 0/0, 1/0, 2/0}
				{
					\draw (\i,\j) -- (\i,\j-1);
				}
				\foreach \i/\j in {1/-1, 3/0}
				{
					\draw (\i,\j) -- (\i-1,\j-1);
				}
		\end{tikzpicture}\end{array}&
		\scalebox{0.95}{$\begin{array}{c}
				\frac{\theta _1 \theta _2 p_1}{\epsilon _1 \epsilon _2}-\frac{\theta _1 \theta _3}{\epsilon _1^2 \epsilon _2}
				
			\end{array}$}&
		{\scriptstyle-\epsilon _2 \left(\epsilon _2-\epsilon _1\right)}&
		{\scriptstyle 4}&
		{\scriptstyle 2}&
		{\scriptstyle \epsilon _1+3 \epsilon _2}&
		{\scriptstyle \epsilon _1+2 \epsilon _2}\\
		\cline{2-8}
		&
		\begin{array}{c}
			\begin{tikzpicture}[scale=0.15]
				\foreach \i/\j in {0/-1, 0/0, 1/-1, 1/0, 2/-1, 2/0}
				{
					\draw (\i,\j) -- (\i+1,\j);
				}
				\foreach \i/\j in {0/0, 1/0, 2/0, 3/0}
				{
					\draw (\i,\j) -- (\i,\j-1);
				}
		\end{tikzpicture}\end{array}&
		\scalebox{0.95}{$\begin{array}{c}
				-\frac{3 p_2 p_1}{\epsilon _1}+\frac{2 p_3}{\epsilon _1^2}+p_1^3
				
			\end{array}$}&
		{\scriptstyle6 \epsilon _1 \epsilon _2^3 \left(\epsilon _2-\epsilon _1\right) \left(2 \epsilon _2-\epsilon _1\right)}&
		{\scriptstyle 3}&
		{\scriptstyle 3}&
		{\scriptstyle 3 \epsilon _2}&
		{\scriptstyle \frac{3 \epsilon _1}{2}+\frac{9 \epsilon _2}{2}}\\
		\hline
		\frac{7}{2} &
		\begin{array}{c}
			\begin{tikzpicture}[scale=0.15]
				\foreach \i/\j in {0/-3, 0/-2, 0/-1, 0/0}
				{
					\draw (\i,\j) -- (\i+1,\j);
				}
				\foreach \i/\j in {0/-3, 0/-2, 0/-1, 0/0, 1/-2, 1/-1, 1/0}
				{
					\draw (\i,\j) -- (\i,\j-1);
				}
				\foreach \i/\j in {1/-3}
				{
					\draw (\i,\j) -- (\i-1,\j-1);
				}
		\end{tikzpicture}\end{array}&
		\scalebox{0.95}{$\begin{array}{c}
				\theta _1 p_1^3-\frac{3 \theta _2 p_1^2}{\epsilon _2}-\frac{3 \theta _1 p_2 p_1}{\epsilon _2}+\frac{6 \theta _3 p_1}{\epsilon _2^2}+\frac{2 \theta _1 p_3}{\epsilon _2^2}+\frac{3 \theta _2 p_2}{\epsilon _2^2}-\frac{6 \theta _4}{\epsilon _2^3}
				
			\end{array}$}&
		{\scriptstyle-6 \epsilon _1^3 \left(\epsilon _1-\epsilon _2\right) \left(2 \epsilon _1-\epsilon _2\right) \left(3 \epsilon _1-\epsilon _2\right)}&
		{\scriptstyle 4}&
		{\scriptstyle 3}&
		{\scriptstyle 6 \epsilon _1}&
		{\scriptstyle \frac{9 \epsilon _1}{2}+\frac{3 \epsilon _2}{2}}\\
		\cline{2-8}
		&
		\begin{array}{c}
			\begin{tikzpicture}[scale=0.15]
				\foreach \i/\j in {0/-3, 0/-2, 0/-1, 0/0, 1/0}
				{
					\draw (\i,\j) -- (\i+1,\j);
				}
				\foreach \i/\j in {0/-2, 0/-1, 0/0, 1/-2, 1/-1, 1/0}
				{
					\draw (\i,\j) -- (\i,\j-1);
				}
				\foreach \i/\j in {2/0}
				{
					\draw (\i,\j) -- (\i-1,\j-1);
				}
		\end{tikzpicture}\end{array}&
		\scalebox{0.95}{$\begin{array}{c}
				\theta _1 p_1^3-\frac{\theta _2 p_1^2}{\epsilon _1}-\frac{3 \theta _1 p_2 p_1}{\epsilon _2}+\frac{2 \theta _3 p_1}{\epsilon _1 \epsilon _2}+\frac{2 \theta _1 p_3}{\epsilon _2^2}+\frac{\theta _2 p_2}{\epsilon _1 \epsilon _2}-\frac{2 \theta _4}{\epsilon _1 \epsilon _2^2}
				
			\end{array}$}&
		{\scriptstyle-2 \epsilon _1^2 \left(\epsilon _1-\epsilon _2\right) \left(2 \epsilon _1-\epsilon _2\right) \epsilon _2 \left(\epsilon _2-3 \epsilon _1\right)}&
		{\scriptstyle 4}&
		{\scriptstyle 3}&
		{\scriptstyle 3 \epsilon _1+\epsilon _2}&
		{\scriptstyle \frac{9 \epsilon _1}{2}+\frac{3 \epsilon _2}{2}}\\
		\cline{2-8}
		&
		\begin{array}{c}
			\begin{tikzpicture}[scale=0.15]
				\foreach \i/\j in {0/-2, 0/-1, 0/0, 1/-1, 1/0}
				{
					\draw (\i,\j) -- (\i+1,\j);
				}
				\foreach \i/\j in {0/-2, 0/-1, 0/0, 1/-1, 1/0, 2/0}
				{
					\draw (\i,\j) -- (\i,\j-1);
				}
				\foreach \i/\j in {1/-2}
				{
					\draw (\i,\j) -- (\i-1,\j-1);
				}
		\end{tikzpicture}\end{array}&
		\scalebox{0.95}{$\begin{array}{c}
				\theta _1 p_1^3-\frac{2 \theta _2 p_1^2}{\epsilon _2}-\frac{\theta _1 p_2 p_1 \left(\epsilon _1+\epsilon _2\right)}{\epsilon _1 \epsilon _2}+\frac{\theta _3 p_1 \left(2 \epsilon _1+\epsilon _2\right)}{\epsilon _1 \epsilon _2^2}+\frac{\theta _1 p_3}{\epsilon _1 \epsilon _2}+\frac{\theta _2 p_2}{\epsilon _1 \epsilon _2}-\frac{2 \theta _4}{\epsilon _1 \epsilon _2^2}
				
			\end{array}$}&
		{\scriptstyle-\epsilon _1^2 \left(2 \epsilon _1-2 \epsilon _2\right) \left(\epsilon _1-\epsilon _2\right) \epsilon _2 \left(\epsilon _2-2 \epsilon _1\right)}&
		{\scriptstyle 4}&
		{\scriptstyle 3}&
		{\scriptstyle 3 \epsilon _1+\epsilon _2}&
		{\scriptstyle \frac{5 \epsilon _1}{2}+\frac{5 \epsilon _2}{2}}\\
		\cline{2-8}
		&
		\begin{array}{c}
			\begin{tikzpicture}[scale=0.15]
				\foreach \i/\j in {0/-2, 0/-1, 0/0, 1/-1, 1/0}
				{
					\draw (\i,\j) -- (\i+1,\j);
				}
				\foreach \i/\j in {0/-1, 0/0, 1/-1, 1/0, 2/0}
				{
					\draw (\i,\j) -- (\i,\j-1);
				}
				\foreach \i/\j in {2/-1}
				{
					\draw (\i,\j) -- (\i-1,\j-1);
				}
		\end{tikzpicture}\end{array}&
		\scalebox{0.95}{$\begin{array}{c}
				\theta _1 p_1^3-\frac{\theta _2 p_1^2 \left(\epsilon _1+\epsilon _2\right)}{\epsilon _1 \epsilon _2}-\frac{\theta _1 p_2 p_1 \left(\epsilon _1+\epsilon _2\right)}{\epsilon _1 \epsilon _2}+\frac{3 \theta _3 p_1}{\epsilon _1 \epsilon _2}+\frac{\theta _1 p_3}{\epsilon _1 \epsilon _2}+\frac{\theta _2 p_2 \left(\epsilon _1^2-\epsilon _2 \epsilon _1+\epsilon _2^2\right)}{\epsilon _1^2 \epsilon _2^2}-\frac{\theta _4 \left(\epsilon _1+\epsilon _2\right)}{\epsilon _1^2 \epsilon _2^2}
				
			\end{array}$}&
		{\scriptstyle\epsilon _1 \left(\epsilon _1-2 \epsilon _2\right) \left(\epsilon _1-\epsilon _2\right) \epsilon _2 \left(\epsilon _2-2 \epsilon _1\right) \left(\epsilon _2-\epsilon _1\right)}&
		{\scriptstyle 4}&
		{\scriptstyle 3}&
		{\scriptstyle 2 \epsilon _1+2 \epsilon _2}&
		{\scriptstyle \frac{5 \epsilon _1}{2}+\frac{5 \epsilon _2}{2}}\\
		\cline{2-8}
		&
		\begin{array}{c}
			\begin{tikzpicture}[scale=0.15]
				\foreach \i/\j in {0/-2, 0/-1, 0/0, 1/-1, 1/0, 2/0}
				{
					\draw (\i,\j) -- (\i+1,\j);
				}
				\foreach \i/\j in {0/-1, 0/0, 1/-1, 1/0, 2/0}
				{
					\draw (\i,\j) -- (\i,\j-1);
				}
				\foreach \i/\j in {3/0}
				{
					\draw (\i,\j) -- (\i-1,\j-1);
				}
		\end{tikzpicture}\end{array}&
		\scalebox{0.95}{$\begin{array}{c}
				\theta _1 p_1^3-\frac{2 \theta _2 p_1^2}{\epsilon _1}-\frac{\theta _1 p_2 p_1 \left(\epsilon _1+\epsilon _2\right)}{\epsilon _1 \epsilon _2}+\frac{\theta _3 p_1 \left(\epsilon _1+2 \epsilon _2\right)}{\epsilon _1^2 \epsilon _2}+\frac{\theta _1 p_3}{\epsilon _1 \epsilon _2}+\frac{\theta _2 p_2}{\epsilon _1 \epsilon _2}-\frac{2 \theta _4}{\epsilon _1^2 \epsilon _2}
				
			\end{array}$}&
		{\scriptstyle-\epsilon _1 \left(\epsilon _1-2 \epsilon _2\right) \epsilon _2^2 \left(\epsilon _2-\epsilon _1\right) \left(2 \epsilon _2-2 \epsilon _1\right)}&
		{\scriptstyle 4}&
		{\scriptstyle 3}&
		{\scriptstyle \epsilon _1+3 \epsilon _2}&
		{\scriptstyle \frac{5 \epsilon _1}{2}+\frac{5 \epsilon _2}{2}}\\
		\cline{2-8}
		&
		\begin{array}{c}
			\begin{tikzpicture}[scale=0.15]
				\foreach \i/\j in {0/-1, 0/0, 1/-1, 1/0, 2/-1, 2/0}
				{
					\draw (\i,\j) -- (\i+1,\j);
				}
				\foreach \i/\j in {0/-1, 0/0, 1/0, 2/0, 3/0}
				{
					\draw (\i,\j) -- (\i,\j-1);
				}
				\foreach \i/\j in {1/-1}
				{
					\draw (\i,\j) -- (\i-1,\j-1);
				}
		\end{tikzpicture}\end{array}&
		\scalebox{0.95}{$\begin{array}{c}
				\theta _1 p_1^3-\frac{\theta _2 p_1^2}{\epsilon _2}-\frac{3 \theta _1 p_2 p_1}{\epsilon _1}+\frac{2 \theta _3 p_1}{\epsilon _1 \epsilon _2}+\frac{2 \theta _1 p_3}{\epsilon _1^2}+\frac{\theta _2 p_2}{\epsilon _1 \epsilon _2}-\frac{2 \theta _4}{\epsilon _1^2 \epsilon _2}
				
			\end{array}$}&
		{\scriptstyle-2 \epsilon _1 \left(\epsilon _1-3 \epsilon _2\right) \epsilon _2^2 \left(\epsilon _2-\epsilon _1\right) \left(2 \epsilon _2-\epsilon _1\right)}&
		{\scriptstyle 4}&
		{\scriptstyle 3}&
		{\scriptstyle \epsilon _1+3 \epsilon _2}&
		{\scriptstyle \frac{3 \epsilon _1}{2}+\frac{9 \epsilon _2}{2}}\\
		\cline{2-8}
		&
		\begin{array}{c}
			\begin{tikzpicture}[scale=0.15]
				\foreach \i/\j in {0/-1, 0/0, 1/-1, 1/0, 2/-1, 2/0, 3/0}
				{
					\draw (\i,\j) -- (\i+1,\j);
				}
				\foreach \i/\j in {0/0, 1/0, 2/0, 3/0}
				{
					\draw (\i,\j) -- (\i,\j-1);
				}
				\foreach \i/\j in {4/0}
				{
					\draw (\i,\j) -- (\i-1,\j-1);
				}
		\end{tikzpicture}\end{array}&
		\scalebox{0.95}{$\begin{array}{c}
				\theta _1 p_1^3-\frac{3 \theta _2 p_1^2}{\epsilon _1}-\frac{3 \theta _1 p_2 p_1}{\epsilon _1}+\frac{6 \theta _3 p_1}{\epsilon _1^2}+\frac{2 \theta _1 p_3}{\epsilon _1^2}+\frac{3 \theta _2 p_2}{\epsilon _1^2}-\frac{6 \theta _4}{\epsilon _1^3}
				
			\end{array}$}&
		{\scriptstyle-6 \epsilon _2^3 \left(\epsilon _2-\epsilon _1\right) \left(2 \epsilon _2-\epsilon _1\right) \left(3 \epsilon _2-\epsilon _1\right)}&
		{\scriptstyle 4}&
		{\scriptstyle 3}&
		{\scriptstyle 6 \epsilon _2}&
		{\scriptstyle \frac{3 \epsilon _1}{2}+\frac{9 \epsilon _2}{2}}\\
		\hline
		4 &
		\begin{array}{c}
			\begin{tikzpicture}[scale=0.15]
				\foreach \i/\j in {0/-4, 0/-3, 0/-2, 0/-1, 0/0}
				{
					\draw (\i,\j) -- (\i+1,\j);
				}
				\foreach \i/\j in {0/-3, 0/-2, 0/-1, 0/0, 1/-3, 1/-2, 1/-1, 1/0}
				{
					\draw (\i,\j) -- (\i,\j-1);
				}
		\end{tikzpicture}\end{array}&
		\scalebox{0.95}{$\begin{array}{c}
				-\frac{6 p_2 p_1^2}{\epsilon _2}+\frac{8 p_3 p_1}{\epsilon _2^2}+\frac{3 p_2^2}{\epsilon _2^2}-\frac{6 p_4}{\epsilon _2^3}+p_1^4
				
			\end{array}$}&
		{\scriptstyle-24 \epsilon _1^4 \left(\epsilon _1-\epsilon _2\right) \left(2 \epsilon _1-\epsilon _2\right) \left(3 \epsilon _1-\epsilon _2\right) \epsilon _2}&
		{\scriptstyle 4}&
		{\scriptstyle 4}&
		{\scriptstyle 6 \epsilon _1}&
		{\scriptstyle 8 \epsilon _1+2 \epsilon _2}\\
		\cline{2-8}
		&
		\begin{array}{c}
			\begin{tikzpicture}[scale=0.15]
				\foreach \i/\j in {0/-3, 0/-2, 0/-1, 0/0, 1/0}
				{
					\draw (\i,\j) -- (\i+1,\j);
				}
				\foreach \i/\j in {0/-3, 0/-2, 0/-1, 0/0, 1/-2, 1/-1, 1/0}
				{
					\draw (\i,\j) -- (\i,\j-1);
				}
				\foreach \i/\j in {1/-3, 2/0}
				{
					\draw (\i,\j) -- (\i-1,\j-1);
				}
		\end{tikzpicture}\end{array}&
		\scalebox{0.95}{$\begin{array}{c}
				\frac{\theta _1 \theta _2 p_1^2}{\epsilon _1 \epsilon _2}-\frac{2 \theta _1 \theta _3 p_1}{\epsilon _1 \epsilon _2^2}-\frac{\theta _1 \theta _2 p_2}{\epsilon _1 \epsilon _2^2}+\frac{2 \theta _1 \theta _4}{\epsilon _1 \epsilon _2^3}
				
			\end{array}$}&
		{\scriptstyle2 \epsilon _1^2 \left(\epsilon _1-\epsilon _2\right) \left(2 \epsilon _1-\epsilon _2\right)}&
		{\scriptstyle 5}&
		{\scriptstyle 3}&
		{\scriptstyle 6 \epsilon _1+\epsilon _2}&
		{\scriptstyle \frac{9 \epsilon _1}{2}+\frac{3 \epsilon _2}{2}}\\
		\cline{2-8}
		&
		\begin{array}{c}
			\begin{tikzpicture}[scale=0.15]
				\foreach \i/\j in {0/-3, 0/-2, 0/-1, 0/0, 1/-1, 1/0}
				{
					\draw (\i,\j) -- (\i+1,\j);
				}
				\foreach \i/\j in {0/-2, 0/-1, 0/0, 1/-2, 1/-1, 1/0, 2/0}
				{
					\draw (\i,\j) -- (\i,\j-1);
				}
		\end{tikzpicture}\end{array}&
		\scalebox{0.95}{$\begin{array}{c}
				-\frac{p_2 p_1^2 \left(3 \epsilon _1+\epsilon _2\right)}{\epsilon _1 \epsilon _2}+\frac{2 p_3 p_1 \left(\epsilon _1+\epsilon _2\right)}{\epsilon _1 \epsilon _2^2}+\frac{p_2^2}{\epsilon _1 \epsilon _2}-\frac{2 p_4}{\epsilon _1 \epsilon _2^2}+p_1^4
				
			\end{array}$}&
		{\scriptstyle-2 \epsilon _1^3 \left(2 \epsilon _1-2 \epsilon _2\right) \left(\epsilon _1-\epsilon _2\right) \epsilon _2^2 \left(\epsilon _2-3 \epsilon _1\right)}&
		{\scriptstyle 4}&
		{\scriptstyle 4}&
		{\scriptstyle 3 \epsilon _1+\epsilon _2}&
		{\scriptstyle 5 \epsilon _1+3 \epsilon _2}\\
		\cline{2-8}
		&
		\begin{array}{c}
			\begin{tikzpicture}[scale=0.15]
				\foreach \i/\j in {0/-2, 0/-1, 0/0, 1/-1, 1/0}
				{
					\draw (\i,\j) -- (\i+1,\j);
				}
				\foreach \i/\j in {0/-2, 0/-1, 0/0, 1/-1, 1/0, 2/0}
				{
					\draw (\i,\j) -- (\i,\j-1);
				}
				\foreach \i/\j in {1/-2, 2/-1}
				{
					\draw (\i,\j) -- (\i-1,\j-1);
				}
		\end{tikzpicture}\end{array}&
		\scalebox{0.95}{$\begin{array}{c}
				\frac{\theta _1 \theta _2 p_1^2}{\epsilon _1 \epsilon _2}-\frac{2 \theta _1 \theta _3 p_1}{\epsilon _1 \epsilon _2^2}+\frac{\theta _1 \theta _2 p_2 \left(\epsilon _1-\epsilon _2\right)}{\epsilon _1^2 \epsilon _2^2}+\frac{\theta _2 \theta _3 \left(2 \epsilon _1-\epsilon _2\right)}{\epsilon _1^2 \epsilon _2^3}+\frac{\theta _1 \theta _4}{\epsilon _1^2 \epsilon _2^2}
				
			\end{array}$}&
		{\scriptstyle-\epsilon _1 \left(2 \epsilon _1-2 \epsilon _2\right) \left(\epsilon _1-\epsilon _2\right) \left(\epsilon _2-2 \epsilon _1\right)}&
		{\scriptstyle 5}&
		{\scriptstyle 3}&
		{\scriptstyle 4 \epsilon _1+2 \epsilon _2}&
		{\scriptstyle \frac{5 \epsilon _1}{2}+\frac{5 \epsilon _2}{2}}\\
		\cline{2-8}
		&
		\begin{array}{c}
			\begin{tikzpicture}[scale=0.15]
				\foreach \i/\j in {0/-2, 0/-1, 0/0, 1/-1, 1/0, 2/0}
				{
					\draw (\i,\j) -- (\i+1,\j);
				}
				\foreach \i/\j in {0/-2, 0/-1, 0/0, 1/-1, 1/0, 2/0}
				{
					\draw (\i,\j) -- (\i,\j-1);
				}
				\foreach \i/\j in {1/-2, 3/0}
				{
					\draw (\i,\j) -- (\i-1,\j-1);
				}
		\end{tikzpicture}\end{array}&
		\scalebox{0.95}{$\begin{array}{c}
				\frac{\theta _1 \theta _2 p_1^2}{\epsilon _1 \epsilon _2}-\frac{\theta _1 \theta _3 p_1 \left(\epsilon _1+\epsilon _2\right)}{\epsilon _1^2 \epsilon _2^2}+\frac{\theta _2 \theta _3}{\epsilon _1^2 \epsilon _2^2}+\frac{\theta _1 \theta _4}{\epsilon _1^2 \epsilon _2^2}
				
			\end{array}$}&
		{\scriptstyle\epsilon _1 \left(\epsilon _1-\epsilon _2\right) \epsilon _2 \left(\epsilon _2-\epsilon _1\right)}&
		{\scriptstyle 5}&
		{\scriptstyle 3}&
		{\scriptstyle 3 \epsilon _1+3 \epsilon _2}&
		{\scriptstyle \frac{5 \epsilon _1}{2}+\frac{5 \epsilon _2}{2}}\\
		\cline{2-8}
		&
		\begin{array}{c}
			\begin{tikzpicture}[scale=0.15]
				\foreach \i/\j in {0/-2, 0/-1, 0/0, 1/-2, 1/-1, 1/0}
				{
					\draw (\i,\j) -- (\i+1,\j);
				}
				\foreach \i/\j in {0/-1, 0/0, 1/-1, 1/0, 2/-1, 2/0}
				{
					\draw (\i,\j) -- (\i,\j-1);
				}
		\end{tikzpicture}\end{array}&
		\scalebox{0.95}{$\begin{array}{c}
				-\frac{2 p_2 p_1^2 \left(\epsilon _1+\epsilon _2\right)}{\epsilon _1 \epsilon _2}+\frac{4 p_3 p_1}{\epsilon _1 \epsilon _2}-\frac{p_4 \left(\epsilon _1+\epsilon _2\right)}{\epsilon _1^2 \epsilon _2^2}+\frac{p_2^2 \left(\epsilon _1^2-\epsilon _2 \epsilon _1+\epsilon _2^2\right)}{\epsilon _1^2 \epsilon _2^2}+p_1^4
				
			\end{array}$}&
		{\scriptstyle4 \epsilon _1^2 \left(\epsilon _1-2 \epsilon _2\right) \left(\epsilon _1-\epsilon _2\right) \epsilon _2^2 \left(\epsilon _2-2 \epsilon _1\right) \left(\epsilon _2-\epsilon _1\right)}&
		{\scriptstyle 4}&
		{\scriptstyle 4}&
		{\scriptstyle 2 \epsilon _1+2 \epsilon _2}&
		{\scriptstyle 4 \epsilon _1+4 \epsilon _2}\\
		\cline{2-8}
		&
		\begin{array}{c}
			\begin{tikzpicture}[scale=0.15]
				\foreach \i/\j in {0/-2, 0/-1, 0/0, 1/-1, 1/0, 2/0}
				{
					\draw (\i,\j) -- (\i+1,\j);
				}
				\foreach \i/\j in {0/-1, 0/0, 1/-1, 1/0, 2/0}
				{
					\draw (\i,\j) -- (\i,\j-1);
				}
				\foreach \i/\j in {2/-1, 3/0}
				{
					\draw (\i,\j) -- (\i-1,\j-1);
				}
		\end{tikzpicture}\end{array}&
		\scalebox{0.95}{$\begin{array}{c}
				\frac{\theta _1 \theta _2 p_1^2}{\epsilon _1 \epsilon _2}-\frac{2 \theta _1 \theta _3 p_1}{\epsilon _1^2 \epsilon _2}-\frac{\theta _1 \theta _2 p_2 \left(\epsilon _1-\epsilon _2\right)}{\epsilon _1^2 \epsilon _2^2}-\frac{\theta _2 \theta _3 \left(\epsilon _1-2 \epsilon _2\right)}{\epsilon _1^3 \epsilon _2^2}+\frac{\theta _1 \theta _4}{\epsilon _1^2 \epsilon _2^2}
				
			\end{array}$}&
		{\scriptstyle-\left(\epsilon _1-2 \epsilon _2\right) \epsilon _2 \left(\epsilon _2-\epsilon _1\right) \left(2 \epsilon _2-2 \epsilon _1\right)}&
		{\scriptstyle 5}&
		{\scriptstyle 3}&
		{\scriptstyle 2 \epsilon _1+4 \epsilon _2}&
		{\scriptstyle \frac{5 \epsilon _1}{2}+\frac{5 \epsilon _2}{2}}\\
		\cline{2-8}
		&
		\begin{array}{c}
			\begin{tikzpicture}[scale=0.15]
				\foreach \i/\j in {0/-2, 0/-1, 0/0, 1/-1, 1/0, 2/-1, 2/0}
				{
					\draw (\i,\j) -- (\i+1,\j);
				}
				\foreach \i/\j in {0/-1, 0/0, 1/-1, 1/0, 2/0, 3/0}
				{
					\draw (\i,\j) -- (\i,\j-1);
				}
		\end{tikzpicture}\end{array}&
		\scalebox{0.95}{$\begin{array}{c}
				-\frac{p_2 p_1^2 \left(\epsilon _1+3 \epsilon _2\right)}{\epsilon _1 \epsilon _2}+\frac{2 p_3 p_1 \left(\epsilon _1+\epsilon _2\right)}{\epsilon _1^2 \epsilon _2}+\frac{p_2^2}{\epsilon _1 \epsilon _2}-\frac{2 p_4}{\epsilon _1^2 \epsilon _2}+p_1^4
				
			\end{array}$}&
		{\scriptstyle-2 \epsilon _1^2 \left(\epsilon _1-3 \epsilon _2\right) \epsilon _2^3 \left(\epsilon _2-\epsilon _1\right) \left(2 \epsilon _2-2 \epsilon _1\right)}&
		{\scriptstyle 4}&
		{\scriptstyle 4}&
		{\scriptstyle \epsilon _1+3 \epsilon _2}&
		{\scriptstyle 3 \epsilon _1+5 \epsilon _2}\\
		\cline{2-8}
		&
		\begin{array}{c}
			\begin{tikzpicture}[scale=0.15]
				\foreach \i/\j in {0/-1, 0/0, 1/-1, 1/0, 2/-1, 2/0, 3/0}
				{
					\draw (\i,\j) -- (\i+1,\j);
				}
				\foreach \i/\j in {0/-1, 0/0, 1/0, 2/0, 3/0}
				{
					\draw (\i,\j) -- (\i,\j-1);
				}
				\foreach \i/\j in {1/-1, 4/0}
				{
					\draw (\i,\j) -- (\i-1,\j-1);
				}
		\end{tikzpicture}\end{array}&
		\scalebox{0.95}{$\begin{array}{c}
				\frac{\theta _1 \theta _2 p_1^2}{\epsilon _1 \epsilon _2}-\frac{2 \theta _1 \theta _3 p_1}{\epsilon _1^2 \epsilon _2}-\frac{\theta _1 \theta _2 p_2}{\epsilon _1^2 \epsilon _2}+\frac{2 \theta _1 \theta _4}{\epsilon _1^3 \epsilon _2}
				
			\end{array}$}&
		{\scriptstyle2 \epsilon _2^2 \left(\epsilon _2-\epsilon _1\right) \left(2 \epsilon _2-\epsilon _1\right)}&
		{\scriptstyle 5}&
		{\scriptstyle 3}&
		{\scriptstyle \epsilon _1+6 \epsilon _2}&
		{\scriptstyle \frac{3 \epsilon _1}{2}+\frac{9 \epsilon _2}{2}}\\
		\cline{2-8}
		&
		\begin{array}{c}
			\begin{tikzpicture}[scale=0.15]
				\foreach \i/\j in {0/-1, 0/0, 1/-1, 1/0, 2/-1, 2/0, 3/-1, 3/0}
				{
					\draw (\i,\j) -- (\i+1,\j);
				}
				\foreach \i/\j in {0/0, 1/0, 2/0, 3/0, 4/0}
				{
					\draw (\i,\j) -- (\i,\j-1);
				}
		\end{tikzpicture}\end{array}&
		\scalebox{0.95}{$\begin{array}{c}
				-\frac{6 p_2 p_1^2}{\epsilon _1}+\frac{8 p_3 p_1}{\epsilon _1^2}+\frac{3 p_2^2}{\epsilon _1^2}-\frac{6 p_4}{\epsilon _1^3}+p_1^4
				
			\end{array}$}&
		{\scriptstyle-24 \epsilon _1 \epsilon _2^4 \left(\epsilon _2-\epsilon _1\right) \left(2 \epsilon _2-\epsilon _1\right) \left(3 \epsilon _2-\epsilon _1\right)}&
		{\scriptstyle 4}&
		{\scriptstyle 4}&
		{\scriptstyle 6 \epsilon _2}&
		{\scriptstyle 2 \epsilon _1+8 \epsilon _2}\\
		\hline
		\frac{9}{2} &
		\begin{array}{c}
			\begin{tikzpicture}[scale=0.15]
				\foreach \i/\j in {0/-4, 0/-3, 0/-2, 0/-1, 0/0}
				{
					\draw (\i,\j) -- (\i+1,\j);
				}
				\foreach \i/\j in {0/-4, 0/-3, 0/-2, 0/-1, 0/0, 1/-3, 1/-2, 1/-1, 1/0}
				{
					\draw (\i,\j) -- (\i,\j-1);
				}
				\foreach \i/\j in {1/-4}
				{
					\draw (\i,\j) -- (\i-1,\j-1);
				}
		\end{tikzpicture}\end{array}&
		\scalebox{0.95}{$\begin{array}{c}
				\theta _1 p_1^4-\frac{6 \theta _1 p_2 p_1^2}{\epsilon _2}+\frac{8 \theta _1 p_3 p_1}{\epsilon _2^2}+\frac{3 \theta _1 p_2^2}{\epsilon _2^2}-\frac{6 \theta _1 p_4}{\epsilon _2^3}-\frac{8 \theta _2 p_3}{\epsilon _2^3}\\
				-\frac{4 \theta _2 p_1^3}{\epsilon _2}+\frac{12 \theta _3 p_1^2}{\epsilon _2^2}+\frac{12 \theta _2 p_2 p_1}{\epsilon _2^2}-\frac{24 \theta _4 p_1}{\epsilon _2^3}-\frac{12 \theta _3 p_2}{\epsilon _2^3}+\frac{24 \theta _5}{\epsilon _2^4}
				
			\end{array}$}&
		{\scriptstyle24 \epsilon _1^4 \left(\epsilon _1-\epsilon _2\right) \left(2 \epsilon _1-\epsilon _2\right) \left(3 \epsilon _1-\epsilon _2\right) \left(4 \epsilon _1-\epsilon _2\right)}&
		{\scriptstyle 5}&
		{\scriptstyle 4}&
		{\scriptstyle 10 \epsilon _1}&
		{\scriptstyle 8 \epsilon _1+2 \epsilon _2}\\
		\cline{2-8}
		&
		\begin{array}{c}
			\begin{tikzpicture}[scale=0.15]
				\foreach \i/\j in {0/-4, 0/-3, 0/-2, 0/-1, 0/0, 1/0}
				{
					\draw (\i,\j) -- (\i+1,\j);
				}
				\foreach \i/\j in {0/-3, 0/-2, 0/-1, 0/0, 1/-3, 1/-2, 1/-1, 1/0}
				{
					\draw (\i,\j) -- (\i,\j-1);
				}
				\foreach \i/\j in {2/0}
				{
					\draw (\i,\j) -- (\i-1,\j-1);
				}
		\end{tikzpicture}\end{array}&
		\scalebox{0.95}{$\begin{array}{c}
				\theta _1 p_1^4-\frac{\theta _2 p_1^3}{\epsilon _1}-\frac{6 \theta _1 p_2 p_1^2}{\epsilon _2}+\frac{8 \theta _1 p_3 p_1}{\epsilon _2^2}+\frac{3 \theta _1 p_2^2}{\epsilon _2^2}-\frac{6 \theta _1 p_4}{\epsilon _2^3}\\
				+\frac{3 \theta _3 p_1^2}{\epsilon _1 \epsilon _2}+\frac{3 \theta _2 p_2 p_1}{\epsilon _1 \epsilon _2}-\frac{6 \theta _4 p_1}{\epsilon _1 \epsilon _2^2}-\frac{2 \theta _2 p_3}{\epsilon _1 \epsilon _2^2}-\frac{3 \theta _3 p_2}{\epsilon _1 \epsilon _2^2}+\frac{6 \theta _5}{\epsilon _1 \epsilon _2^3}
				
			\end{array}$}&
		{\scriptstyle6 \epsilon _1^3 \left(\epsilon _1-\epsilon _2\right) \left(2 \epsilon _1-\epsilon _2\right) \left(3 \epsilon _1-\epsilon _2\right) \epsilon _2 \left(\epsilon _2-4 \epsilon _1\right)}&
		{\scriptstyle 5}&
		{\scriptstyle 4}&
		{\scriptstyle 6 \epsilon _1+\epsilon _2}&
		{\scriptstyle 8 \epsilon _1+2 \epsilon _2}\\
		\cline{2-8}
		&
		\begin{array}{c}
			\begin{tikzpicture}[scale=0.15]
				\foreach \i/\j in {0/-3, 0/-2, 0/-1, 0/0, 1/-1, 1/0}
				{
					\draw (\i,\j) -- (\i+1,\j);
				}
				\foreach \i/\j in {0/-3, 0/-2, 0/-1, 0/0, 1/-2, 1/-1, 1/0, 2/0}
				{
					\draw (\i,\j) -- (\i,\j-1);
				}
				\foreach \i/\j in {1/-3}
				{
					\draw (\i,\j) -- (\i-1,\j-1);
				}
		\end{tikzpicture}\end{array}&
		\scalebox{0.95}{$\begin{array}{c}
				\theta _1 p_1^4-\frac{\theta _1 p_2 p_1^2 \left(3 \epsilon _1+\epsilon _2\right)}{\epsilon _1 \epsilon _2}+\frac{2 \theta _1 p_3 p_1 \left(\epsilon _1+\epsilon _2\right)}{\epsilon _1 \epsilon _2^2}+\frac{\theta _1 p_2^2}{\epsilon _1 \epsilon _2}-\frac{2 \theta _1 p_4}{\epsilon _1 \epsilon _2^2}-\frac{2 \theta _2 p_3}{\epsilon _1 \epsilon _2^2}\\
				-\frac{3 \theta _2 p_1^3}{\epsilon _2}+\frac{\theta _3 p_1^2 \left(6 \epsilon _1+\epsilon _2\right)}{\epsilon _1 \epsilon _2^2}+\frac{\theta _2 p_2 p_1 \left(3 \epsilon _1+2 \epsilon _2\right)}{\epsilon _1 \epsilon _2^2}-\frac{2 \theta _4 p_1 \left(3 \epsilon _1+2 \epsilon _2\right)}{\epsilon _1 \epsilon _2^3}-\frac{3 \theta _3 p_2}{\epsilon _1 \epsilon _2^2}+\frac{6 \theta _5}{\epsilon _1 \epsilon _2^3}
				
			\end{array}$}&
		{\scriptstyle2 \epsilon _1^3 \left(3 \epsilon _1-2 \epsilon _2\right) \left(\epsilon _1-\epsilon _2\right) \left(2 \epsilon _1-\epsilon _2\right) \epsilon _2 \left(\epsilon _2-3 \epsilon _1\right)}&
		{\scriptstyle 5}&
		{\scriptstyle 4}&
		{\scriptstyle 6 \epsilon _1+\epsilon _2}&
		{\scriptstyle 5 \epsilon _1+3 \epsilon _2}\\
		\cline{2-8}
		&
		\begin{array}{c}
			\begin{tikzpicture}[scale=0.15]
				\foreach \i/\j in {0/-3, 0/-2, 0/-1, 0/0, 1/-1, 1/0}
				{
					\draw (\i,\j) -- (\i+1,\j);
				}
				\foreach \i/\j in {0/-2, 0/-1, 0/0, 1/-2, 1/-1, 1/0, 2/0}
				{
					\draw (\i,\j) -- (\i,\j-1);
				}
				\foreach \i/\j in {2/-1}
				{
					\draw (\i,\j) -- (\i-1,\j-1);
				}
		\end{tikzpicture}\end{array}&
		\scalebox{0.95}{$\begin{array}{c}
				\theta _1 p_1^4-\frac{\theta _2 p_1^3 \left(\epsilon _1+\epsilon _2\right)}{\epsilon _1 \epsilon _2}-\frac{\theta _1 p_2 p_1^2 \left(3 \epsilon _1+\epsilon _2\right)}{\epsilon _1 \epsilon _2}+\frac{2 \theta _1 p_3 p_1 \left(\epsilon _1+\epsilon _2\right)}{\epsilon _1 \epsilon _2^2}+\frac{\theta _1 p_2^2}{\epsilon _1 \epsilon _2}-\frac{2 \theta _1 p_4}{\epsilon _1 \epsilon _2^2}\\
				+\frac{4 \theta _3 p_1^2}{\epsilon _1 \epsilon _2}+\frac{\theta _2 p_2 p_1 \left(3 \epsilon _1^2+\epsilon _2^2\right)}{\epsilon _1^2 \epsilon _2^2}-\frac{\theta _4 p_1 \left(5 \epsilon _1+\epsilon _2\right)}{\epsilon _1^2 \epsilon _2^2}-\frac{\theta _2 p_3 \left(2 \epsilon _1^2-\epsilon _2 \epsilon _1+\epsilon _2^2\right)}{\epsilon _1^2 \epsilon _2^3}-\frac{\theta _3 p_2 \left(\epsilon _1+\epsilon _2\right)}{\epsilon _1^2 \epsilon _2^2}+\frac{2 \theta _5 \left(\epsilon _1+\epsilon _2\right)}{\epsilon _1^2 \epsilon _2^3}
				
			\end{array}$}&
		{\scriptstyle-\epsilon _1^2 \left(2 \epsilon _1-2 \epsilon _2\right) \left(\epsilon _1-\epsilon _2\right){}^2 \epsilon _2 \left(\epsilon _2-3 \epsilon _1\right) \left(\epsilon _2-2 \epsilon _1\right)}&
		{\scriptstyle 5}&
		{\scriptstyle 4}&
		{\scriptstyle 4 \epsilon _1+2 \epsilon _2}&
		{\scriptstyle 5 \epsilon _1+3 \epsilon _2}\\
		\cline{2-8}
		&
		\begin{array}{c}
			\begin{tikzpicture}[scale=0.15]
				\foreach \i/\j in {0/-3, 0/-2, 0/-1, 0/0, 1/-1, 1/0, 2/0}
				{
					\draw (\i,\j) -- (\i+1,\j);
				}
				\foreach \i/\j in {0/-2, 0/-1, 0/0, 1/-2, 1/-1, 1/0, 2/0}
				{
					\draw (\i,\j) -- (\i,\j-1);
				}
				\foreach \i/\j in {3/0}
				{
					\draw (\i,\j) -- (\i-1,\j-1);
				}
		\end{tikzpicture}\end{array}&
		\scalebox{0.95}{$\begin{array}{c}
				\theta _1 p_1^4-\frac{2 \theta _2 p_1^3}{\epsilon _1}-\frac{\theta _1 p_2 p_1^2 \left(3 \epsilon _1+\epsilon _2\right)}{\epsilon _1 \epsilon _2}+\frac{2 \theta _1 p_3 p_1 \left(\epsilon _1+\epsilon _2\right)}{\epsilon _1 \epsilon _2^2}+\frac{\theta _1 p_2^2}{\epsilon _1 \epsilon _2}-\frac{2 \theta _1 p_4}{\epsilon _1 \epsilon _2^2}\\
				+\frac{2 \theta _3 p_1^2 \left(\epsilon _1+\epsilon _2\right)}{\epsilon _1^2 \epsilon _2}+\frac{4 \theta _2 p_2 p_1}{\epsilon _1 \epsilon _2}-\frac{2 \theta _4 p_1 \left(\epsilon _1+2 \epsilon _2\right)}{\epsilon _1^2 \epsilon _2^2}-\frac{2 \theta _2 p_3}{\epsilon _1 \epsilon _2^2}-\frac{2 \theta _3 p_2}{\epsilon _1^2 \epsilon _2}+\frac{4 \theta _5}{\epsilon _1^2 \epsilon _2^2}
				
			\end{array}$}&
		{\scriptstyle2 \epsilon _1^2 \left(2 \epsilon _1-2 \epsilon _2\right) \left(\epsilon _1-\epsilon _2\right) \epsilon _2^2 \left(\epsilon _2-\epsilon _1\right) \left(2 \epsilon _2-3 \epsilon _1\right)}&
		{\scriptstyle 5}&
		{\scriptstyle 4}&
		{\scriptstyle 3 \epsilon _1+3 \epsilon _2}&
		{\scriptstyle 5 \epsilon _1+3 \epsilon _2}\\
		\cline{2-8}
		&
		\begin{array}{c}
			\begin{tikzpicture}[scale=0.15]
				\foreach \i/\j in {0/-2, 0/-1, 0/0, 1/-2, 1/-1, 1/0}
				{
					\draw (\i,\j) -- (\i+1,\j);
				}
				\foreach \i/\j in {0/-2, 0/-1, 0/0, 1/-1, 1/0, 2/-1, 2/0}
				{
					\draw (\i,\j) -- (\i,\j-1);
				}
				\foreach \i/\j in {1/-2}
				{
					\draw (\i,\j) -- (\i-1,\j-1);
				}
		\end{tikzpicture}\end{array}&
		\scalebox{0.95}{$\begin{array}{c}
				\theta _1 p_1^4-\frac{2 \theta _1 p_2 p_1^2 \left(\epsilon _1+\epsilon _2\right)}{\epsilon _1 \epsilon _2}+\frac{4 \theta _1 p_3 p_1}{\epsilon _1 \epsilon _2}-\frac{\theta _1 p_4 \left(\epsilon _1+\epsilon _2\right)}{\epsilon _1^2 \epsilon _2^2}+\frac{\theta _1 p_2^2 \left(\epsilon _1^2-\epsilon _2 \epsilon _1+\epsilon _2^2\right)}{\epsilon _1^2 \epsilon _2^2}-\frac{2 \theta _2 p_3}{\epsilon _1 \epsilon _2^2}\\
				-\frac{2 \theta _2 p_1^3}{\epsilon _2}+\frac{2 \theta _3 p_1^2 \left(\epsilon _1+\epsilon _2\right)}{\epsilon _1 \epsilon _2^2}+\frac{2 \theta _2 p_2 p_1 \left(\epsilon _1+\epsilon _2\right)}{\epsilon _1 \epsilon _2^2}-\frac{6 \theta _4 p_1}{\epsilon _1 \epsilon _2^2}-\frac{2 \theta _3 p_2 \left(\epsilon _1^2-\epsilon _2 \epsilon _1+\epsilon _2^2\right)}{\epsilon _1^2 \epsilon _2^3}+\frac{2 \theta _5 \left(\epsilon _1+\epsilon _2\right)}{\epsilon _1^2 \epsilon _2^3}
				
			\end{array}$}&
		{\scriptstyle-2 \epsilon _1^2 \left(\epsilon _1-2 \epsilon _2\right) \left(2 \epsilon _1-2 \epsilon _2\right) \left(\epsilon _1-\epsilon _2\right) \epsilon _2 \left(\epsilon _2-2 \epsilon _1\right) \left(\epsilon _2-\epsilon _1\right)}&
		{\scriptstyle 5}&
		{\scriptstyle 4}&
		{\scriptstyle 4 \epsilon _1+2 \epsilon _2}&
		{\scriptstyle 4 \epsilon _1+4 \epsilon _2}\\
		\cline{2-8}
		&
		\begin{array}{c}
			\begin{tikzpicture}[scale=0.15]
				\foreach \i/\j in {0/-2, 0/-1, 0/0, 1/-1, 1/0, 2/0}
				{
					\draw (\i,\j) -- (\i+1,\j);
				}
				\foreach \i/\j in {0/-2, 0/-1, 0/0, 1/-1, 1/0, 2/0}
				{
					\draw (\i,\j) -- (\i,\j-1);
				}
				\foreach \i/\j in {1/-2, 2/-1, 3/0}
				{
					\draw (\i,\j) -- (\i-1,\j-1);
				}
		\end{tikzpicture}\end{array}&
		\scalebox{0.95}{$\begin{array}{c}
				\frac{\theta _1 \theta _2 \theta _3}{\epsilon _1^3 \epsilon _2^3}
				
			\end{array}$}&
		{\scriptstyle1}&
		{\scriptstyle 6}&
		{\scriptstyle 3}&
		{\scriptstyle 4 \epsilon _1+4 \epsilon _2}&
		{\scriptstyle \frac{5 \epsilon _1}{2}+\frac{5 \epsilon _2}{2}}\\
		\cline{2-8}
		&
		\begin{array}{c}
			\begin{tikzpicture}[scale=0.15]
				\foreach \i/\j in {0/-2, 0/-1, 0/0, 1/-1, 1/0, 2/-1, 2/0}
				{
					\draw (\i,\j) -- (\i+1,\j);
				}
				\foreach \i/\j in {0/-2, 0/-1, 0/0, 1/-1, 1/0, 2/0, 3/0}
				{
					\draw (\i,\j) -- (\i,\j-1);
				}
				\foreach \i/\j in {1/-2}
				{
					\draw (\i,\j) -- (\i-1,\j-1);
				}
		\end{tikzpicture}\end{array}&
		\scalebox{0.95}{$\begin{array}{c}
				\theta _1 p_1^4-\frac{2 \theta _2 p_1^3}{\epsilon _2}-\frac{\theta _1 p_2 p_1^2 \left(\epsilon _1+3 \epsilon _2\right)}{\epsilon _1 \epsilon _2}+\frac{2 \theta _1 p_3 p_1 \left(\epsilon _1+\epsilon _2\right)}{\epsilon _1^2 \epsilon _2}+\frac{\theta _1 p_2^2}{\epsilon _1 \epsilon _2}-\frac{2 \theta _1 p_4}{\epsilon _1^2 \epsilon _2}\\
				+\frac{2 \theta _3 p_1^2 \left(\epsilon _1+\epsilon _2\right)}{\epsilon _1 \epsilon _2^2}+\frac{4 \theta _2 p_2 p_1}{\epsilon _1 \epsilon _2}-\frac{2 \theta _4 p_1 \left(2 \epsilon _1+\epsilon _2\right)}{\epsilon _1^2 \epsilon _2^2}-\frac{2 \theta _2 p_3}{\epsilon _1^2 \epsilon _2}-\frac{2 \theta _3 p_2}{\epsilon _1 \epsilon _2^2}+\frac{4 \theta _5}{\epsilon _1^2 \epsilon _2^2}
				
			\end{array}$}&
		{\scriptstyle2 \epsilon _1^2 \left(2 \epsilon _1-3 \epsilon _2\right) \left(\epsilon _1-\epsilon _2\right) \epsilon _2^2 \left(\epsilon _2-\epsilon _1\right) \left(2 \epsilon _2-2 \epsilon _1\right)}&
		{\scriptstyle 5}&
		{\scriptstyle 4}&
		{\scriptstyle 3 \epsilon _1+3 \epsilon _2}&
		{\scriptstyle 3 \epsilon _1+5 \epsilon _2}\\
		\cline{2-8}
		&
		\begin{array}{c}
			\begin{tikzpicture}[scale=0.15]
				\foreach \i/\j in {0/-2, 0/-1, 0/0, 1/-2, 1/-1, 1/0, 2/0}
				{
					\draw (\i,\j) -- (\i+1,\j);
				}
				\foreach \i/\j in {0/-1, 0/0, 1/-1, 1/0, 2/-1, 2/0}
				{
					\draw (\i,\j) -- (\i,\j-1);
				}
				\foreach \i/\j in {3/0}
				{
					\draw (\i,\j) -- (\i-1,\j-1);
				}
		\end{tikzpicture}\end{array}&
		\scalebox{0.95}{$\begin{array}{c}
				\theta _1 p_1^4-\frac{2 \theta _2 p_1^3}{\epsilon _1}-\frac{2 \theta _1 p_2 p_1^2 \left(\epsilon _1+\epsilon _2\right)}{\epsilon _1 \epsilon _2}+\frac{4 \theta _1 p_3 p_1}{\epsilon _1 \epsilon _2}-\frac{\theta _1 p_4 \left(\epsilon _1+\epsilon _2\right)}{\epsilon _1^2 \epsilon _2^2}+\frac{\theta _1 p_2^2 \left(\epsilon _1^2-\epsilon _2 \epsilon _1+\epsilon _2^2\right)}{\epsilon _1^2 \epsilon _2^2}\\
				+\frac{2 \theta _3 p_1^2 \left(\epsilon _1+\epsilon _2\right)}{\epsilon _1^2 \epsilon _2}+\frac{2 \theta _2 p_2 p_1 \left(\epsilon _1+\epsilon _2\right)}{\epsilon _1^2 \epsilon _2}-\frac{6 \theta _4 p_1}{\epsilon _1^2 \epsilon _2}-\frac{2 \theta _2 p_3}{\epsilon _1^2 \epsilon _2}-\frac{2 \theta _3 p_2 \left(\epsilon _1^2-\epsilon _2 \epsilon _1+\epsilon _2^2\right)}{\epsilon _1^3 \epsilon _2^2}+\frac{2 \theta _5 \left(\epsilon _1+\epsilon _2\right)}{\epsilon _1^3 \epsilon _2^2}
				
			\end{array}$}&
		{\scriptstyle-2 \epsilon _1 \left(\epsilon _1-2 \epsilon _2\right) \left(\epsilon _1-\epsilon _2\right) \epsilon _2^2 \left(\epsilon _2-2 \epsilon _1\right) \left(\epsilon _2-\epsilon _1\right) \left(2 \epsilon _2-2 \epsilon _1\right)}&
		{\scriptstyle 5}&
		{\scriptstyle 4}&
		{\scriptstyle 2 \epsilon _1+4 \epsilon _2}&
		{\scriptstyle 4 \epsilon _1+4 \epsilon _2}\\
		\cline{2-8}
		&
		\begin{array}{c}
			\begin{tikzpicture}[scale=0.15]
				\foreach \i/\j in {0/-2, 0/-1, 0/0, 1/-1, 1/0, 2/-1, 2/0}
				{
					\draw (\i,\j) -- (\i+1,\j);
				}
				\foreach \i/\j in {0/-1, 0/0, 1/-1, 1/0, 2/0, 3/0}
				{
					\draw (\i,\j) -- (\i,\j-1);
				}
				\foreach \i/\j in {2/-1}
				{
					\draw (\i,\j) -- (\i-1,\j-1);
				}
		\end{tikzpicture}\end{array}&
		\scalebox{0.95}{$\begin{array}{c}
				\theta _1 p_1^4-\frac{\theta _2 p_1^3 \left(\epsilon _1+\epsilon _2\right)}{\epsilon _1 \epsilon _2}-\frac{\theta _1 p_2 p_1^2 \left(\epsilon _1+3 \epsilon _2\right)}{\epsilon _1 \epsilon _2}+\frac{2 \theta _1 p_3 p_1 \left(\epsilon _1+\epsilon _2\right)}{\epsilon _1^2 \epsilon _2}+\frac{\theta _1 p_2^2}{\epsilon _1 \epsilon _2}-\frac{2 \theta _1 p_4}{\epsilon _1^2 \epsilon _2}\\
				+\frac{4 \theta _3 p_1^2}{\epsilon _1 \epsilon _2}+\frac{\theta _2 p_2 p_1 \left(\epsilon _1^2+3 \epsilon _2^2\right)}{\epsilon _1^2 \epsilon _2^2}-\frac{\theta _4 p_1 \left(\epsilon _1+5 \epsilon _2\right)}{\epsilon _1^2 \epsilon _2^2}-\frac{\theta _2 p_3 \left(\epsilon _1^2-\epsilon _2 \epsilon _1+2 \epsilon _2^2\right)}{\epsilon _1^3 \epsilon _2^2}-\frac{\theta _3 p_2 \left(\epsilon _1+\epsilon _2\right)}{\epsilon _1^2 \epsilon _2^2}+\frac{2 \theta _5 \left(\epsilon _1+\epsilon _2\right)}{\epsilon _1^3 \epsilon _2^2}
				
			\end{array}$}&
		{\scriptstyle-\epsilon _1 \left(\epsilon _1-3 \epsilon _2\right) \left(\epsilon _1-2 \epsilon _2\right) \epsilon _2^2 \left(\epsilon _2-\epsilon _1\right){}^2 \left(2 \epsilon _2-2 \epsilon _1\right)}&
		{\scriptstyle 5}&
		{\scriptstyle 4}&
		{\scriptstyle 2 \epsilon _1+4 \epsilon _2}&
		{\scriptstyle 3 \epsilon _1+5 \epsilon _2}\\
		\cline{2-8}
		&
		\begin{array}{c}
			\begin{tikzpicture}[scale=0.15]
				\foreach \i/\j in {0/-2, 0/-1, 0/0, 1/-1, 1/0, 2/-1, 2/0, 3/0}
				{
					\draw (\i,\j) -- (\i+1,\j);
				}
				\foreach \i/\j in {0/-1, 0/0, 1/-1, 1/0, 2/0, 3/0}
				{
					\draw (\i,\j) -- (\i,\j-1);
				}
				\foreach \i/\j in {4/0}
				{
					\draw (\i,\j) -- (\i-1,\j-1);
				}
		\end{tikzpicture}\end{array}&
		\scalebox{0.95}{$\begin{array}{c}
				\theta _1 p_1^4-\frac{3 \theta _2 p_1^3}{\epsilon _1}-\frac{\theta _1 p_2 p_1^2 \left(\epsilon _1+3 \epsilon _2\right)}{\epsilon _1 \epsilon _2}+\frac{2 \theta _1 p_3 p_1 \left(\epsilon _1+\epsilon _2\right)}{\epsilon _1^2 \epsilon _2}+\frac{\theta _1 p_2^2}{\epsilon _1 \epsilon _2}-\frac{2 \theta _1 p_4}{\epsilon _1^2 \epsilon _2}\\
				+\frac{\theta _3 p_1^2 \left(\epsilon _1+6 \epsilon _2\right)}{\epsilon _1^2 \epsilon _2}+\frac{\theta _2 p_2 p_1 \left(2 \epsilon _1+3 \epsilon _2\right)}{\epsilon _1^2 \epsilon _2}-\frac{2 \theta _4 p_1 \left(2 \epsilon _1+3 \epsilon _2\right)}{\epsilon _1^3 \epsilon _2}-\frac{2 \theta _2 p_3}{\epsilon _1^2 \epsilon _2}-\frac{3 \theta _3 p_2}{\epsilon _1^2 \epsilon _2}+\frac{6 \theta _5}{\epsilon _1^3 \epsilon _2}
				
			\end{array}$}&
		{\scriptstyle2 \epsilon _1 \left(\epsilon _1-3 \epsilon _2\right) \epsilon _2^3 \left(\epsilon _2-\epsilon _1\right) \left(2 \epsilon _2-\epsilon _1\right) \left(3 \epsilon _2-2 \epsilon _1\right)}&
		{\scriptstyle 5}&
		{\scriptstyle 4}&
		{\scriptstyle \epsilon _1+6 \epsilon _2}&
		{\scriptstyle 3 \epsilon _1+5 \epsilon _2}\\
		\cline{2-8}
		&
		\begin{array}{c}
			\begin{tikzpicture}[scale=0.15]
				\foreach \i/\j in {0/-1, 0/0, 1/-1, 1/0, 2/-1, 2/0, 3/-1, 3/0}
				{
					\draw (\i,\j) -- (\i+1,\j);
				}
				\foreach \i/\j in {0/-1, 0/0, 1/0, 2/0, 3/0, 4/0}
				{
					\draw (\i,\j) -- (\i,\j-1);
				}
				\foreach \i/\j in {1/-1}
				{
					\draw (\i,\j) -- (\i-1,\j-1);
				}
		\end{tikzpicture}\end{array}&
		\scalebox{0.95}{$\begin{array}{c}
				\theta _1 p_1^4-\frac{\theta _2 p_1^3}{\epsilon _2}-\frac{6 \theta _1 p_2 p_1^2}{\epsilon _1}+\frac{8 \theta _1 p_3 p_1}{\epsilon _1^2}+\frac{3 \theta _1 p_2^2}{\epsilon _1^2}-\frac{6 \theta _1 p_4}{\epsilon _1^3}\\
				+\frac{3 \theta _3 p_1^2}{\epsilon _1 \epsilon _2}+\frac{3 \theta _2 p_2 p_1}{\epsilon _1 \epsilon _2}-\frac{6 \theta _4 p_1}{\epsilon _1^2 \epsilon _2}-\frac{2 \theta _2 p_3}{\epsilon _1^2 \epsilon _2}-\frac{3 \theta _3 p_2}{\epsilon _1^2 \epsilon _2}+\frac{6 \theta _5}{\epsilon _1^3 \epsilon _2}
				
			\end{array}$}&
		{\scriptstyle6 \epsilon _1 \left(\epsilon _1-4 \epsilon _2\right) \epsilon _2^3 \left(\epsilon _2-\epsilon _1\right) \left(2 \epsilon _2-\epsilon _1\right) \left(3 \epsilon _2-\epsilon _1\right)}&
		{\scriptstyle 5}&
		{\scriptstyle 4}&
		{\scriptstyle \epsilon _1+6 \epsilon _2}&
		{\scriptstyle 2 \epsilon _1+8 \epsilon _2}\\
		\cline{2-8}
		&
		\begin{array}{c}
			\begin{tikzpicture}[scale=0.15]
				\foreach \i/\j in {0/-1, 0/0, 1/-1, 1/0, 2/-1, 2/0, 3/-1, 3/0, 4/0}
				{
					\draw (\i,\j) -- (\i+1,\j);
				}
				\foreach \i/\j in {0/0, 1/0, 2/0, 3/0, 4/0}
				{
					\draw (\i,\j) -- (\i,\j-1);
				}
				\foreach \i/\j in {5/0}
				{
					\draw (\i,\j) -- (\i-1,\j-1);
				}
		\end{tikzpicture}\end{array}&
		\scalebox{0.95}{$\begin{array}{c}
				\theta _1 p_1^4-\frac{6 \theta _1 p_2 p_1^2}{\epsilon _1}+\frac{8 \theta _1 p_3 p_1}{\epsilon _1^2}+\frac{3 \theta _1 p_2^2}{\epsilon _1^2}-\frac{6 \theta _1 p_4}{\epsilon _1^3}-\frac{8 \theta _2 p_3}{\epsilon _1^3}\\
				-\frac{4 \theta _2 p_1^3}{\epsilon _1}+\frac{12 \theta _3 p_1^2}{\epsilon _1^2}+\frac{12 \theta _2 p_2 p_1}{\epsilon _1^2}-\frac{24 \theta _4 p_1}{\epsilon _1^3}-\frac{12 \theta _3 p_2}{\epsilon _1^3}+\frac{24 \theta _5}{\epsilon _1^4}
				
			\end{array}$}&
		{\scriptstyle24 \epsilon _2^4 \left(\epsilon _2-\epsilon _1\right) \left(2 \epsilon _2-\epsilon _1\right) \left(3 \epsilon _2-\epsilon _1\right) \left(4 \epsilon _2-\epsilon _1\right)}&
		{\scriptstyle 5}&
		{\scriptstyle 4}&
		{\scriptstyle 10 \epsilon _2}&
		{\scriptstyle 2 \epsilon _1+8 \epsilon _2}\\
		\hline
	\end{longtable}
	\endgroup

\end{landscape}

\end{document}